\documentclass[aps, amsmath, amssymb, twocolumn, 10pt, 
superscriptaddress, longbibliography, nofootinbib]{revtex4-2}

\usepackage{textcomp}
\usepackage{amsmath}
\usepackage{amsfonts}
\usepackage{calc}
\usepackage{mathrsfs}
\usepackage{array}
\usepackage{xcolor}
\usepackage{bm}
\usepackage{graphicx}
\usepackage{hyperref}
\usepackage{tensor}
\usepackage{booktabs}
\usepackage{mathtools}
\usepackage[caption=false]{subfig}
\usepackage{float}
\usepackage{orcidlink}
\usepackage[normalem]{ulem}
\usepackage{placeins}

\graphicspath{{Figures/}}

\renewcommand{\P}{\mathcal{P}}
\newcommand{\res}{\mathcal{R}}

\newcommand{\diff}[2]  {\frac{d #1}{d #2}}
\newcommand{\sdiff}[2]  {\frac{d^2 #1}{d #2^2}}

\newcommand{\psdiff}[2]  {\frac{\partial^2 #1}{\partial #2^2}}
\newcommand{\eqn}[1] {Eq.~\eqref{eq:#1}}
\newcommand{\fig}[1] {Fig.~\ref{fig:#1}}

\newcommand{\secref}[1] {Sec.~\ref{sec:#1}}

\newcommand{\nn}{\nonumber}

\renewcommand{\l}{\left(}
\renewcommand{\r}{\right)}

\newcommand{\lmn}  {\ell m n}
\newcommand{\lm}  {\ell m}
\newcommand{\fdbox}  {\Box_{\lmn}}
\newcommand{\retfield}  {\psi_{\lmn}}
\newcommand{\fdretfield}  {\psi_{\lmn}}
\newcommand{\fdretfieldplus}  {\psi^+_{\lmn}}
\newcommand{\fdregfield}  {\psi^{\mathcal{R}}_{\lmn}}
\newcommand{\fdregfieldplus}  {\psi^{\mathcal{R},+}_{\lmn}}

\newcommand{\tdresfield} {\psi^{\mathcal{R}}_{\lm}}
\newcommand{\tdresfieldEHS} {\psi^{\mathcal{R},{\rm EHS}}_{\lm}}
\newcommand{\tdresfieldplus} {\psi^{\mathcal{R},+}_{\lm}}
\newcommand{\tdresfieldpm} {\psi^{\mathcal{R},\pm}_{\lm}}

\newcommand{\tdresfieldminus} {\psi^{\mathcal{R},-}_{\lm}}

\newcommand{\tdsingfield}  {\psi_{\lm}^{\P}}
\newcommand{\tdsingfieldplus}  {\psi_{\lm}^{\P,+}}
\newcommand{\tdsingfieldminus}  {\psi_{\lm}^{\P,-}}
\newcommand{\tdsingfieldpm}  {\psi_{\lm}^{\P,\pm}}

\newcommand{\fdsingfield}  {\psi^{\mathcal{P}}_{\lmn}}
\newcommand{\fdsingfieldplus}  {\psi^{\mathcal{P},+}_{\lmn}}

\newcommand{\fdinhfield}  {\psi^{\text{inh}}_{\lmn}}
\newcommand{\fdinhfieldplus}  {\psi^{\text{inh},+}_{\lmn}}

\newcommand{\fdregfieldminus}  {\psi^{\mathcal{R},-}_{\lmn}}
\newcommand{\fdretfieldminus}  {\psi^-_{\lmn}}
\newcommand{\fdsingfieldminus}  {\psi^{\mathcal{P},-}_{\lmn}}

\newcommand{\fdregfieldpm}  {\psi^{\mathcal{R},\pm}_{\lmn}}

\newcommand{\fdsingfieldpm}  {\psi^{\mathcal{P},\pm}_{\lmn}}

\newcommand{\fdJ}  {J_{\lmn}}

\newcommand{\Seff}  {S^{\text{eff}}_{\lm}}
\newcommand{\fdSeff}  {S^{\text{eff}}_{\lmn}}
\newcommand{\tdSeff}  {S^{\text{eff}}_{\lm}}
\newcommand{\fdSeffplus}  {S^{\text{eff}, +}_{\lmn}}
\newcommand{\fdSeffminus}  {S^{\text{eff}, -}_{\lmn}}
\newcommand{\fdSeffpm}  {S^{\text{eff}, \pm}_{\lmn}}
\newcommand{\tdSeffplus}  {S^{\text{eff}, +}_{\lm}}
\newcommand{\tdSeffminus}  {S^{\text{eff}, -}_{\lm}}
\newcommand{\tdSeffpm}  {S^{\text{eff}, \pm}_{\lm}}
\newcommand{\omn} {\omega_{m n}}
\newcommand{\f} {f(r)}
\newcommand{\rmin} {r_{\text{min}}}
\newcommand{\rmax} {r_{\text{max}}}
\newcommand{\rin} {r_{\text{in}}}
\newcommand{\rout} {r_{\text{out}}}

\newcommand{\fdh}  { \psi^{\infty/h}_{\lmn}}
\newcommand{\fdhplus}  { \psi^{\infty}_{\lmn}}
\newcommand{\fdhminus}  { \psi^{h}_{\lmn}}
\newcommand{\T} {T_{r}}
\newcommand{\nsum} {\sum^{\infty}_{n = -\infty}}
\newcommand{\etplus} {e^{i \omn t}}
\newcommand{\etminus} {e^{-i \omn t}}
\newcommand{\etortplus} {e^{i \omn r_{*}}}
\newcommand{\etortminus} {e^{-i \omn r_{*}}}
\newcommand{\rp} {r_{p}(t)}
\newcommand{\tp} {t_{p}}
\newcommand{\phip} {\varphi_{p}}

\newcommand{\eesplus} {\psi^{{\cal R},+}_{\lmn}}
\newcommand{\eesminus} {\psi^{{\cal R},-}_{\lmn}}

\newcommand{\rstar} {r_{*}}

\DeclarePairedDelimiterX\braket[1]{\langle}{\rangle}{#1}

\newcommand{\beq}{\begin{equation}}
\newcommand{\eeq}{\end{equation}}
\newcommand{\bec}{\begin{cases}}
\newcommand{\eec}{\end{cases}}

\definecolor{colour1}{HTML}{0571b0} 
\definecolor{colour2}{HTML}{92c5de} 
\definecolor{colour3}{HTML}{f4a582} 
\definecolor{colour4}{HTML}{ca0020} 
\definecolor{colour5}{HTML}{fe4a49} 

\hypersetup{colorlinks=true, linkcolor=colour1, citecolor=colour1,
filecolor=colour1, urlcolor=colour1}

\begin{document}
\title{Applying the effective-source approach to frequency-domain self-force
calculations for eccentric orbits}

\author{Benjamin Leather\,\orcidlink{0000-0001-6186-7271},} 
\affiliation{Max Planck Institute for Gravitational Physics (Albert Einstein
Institute), Am M\"{u}hlenberg 1, 14476 Potsdam, Germany}
\affiliation{School of Mathematics and Statistics, University College Dublin,
Belfield, Dublin 4, Ireland}
\author{Niels Warburton\,\orcidlink{0000-0003-0914-8645},}
\affiliation{School of Mathematics and Statistics, University College Dublin,
Belfield, Dublin 4, Ireland}
\date{\today}


\begin{abstract}
Extreme mass-ratio inspirals (EMRIs) are expected to have considerable eccentricity when emitting gravitational waves (GWs) in the LISA band.
Developing GW templates that remain phase accurate over these long inspirals requires the use of second-order self-force theory and practical second-order self-force calculations are now emerging for quasi-circular EMRIs.
These calculations rely on effective-source regularization techniques in the frequency domain that presently are specialized to circular orbits.
Here we make a first step towards more generic second-order calculations by extending the frequency domain effective-source approach to eccentric orbits.
In order to overcome the slow convergence of the Fourier sum over radial modes, we develop a new \textit{extended effective-sources} approach which builds upon the method of extended particular solutions.
To demonstrate our new computational technique we apply it a toy scalar-field problem which is conceptually similar to the gravitational case.

\end{abstract}

\maketitle

\section{Introduction}
\label{sec:introduction}

Direct observations of gravitational waves from binary black hole mergers \cite{LIGOScientific:2016vbw, LIGOScientific:2016sjg, LIGOScientific:2017bnn, LIGOScientific:2017vox, LIGOScientific:2017ycc} and more recently binary neutron star mergers \cite{LIGOScientific:2017vwq, LIGOScientific:2020aai, LIGOScientific:2021qlt} signified a turning point in astronomy.  
Gravitational waves are now an observable science rather than interesting theoretical concepts. 

The increasing maturity of space-based gravitational wave detector designs has motivated much recent work to calculate gravitational wave emission from extreme-mass-ratio inspirals (EMRIs). 
There is also considerable theoretical interest in the problem of motion of a point mass in a background geometry in general relativity, influenced by its own self-force.

An EMRI system consists of a compact object of mass $\mu \sim 1-100\,M_{\odot}$ (e.g., a neutron star or black hole) moving on a decaying orbit about, and ultimately into, a massive black hole of mass $M \sim 10^{5}-10^{7}\,M_{\odot}$. 
During their inspiral EMRIs evolve adiabatically due to the back-reaction from the gravitational perturbation sourced by the smaller body.
Each EMRI emits tens to hundreds of thousands of gravitational wave cycles in the millihertz frequency range of the LISA mission \cite{Amaro-Seoane:2014ela, Barack:2018yly} as the smaller body spends months to years orbiting in the strong-field of the massive black hole.
During this time, the associated waveform will encode detailed information of the surrounding spacetime geometry.
Detection of EMRI waveforms will therefore enable high-precision measurements of the central object's mass and spin, test the Kerr hypothesis, and allow tests of proposed alternate theories of gravity \cite{Barack:2006pq}.

The inherent problem with detecting and characterising EMRI waveforms is that the instantaneous signal-to-noise-ratio (SNR) is very small for the vast majority of signals \cite{Gair:2004iv}.
Therefore if one is to extract EMRI signals through matched filtering techniques we require accurate theoretical waveform templates.

The natural treatment of the EMRI problem is through the \emph{gravitational self-force} (GSF) approach.  
Here the smaller, compact body sources a perturbation to the metric of the larger black hole.  
The perturbation is expressed a power series expansion in the small mass ratio, $\epsilon := \mu/M$, such that, at leading order, the small body moves along a geodesic of the background spacetime.  
At subsequent sub-leading orders the back-reaction due to the metric perturbation from the smaller body accelerates the body away from geodesic motion in the background spacetime.

Since SF theory is based on a perturbative expansion the accuracy of the result depends upon the order to which the expansion is carried out. 
Producing waveform models that accurately track to the waveform phase to within a fraction of radian requires carrying out this expansion through $\epsilon^2$ \cite{Hinderer:2008dm}.  
These calculations also have to be carried out over a large parameter space for generic binaries that are both highly eccentric, and precessing due to the spins on their component masses \cite{LISA:2022yao}.

First-order calculations have reached this goal and the first-order GSF can be calculated for an object on a generic orbit around a spinning (Kerr) primary \cite{vandeMeent:2017bcc}.
Recently, corrections for generic orbits due to the spin of secondary were also computed \cite{Skoupy:2023lih}.  
A range of works have also calculated the associated inspirals around a Schwarzschild black hole \cite{Warburton:2011fk, Osburn:2015duj, Warburton:2017sxk, VanDeMeent:2018cgn} or Kerr black hole
\cite{Fujita:2020zxe,Lynch:2021ogr,Hughes:2021exa,Lynch:2023gpu}.

Calculations at second-order in the mass ratio are now emerging for quantities such the binding energy \cite{Pound:2019lzj}, the gravitational wave flux \cite{Warburton:2021kwk}, and recently the waveform \cite{Wardell:2021fyy}.
Despite being a perturbative expansion in the mass ratio, the latter two have shown remarkable agreement with numerical relativity simulations, even for near-comparable mass binaries with $\epsilon = 1/10$.
This suggests that GSF results can be used to model intermediate-mass-ratio inspirals (IMRIs) \cite{Warburton:2021kwk}.
This is notable in light of the recent GW observations of binaries with $\epsilon\sim1/30$ \cite{LIGOScientific:2020stg, LIGOScientific:2020zkf}.
So far all second-order calculations have focused on the quasi-circular case.
For EMRIs there is a strong motivation to push these calculations to eccentric orbits \cite{Hopman:2005vr} and this requires the development of new calculational techniques.
The focus of this paper is to develop such a computational approach to second-order perturbations for a compact object moving on an eccentric orbit.

\subsection{Regularization and the effective-source approach}
The point-particle model of EMRIs within SF-theory leads to distributional sources and their associated singular fields.
Consequently, an essential component of SF computations is regularization of the retarded metric perturbation \cite{Mino:1996nk,Quinn:1996am,Gralla:2008fg, Pound:2009sm}.

The first, and most commonly used technique for first-order calculations, is \emph{mode-sum regularization}.
Introduced in \cite{Barack:1999wf}, the technique is based on the observation that a spherical-harmonic decomposition of the first-order metric perturbation yields individual multipole modes that are finite on the worldline.
Thus it is possible to subtract the singular contribution to retarded field on a mode-by-mode basis. 
This procedure then gives the regular contribution to the field from which one can compute the self-force which drives the inspiral.
To date, second-order calculations have been carried out in the Lorenz gauge \cite{Miller:2020bft,Pound:2019lzj} and here the mode-sum prescription is not tractable since the individual multipole modes of the retarded metric perturbation are logarithmically divergent at the location of the particle \cite{Pound:2014xva}.  
There has been some development in formulating a \emph{highly regular gauge} that would render the multipole modes of the retarded field finite and allow a mode-sum procedure to be implemented at second-order \cite{Upton:2021oxf}, but this has not yet been used for practical calculations.  
In this work we shall thus focus an alternative, effective-source treatment.

The effective-source method was pioneered in Refs.~\cite{Barack:2007jh} and \cite{Vega:2007mc}.  
It was designed for situations wherein the modes of the retarded field would diverge at the worldline, meaning it is crucial to second-order formalism as well as 2+1D and 3+1D calculations.  
In this method one splits the retarded field into its singular and regular contributions and reformulates the field equations to solve directly for the residual field using an (effective) regular source term.  
At first order, effective-source schemes have been implemented in time domain (TD) \cite{Vega:2007mc, Dolan:2011dx, Dolan:2012jg, Diener:2011cc} and frequency domain (FD) calculations for quasi-circular orbits \cite{Warburton:2013lea,Wardell:2015ada}.  
Moreover, the approach is the workhorse of current second-order calculations.

\subsection{The challenge of eccentric orbits}
Extending the second-order calculation to eccentric orbits presents numerous theoretical and computational challenges. 
Second-order calculations are presently carried out within a two-timescale framework \cite{Miller:2020bft} which allows for a frequency domain approach where one works with ordinary differential equations (ODEs) for the metric perturbation.
In the time-domain, the second-order effective-source will be finitely differentiable at the particle's worldline.  
In contrast to circular orbits, for a particle in an eccentric orbit, the derivatives of the effective-source will be discontinuous functions of $t$ at a fixed value of $r$.  
Thus, one will have to reconstruct (into the time domain) a discontinuous function of $r$ and $t$ from its Fourier components and this will be tormented by the well known Gibbs phenomenon. 
The standard Fourier transform of the source is therefore very slowly convergent and worse yet, the resultant regular field and its derivatives may not converge at all.

These Gibbs phenomenon issues affect self-force calculations at both first- and second-order and a variety of techniques have been developed to over come the problem of slow convergence of the Fourier sum.
For a distributional source the method of extended homogeneous solutions (EHS) was developed \cite{Barack:2008ms}.
In this approach the frequency domain solutions that are valid on either side of the particle's libration region are analytically extended through the source region.
It is then found that the sum over the extended solutions convergenes exponentially to the correct time-domain solution.
This approach relies crucially on the spacetime away from the particle's worldline being a vacuum solution in the time domain.
The method was first developed and implemented for a single mode of a scalar field, but was later extended to full self-force calculations for scalar \cite{Warburton:2011hp,Nasipak:2019hxh} and gravitational perturbations \cite{Akcay:2013wfa,Osburn:2014hoa,vandeMeent:2015lxa,vandeMeent:2017bcc}.

In the second-order problem the effective-source is non-zero in a finite region around the worldline, and furthermore the full second-order source has unbounded support \cite{Miller:2020bft}.
With no vacuum region around the particle, or indeed anywhere in the spacetime, the method of extended homogeneous solutions cannot be applied.
Fortunately, Hopper and Evens developed the method of extended particular solutions (EPS) to handle this case \cite{Hopper:2012ty}.
In this approach a particular solution to the field equations is sought and then the retarded solution is constructed by comparison with the true solution outside the libration region where the Fourier sum convergences rapidly.

In this paper we develop a version of this method which applies to the effective-source problem where the particle's worldline is an eccentric geodesic.
We call this modified approach the method of \emph{extended effective-sources} (EES).
In lieu of the full second-order source we will consider a scalar self-force analogue to develop the method. 
In our setup we consider a scalar point charge on an eccentric orbit of a non-rotating black hole.
The effective-source we will construct will have support within some finite worldtube and be zero outside.
Although we do not consider a source with unbounded support, as appears in second-order gravitational calculations, the same method can be applicable for such sources.

The format of this paper is as follows.  
We begin in Sec.~\ref{sec:bound_orbits} where we review eccentric geodesic motion around a Schwarzschild black hole.
In Sec.~\ref{sec:scalar_sf} we describe a perturbation due to an orbiting scalar charge, outline the frequency domain decomposition, and discussion regularization approaches.
In Sec.~\ref{sec:standard_effective_source} we discuss the construction of the standard effective-source.
In Sec.~\ref{sec:standard_res_field} we give a worldtube method and use it to compute the regular field using the standard effective-source.
We show here that this results in a very slowly converging Fourier sum for the regular field.
In Sec.~\ref{sec:ees_method} we give our extended effective-sources method that restores exponential convergence to the Fourier sum.
We outline our implementation of the method and present numerical results in Sec.~\ref{sec:numerical_scheme}.
We then give a few concluding remarks in Sec.~\ref{sec:conclusion}.
Additional details are provided in the appendices.

Throughout this work we use geometrized units such that $G = c = 1$ and adopt the metric signature $(- + + +)$.  When shall denote the central black hole's mass as $M$ and use standard Schwarzschild coordinates $\tensor*{x}{^{\mu}} = (t, r, \theta, \varphi)$.
Within these coordinates, the Schwarzschild metric is given by $g_{\alpha\beta}={\rm diag}\left(-f,f^{-1},r^2,r^2\sin^2\theta\right)$, where $f(r):=1-2M/r$.

\section{Bound eccentric orbits on a Schwarzschild black hole}
\label{sec:bound_orbits}

Let us consider a test particle of mass $\mu$ on a bound timelike geodesic
around a Schwarzschild black hole.  We shall denote the worldline of the body by
$\tensor*{x}{^{\alpha}} = \tensor*{x}{^{\alpha}_{p}}(\tau) = [\tp(\tau),
r_{p}(\tau), \theta_{p}(\tau), \phip(\tau)]$ and its associated tangent
four velocity by $\tensor*{u}{^{\alpha}} = d\tensor*{x}{^{\alpha}_{p}}/d\tau$
,where $\tau$ is the particle's propertime.
Here, and throughout this paper, we shall use a subscript $p$ to
indicate evaluation of a quantity at the worldline of the particle.  
The motion of the timelike test body will obey the geodesic equation given by
\beq
	\mu \tensor*{u}{^{\beta}}\tensor*{\nabla}{_{\beta}}(u^{\alpha}) = 0,
	\label{eq:geodesic_eqn}
\eeq 
where the covariant derivative, $\tensor*{\nabla}{_{\beta}}$, is take with respect to the Schwarzschild metric.  
Without loss of generality we specify an equatorial orbit by taking $\theta_{p}(\tau) = \pi / 2$.  
The symmetries of the Schwarzschild metric lead to first integrals: $\mathcal{E} =
-\tensor*{u}{_{t}}$ and $\mathcal{L} = \tensor*{u}{_{\varphi}}$ that can used
to write Eq.~\eqref{eq:geodesic_eqn} in first order form:
\begin{align}
	\diff{\tp}{\tau} &= \tensor*{u}{^{t}} = \frac{\mathcal{E}}{f_{p}}, \quad\quad
	\diff{\phip}{\tau} = \tensor*{u}{^{\varphi}} =
	\frac{\mathcal{L}}{r^{2}_{p}},\\
	&\left( \diff{r_{p}}{\tau} \right)^{2} = (\tensor*{u}{^{r}})^{2} = \mathcal{E}^{2} - U(r_p;\mathcal{L}),
	\label{eq:fourvelocity_eqns}
\end{align}
where $U(r;\mathcal{L})$ is an effective radial potential given by
\beq
	U(r;\mathcal{L}) := \f \left( 1 + \frac{\mathcal{L}^{2}}{r^{2}} \right).
\eeq

Inspired by Newtonian mechanics, we choose an alternative, more gemoetric,
orbital
parameterisation by defining the semi-latus rectum $p$ and the eccentricity $e$
as
orbital parameters \cite{Cutler:1994pb}.  Defining the periapsis and apapsis
by $\rmin$ and $\rmax$ respectively we find these parameters are related to
each other by 
\beq
	p := \frac{2\rmax\rmin}{M(\rmax + \rmin)}, \quad\quad
	e := \frac{\rmax - \rmin}{\rmax + \rmin},
\eeq
or inversely
\beq
	\rmax = \frac{p M}{1 - e}, \quad\quad
	\rmin = \frac{p M}{1 + e}.
\eeq
The specific energy and angular momentum are themselves related to $p$ and $e$ by
\beq
	\mathcal{E}^{2} = \frac{(p - 2 - 2e)(p - 2 + 2e)}{p(p - 3 - e^{2})},
	\quad\quad
	\mathcal{L}^{2} = \frac{p^{2}M^{2}}{p - 3 - e^{2}}.
\eeq
Bound eccentric orbits satisfy the conditions $\mathcal{E} < 1$ and $\mathcal{L} > 2\sqrt{3}M$, or alternatively for $p \geq p_s$ and $0 \le e < 1$, where the separatrix between bound and plunging orbits is given by $p_s = 6+2e$ \cite{Cutler:1994pb}. 
 
For the purposes of later numerical integration, it is useful to introduce the Darwin phase \cite{Darwin:1959} via
\beq
	r_{p}(\chi) = \frac{pM}{1 + e\cos\chi}.
	\label{eq:rp_chi}
\eeq 
If one uses the phase angle, $\chi$, as our orbital parameter instead of the proper time, $\tau$, this circumvents singularities at the orbital turning points in the integrands we shall encounter later \cite{Cutler:1994pb}.  
The orbit of the test particle goes through one full radial libration for a change of the parameter, $\Delta \chi = 2\pi$.  
The periapsis, $\rmin$, occurs at $\chi = 0$ and the apapsis,  $\rmax$, corresponds to $\chi = \pi$.  
We shall assume, without any loss of generality, that at the initial periapsis passage $t = \varphi = 0$.  
The relevant differential equations in terms of $\chi$ are
\begin{align}
	&\diff{\tp}{\chi} = \frac{p^{2}M}{(p - 2 - 2e\cos\chi)(1 +
	e\cos\chi)^{2}}\sqrt{\frac{(p - 2)^{2} - 4e^{2}}{p - 6 - 2e\cos\chi}},
	\label{eq:coordinate_time_chi}\\
	&\diff{\phip}{\chi} = \sqrt{\frac{p}{p - 6 - 2e\cos\chi}},
	\label{eq:azimuthal_angle_chi}
\end{align}
Equation \eqref{eq:azimuthal_angle_chi} for the azimuthal motion admits the
analytical solution
\beq
	\varphi_{p}(\chi) = \sqrt{\frac{4p}{p - 6 - 2e}}\ F\left( \frac{\chi}{2}
	\bigg|
	-\frac{4e}{p - 6 - 2e} \right),
\eeq
where $F(w | m) = \int^{w}_{0} (1 - m^{2}\sin^{2}x)^{-1/2}dx$ is the incomplete elliptic integral of the first kind \cite{Zwillinger2014}.  
The other two equations are usually integrated numerically to compute the period of radial motion and the associated fundamental frequencies
\beq
	\T := \int^{2\pi}_{0}\left( \diff{\tp}{\chi} \right)d\chi, \quad\quad
\Omega_{r} := \frac{2\pi}{\T}.  
	\label{eq:omega_r}
\eeq
Here $\T$ represents one radial libration in coordinate time.  
One can similarly define the fundamental azimuthal frequency by averaging the angular frequency over one radial libration,
\beq
\Omega_{\varphi} := \frac{1}{\T} \int^{\T}_{0}\left( \diff{\phip}{t}
\right)dt = \frac{\phip(2\pi)}{\T}.
	\label{eq:omega_phi}
\eeq

For a particle in a bound eccentric orbit around a Schwarzschild black hole, the motion is not strictly periodic owing to the fact the orbits are generally not closed.  
Whilst the radial motion is periodic with fundamental frequency $\Omega_{r}$, the azimuthal motion of the particle, $\phip(t)$, is monotonically increasing.  
It is useful to note that azimuthal motion can be expressed as
\beq\label{eq:phi_avg_plus_oscillation}
	\phip(t) = \Omega_{\varphi}t + \Delta\phip(t),
\eeq
where the mean azimuthal advance is modulated by the function, $\Delta\phip(t)$, which is periodic with fundamental frequency $\Omega_{r}$ \cite{Cutler:1994pb}.

\section{Scalar field self-force}
\label{sec:scalar_sf}

The challenges of computing the gravitational self-force has motivated studies
of the ``toy model'' scalar-field self-force problem
\cite{Burko:2000xx, Diaz-Rivera:2004nim, Warburton:2010eq,
Warburton:2011hp, Warburton:2014bya, Nasipak:2019hxh, Nasipak:2021qfu,
PanossoMacedo:2022fdi}.
These studies capture much of the computational complexity of gravitational self-force calculations whilst avoiding tedious bookkeeping of solving for the metric perturbation
\cite{Barack:2007tm, Barack:2010tm, Akcay:2013wfa, Osburn:2014hoa, 
vandeMeent:2015lxa, vandeMeent:2017bcc} or subtle gauge issues
\cite{Barack:2001ph,Pound:2013faa}.
In this work we develop a new computational approach and so once again return to the scalar-field problem to elucidate the method.
We describe our scalar-field setup and its decomposition into spherical harmonics below. 
We then consider the Fourier domain decomposition, boundary conditions, and regularization approaches.

\subsection{Field equation and multipole decomposition}
\label{sec:field_equation_decomposition}
Consider a particle of mass $\mu$ carrying a scalar charge, $q$, moving on an eccentric geodesic about a Schwarzschild black hole as described in \secref{bound_orbits}. 
In this work, we ignore the effects of the particle's gravitational self-interaction and focus on the scalar-field self-force (SSF) that arises from the particle's interaction with its own scalar field.  
We develop a method to compute the SSF, and do not compute how this back-reacts on the orbital motion. 
Here, we prescribe that the particle's scalar field, $\Phi$, obeys the a minimally coupled Klein-Gordon equation
\beq
\Box \Phi := \tensor*{\nabla}{^{\alpha}}\tensor*{\nabla}{_{\alpha}} \Phi
= -4\pi\rho,
	\label{eq:klein_gordon_eqn}
\eeq
where $\rho$ is the particle's scalar charge density.  
For a point particle the scalar charge density can be modelled as
\begin{align}
	\rho(t, r, \theta, \varphi) &= q \int \frac{\delta^{4}[\tensor*{x}{^{\mu}} -
	\tensor*{x}{^{\mu}_{p}}(\tau)]}{\sqrt{-g}} d\tau \nn\\ 
	&= \frac{q}{r^{2}_{p}
	\tensor*{u}{^{t}}}{\delta[r - \rp]\delta[\varphi -
	\varphi_{p}(t)]\delta[\theta - \pi/2]}.
	\label{eq:scalar_charge_density}
\end{align}
The $t$-component of the four-velocity, $\tensor*{u}{^{t}}$, is given in
\eqn{fourvelocity_eqns}.

The scalar wave-equation given in \eqn{klein_gordon_eqn} is amenable to a
solution via seperation of variables,
\beq
	\Phi(t, r, \theta, \varphi) = \sum^{\infty}_{\ell = 0} \sum^{\ell}_{m = -\ell}
	\frac{\psi_{\lm}(t, r)}{r} Y_{\lm}(\theta, \varphi),
	\label{eq:field_decomposition} 
\eeq
where $Y_{\lm}(\theta, \varphi) = \hat{c}_{\lm} P^{m}_{\ell}(\cos\theta)e^{i m
\varphi}$ are the standard spherical harmonics with the associated
Legendre polynomial $P^{m}_{\ell}$ and $\hat{c}_{\lm} = \sqrt{\frac{2\ell +
1}{4\pi}\frac{(\ell - m)!}{(\ell + m)!}}$. The extra factor of $1/r$ included
here is added for later convenience. We define the spherical
harmonics to be normalized such that
\beq
	\oint Y_{\lm}(\theta, \varphi) Y^{*}_{\ell^{\prime}m^{\prime}}(\theta, \varphi) d\Omega =
	\delta^{\ell^{\prime}}_{\ell} \delta^{m^{\prime}}_{m},
	\label{eq:orthogonality_relation}
\eeq
where $d\Omega = \sin\theta d\theta d\varphi$, $\delta^{n_{2}}_{n_{1}}$ is the
usual Kronecker delta, and $*$ denotes complex conjugation.  
The charge density, $\rho$, in \eqn{scalar_charge_density} can be decomposed analogously as
\beq
	\rho(t, r, \theta, \varphi) = \sum^{\infty}_{\ell = 0} \sum^{\ell}_{m = -\ell}
	\rho_{\lm}(t, r)
	Y_{\lm}(\theta, \varphi).
	\label{eq:scalar_charge_decomposition}
\eeq
Using the orthonormality relation specified in \eqn{orthogonality_relation} we
find the multipole modes, $\rho_{\lm}(t, r)$, are given by
\begin{align}
	\rho_{\lm}(t, r) &= \oint \, \hat{c}_{\lm} P^{m}_{\ell}(\cos\theta) d\Omega,
	\nn \\
	& = \frac{q\, \hat{c}_{\lm} P^{m}_{\ell}(0)}{\rp^{2} \tensor{u}{^{t}}}
		\delta[r -\rp] e^{- i m\phip(t)}.
	\label{eq:scalar_charge_mulitpole_modes}
\end{align}

Substituting the field decomposition given in \eqn{field_decomposition}
into \eqn{klein_gordon_eqn} and dividing through by $f(r)$ we obtain
\beq
	\Box_{\lm} \psi_{\lm}(t, r) = -4\pi r \rho_{\lm}(t, r),
	\label{eq:td_retarded_field_eqn}
\eeq
where our wave-operator, $\Box_{\lm}$, is defined to be
\beq
	\Box_{\lm} := \frac{1}{f(r)} \left( - \psdiff{\psi_{\lm}}{t}
	+ \psdiff{\psi_{\lm}}{\rstar} \right) - V_{\ell}(r)\, \psi_{\lm}.
	\label{td_box}
\eeq
Here we have introduced the tortoise coordinate defined via $d\rstar/dr =
f(r)^{-1}$.  
Choosing a particular integration constant, the tortoise coordinate is given by
$\rstar = r + 2M \ln \left( r/2M - 1 \right)$.  
The radial potential, $V_{\ell}(r)$, is given by
\beq
	V_{\ell}(r) = \frac{2M}{r^{3}} + \frac{\ell(\ell + 1)}{r^{2}}.
	\label{eq:potential}
\eeq 
One approach to solving the wave equation \eqref{eq:td_retarded_field_eqn} is via time evolution on a $1+1$ dimensional grid \cite{Haas:2007kz,Barack:2022pde}.

\subsection{Frequency domain decomposition}

In this work we focus on the FD approach, where the field $\psi_{\lm}(t, r)$ and the charge density $\rho_{\lm}(t, r)$ are further decomposed into the Fourier frequency modes $\psi_{\lmn}(r)$ and $J_{\lmn}(r)$ respectively.
For a given multipole mode, the frequency spectrum is formed of discrete overtones of the two fundamental frequencies such that
\beq
	\omn := m\Omega_{\varphi} + n\Omega_{r}, \quad\quad m, n \in \mathbb{Z}.
\eeq
The FT of the multipole modes of the field, $\psi_{\lm}(t, r)$, is therefore
given by
\beq
	\psi_{\lmn}(r) = \frac{1}{\T}\int^{\T}_{0}\psi_{\lm}(t, r)\etplus dt,
	\label{eq:fourier_transform_field}
\eeq
where the inverse FT to reconstruct the TD solution reduces to a summation over $n$,
\beq
	\psi_{\lm}(t, r) = \sum^{\infty}_{n = -\infty} \psi_{\lmn}(r) e^{-i \omn t}. 
	\label{eq:psi_TD_from_sum}
\eeq
Similarly, the multipole modes of the charge density also admit a discrete Fourier decomposition of the form
\beq
	\rho_{\lm}(t, r) = -\frac{1}{4\pi r}\sum^{\infty}_{n = -\infty} J_{\lmn}(r)
	e^{-i \omn t}.
\eeq
We note that the multipole decomposition of the charge density, $\rho_{\lm}(t, r)$, given in Eq.~\eqref{eq:scalar_charge_mulitpole_modes}, is proportional to $e^{-i m\phip(t)}$ and so it is not a periodic function with frequency $\Omega_{r}$.  
However, following Eq.~\eqref{eq:phi_avg_plus_oscillation}, if we express this exponential factor as $e^{-i m\Omega_{\varphi}t}e^{-i m\Delta\phip(t)}$ then $\rho_{\lm}(t, r)\,e^{i m\Omega_{\varphi}t}$ is periodic.  
Thus the source term and, by extension, the field $\psi_{\lm}(t, r)$ can be expressed as a Fourier series when multiplied by this appropriate phase factor.  
This is equivalent to being in a frame with angular velocity $\Omega_{\varphi}$ and to an asymptotic observer within this frame the radial and azimuthal motion would appear periodic \cite{Cutler:1994pb}.  
\begin{widetext}
Taking the Fourier transform of \eqn{scalar_charge_mulitpole_modes} we find the Fourier source term given by
\beq
	J_{\lmn}(r) = \frac{q\, \hat{c}_{\lm} P^{m}_{\ell}(0)}{\T} \int^{\T}_{0}\frac{\delta[r - \rp]}{\rp^{2} \tensor{u}{^{t}}} e^{i[\omn t - m\phip(t)]} dt.
\eeq
Our source, $J_{\lmn}$, is compact with support within the libration region $ \rmin \leq r \leq \rmax$.  
The integration over the delta function can be carried out by changing the integration variable from $t$ to $r_{p}$.  
This gives the result
	\beq
		J_{\lmn}(r) = \frac{2q\, \hat{c}_{\lm} P^{m}_{\ell}(0)}{\T r
		|\tensor{u}{^{r}}(r)|f(r)^{2}}\cos[\omn
		t_{p}(r) - m\phip(r)]\times \Theta[r - \rmin]\times \Theta[\rmax - r].
		\label{eq:ft_delta_source}
	\eeq
Here $\Theta$ is the standard Heaviside step function, $\tensor*{u}{^{r}}$ is the $r$-component of the particle's four velocity and the functions $t_{p}(r)$ and $\phip(r)$ are obtained by formally inverting $r_{p}(\chi)$ in the range $0 \leq \chi \leq \pi$.
Using Eq.~\eqref{eq:psi_TD_from_sum} and \eqref{eq:td_retarded_field_eqn} we find the field equation for $\psi_{\lmn}(r)$ is given by
\begin{align}
	\fdbox \fdretfield(r)
	:= \sdiff{\fdretfield}{r} + \frac{2 M}{\f r^3}\l r
	\frac{d\fdretfield}{dr} - \fdretfield(r)\r + \frac{1}{\f}\l\frac{\omn^2}{\f}
	- \frac{\ell(\ell + 1)}{r^2}\r \fdretfield(r) = \fdJ(r),
 	\label{eq:retarded_field_eqn}
\end{align}
\end{widetext}
where we have re-written our differential operator in terms of $r$ instead of $\rstar$ as later this will be our numerical integration variable.

\subsection{Radial boundary conditions}

The physical solutions to \eqn{retarded_field_eqn} are uniquely determined once appropriate boundary conditions are specified at spatial infinity ($r \longrightarrow \infty$) and at the horizon ($r \longrightarrow 2M$).
Here we select the retarded solution which corresponds to outgoing waves at null infinity and purely ingoing waves at the event horizon.  
Formally, let $\fdhplus$ and $\fdhminus$ be two homogeneous solutions to \eqn{retarded_field_eqn},  which are determined by their asymptotic boundary conditions at spatial infinity and the horizon respectively such that
\begin{align}
	\fdhplus(\rstar \longrightarrow \infty) &\sim \etortplus, \nn\\
	\fdhminus(\rstar \longrightarrow -\infty) &\sim \etortminus.
\end{align}

The construction of practical numerical boundary conditions and the computation of the (non-radiative) static modes are discussed in Appendix A of \cite{Warburton:2013lea}.

\subsection{Regularization}
One of the main challenges when computing the SF is that the field of the
particle diverges at the location of the particle.
It is known though that a particular divergent contribution to the field near the particle does not contribute to the orbital evolution \cite{Quinn:1996am,Poisson:2011nh}.
Instead the evolution of the inspiral is driven by a (self-)force that can be
computed from an appropriate regular contribution to the particle's field at
the particle's location.
Extracting this regular field requires applying a regularization technique to
either the retarded field or the field equations themselves.

Follow on from the original formulation \cite{Quinn:1996am},  Detweiler and Whiting \cite{Detweiler:2003} recast the regularization scheme such that self-force is computed from the regular field, $\Phi^{R}$, with
\beq
	\tensor*{F}{^{\text{self}}_{\alpha}}(x_{p}) = q
	\tensor{\nabla}{_{\alpha}}\tensor{\Phi}{^{R}}(x_{p}).
	\label{eq:self_force}
\eeq
The regular field is defined by
\beq
	\tensor*{\Phi}{^{R}}(x_{p}) = \lim_{x \rightarrow x_{p}} [
	\tensor*{\Phi}{^{\text{ret}}}(x) - \tensor*{\Phi}{^{S}}(x) ],
	\label{eq:regular_field}
\eeq
where $\tensor{\Phi}{^{\text{ret}}}$ is the usual retarded field and $\tensor{\Phi}{^{S}}$ is an appropriately constructed singular field \cite{Detweiler:2003,Poisson:2011nh} .  
Here the argument ``$x$" represents a field point in the normal neighbourhood of the particle's worldline.  
The retarded and singular fields obey the inhomogeneous field equation, \eqn{klein_gordon_eqn}:
\beq
	\Box \tensor{\Phi}{^{\text{ret}/S}} = -4\pi\rho,
\eeq
whilst the regular field obeys the homogeneous version of the same field
equation,
\beq
	\Box \tensor{\Phi}{^{R}} = 0.
\eeq
Using this split, one finds the self-force \eqn{self_force} can be written as as
\begin{align}
	\tensor*{F}{^{\text{self}}_{\alpha}}(x_{p}) &= q \lim_{x \rightarrow x_{p}} [
	\tensor{\nabla}{_{\alpha}} \left( \tensor*{\Phi}{^{\text{ret}}}(x) -
	\tensor*{\Phi}{^{S}}(x)\right)
]
	\nn\\
	& = \lim_{x \rightarrow x_{p}} [ \tensor*{F}{^{\text{ret}}_{\alpha}}(x) -
	\tensor*{F}{^{S}_{\alpha}}(x) ],
	\label{eq:sf_split}
\end{align}
where we have defined
\beq
	\tensor*{F}{^{\text{ret}/S}_{\alpha}}(x) := q
	\tensor{\nabla}{_{\alpha}}\tensor*{\Phi}{^{\text{ret}/S}}(x).
\eeq
The singular field is not global defined and in practice it is approximated by a puncture field, $\tensor*{\Phi}{^{\mathcal{P}}}$, which is computed by taking a local expansion of the singular field and truncating at a certain order
\cite{Barack:1999wf,Heffernan:2012su}.  
The puncture field is defined such that
\begin{align}
	\lim_{x \rightarrow x_{p}} [ \tensor*{\Phi}{^{\mathcal{P}}}(x) &-
	\tensor*{\Phi}{^{S}}(x) ] = 0, \nn\\
	\lim_{x \rightarrow x_{p}} [ \tensor{\nabla}{_{\alpha}}
	\tensor*{\Phi}{^{\mathcal{P}}}(x) &-
	\tensor{\nabla}{_{\alpha}} \tensor*{\Phi}{^{S}}(x) ] =
	0.
\end{align}
Analogous to the regular field, a ``residual" field is then defined via
\beq
	\tensor*{\Phi}{^{\mathcal{R}}} := \tensor*{\Phi}{^{\text{ret}}} - \tensor*{\Phi}{^{\mathcal{P}}} \sim \tensor*{\Phi}{^{R}}.
	\label{eq:residual_field}
\eeq
As such, the smoothness of the residual field on the worldline is determined by the order of the approximation of the puncture to the singular field. 
So long as the puncture field approximates the singular field to high enough order to make the residual field $C^1$ differentiable on the worldline, the self-force can be computed via
\beq
	\tensor*{F}{^{\text{self}}_{\alpha}}(x_{p}) = \lim_{x \rightarrow x_{p}} \mu
	\tensor{\nabla}{_{\alpha}} \tensor*{\Phi}{^{\mathcal{R}}}(x).
	\label{eq:sf_residual}
\eeq

If we approach Eq.~\eqref{eq:sf_residual} from a computational standpoint, however, we encounter  significant difficulties.
In particular if we try to calculate the residual field via Eq.~\eqref{eq:residual_field} we have to subtract one diverging quantity from another before taking the limit to the worldline of the particle.
Numerically, this would be extremely challenging.
The \emph{mode-sum} scheme and the \emph{effective-source} approach are two ways of circumventing this issue.

If the field near the particle diverges as $1/(\Delta r)$, where $\Delta r := r - \rp$ is defined as the distance from the worldline, then a decomposition of the perturbation into spherical harmonic $\ell m$-modes will render each $\ell m$ mode of the field finite at the location of the particle.
This means the subtraction in Eq.~\eqref{eq:residual_field} can be carried out mode-by-mode in a procedure known as mode-sum regularization \cite{Barack:1999wf,Barack:2001bw}.
The scalar model considered in this work, and first-order in the mass-ratio gravitational perturbations in, e.g., Lorenz, radiation and Regge-Wheeler
gauges, can be regularised in this manner \cite{Burko:2000xx,Diaz-Rivera:2004nim,Warburton:2010eq}.
In general, gravitational perturbations at second-order in the mass ratio diverge more strongly around the particle and so the individual $\ell m$-modes
of the perturbation diverge at the location of the particle \cite{Pound:2014xva}.\footnote{There is a special class of gauges where the divergence near
the particle is weak enough that a mode decomposition renders the individual
modes finite at the particle \cite{Upton:2021oxf}. A significant amount of new
theoretical and computational infrastructure is needed before this approach can be used in practice.}
We thus need an alternative approach.

The effective-source method sides steps the issue of the divergence in the retarded and puncture fields by directly solving for the residual field. 
In terms of the SSF model, one can rewrite the retarded field, $\tensor*{\Phi}{^{\text{ret}}}$, in terms of the residual field,
$\tensor*{\Phi}{^{\cal R}}$, and the puncture field, $\tensor*{\Phi}{^{\cal P}}$, 
using \eqn{residual_field}:
\begin{align}
	\Box \tensor*{\Phi}{^{\cal R}} &= \Box \left(
	\tensor*{\Phi}{^{\text{ret}}}
	-
	\tensor*{\Phi}{^{\cal P}} \right) \nn\\
	&= -4\pi\rho - \Box \tensor*{\Phi}{^{\cal P}} := S^{\text{eff}}.
\end{align}
The effective-source, $S^{\text{eff}}$, defined here will not contain a delta-function since, by construction, the
$\delta$-function term within $\rho$ will exactly cancel with a term that
arises from $\Box \tensor*{\Phi}{^{\mathcal{P}}}$ and hence we are left with
a non-distributional remainder.

In computations using an effective-source, one must restrict its support to the vicinity of the particle's worldline as the puncture field is not defined
outside the particle's normal neighbourhood.  
Practically, this can be done in two different ways: a \emph{window function} \cite{Vega:2007mc} or via a \emph{worldtube}
\cite{Barack:2007jh}.  
These two approaches were shown to be equivalent in Ref.~\cite{Warburton:2013lea} so we focus on the latter which we find easier to implement in practice.
The latter method involves constructing a worldtube such that one solves for $\tensor*{\Phi}{^{\mathcal{R}}}$ inside the worldtube and the physical perturbation $\tensor*{\Phi}{^{\text{ret}}}$ outside. 
Jump conditions, determined by the puncture field, are then supplied at the boundaries of the worldtube.  

We can apply the effective-source approach at the level of $\lm$-modes.
Writing $\psi_{\lm} = \psi^{\res}_{\lm} + \psi^{\P}_{\lm}$ and using Eq.~\eqref{eq:td_retarded_field_eqn} we can write
\begin{align}
\label{eq:time_domain_seff}
	\Box_{\lm} \psi^{\res}_{\lm} &= -4\pi r\rho_{\lm}(t,r) - \Box_{\lm} \tdsingfield \nn \\
					       &:= S^{\rm eff}_{\lm}(t,r)
\end{align}
The explicit form of the puncture we use in this work is given in Appendix
\ref{apdx:scalar_field_puncture}.

\begin{figure}[!t]
	 \centering
         \includegraphics[width=0.48\textwidth]{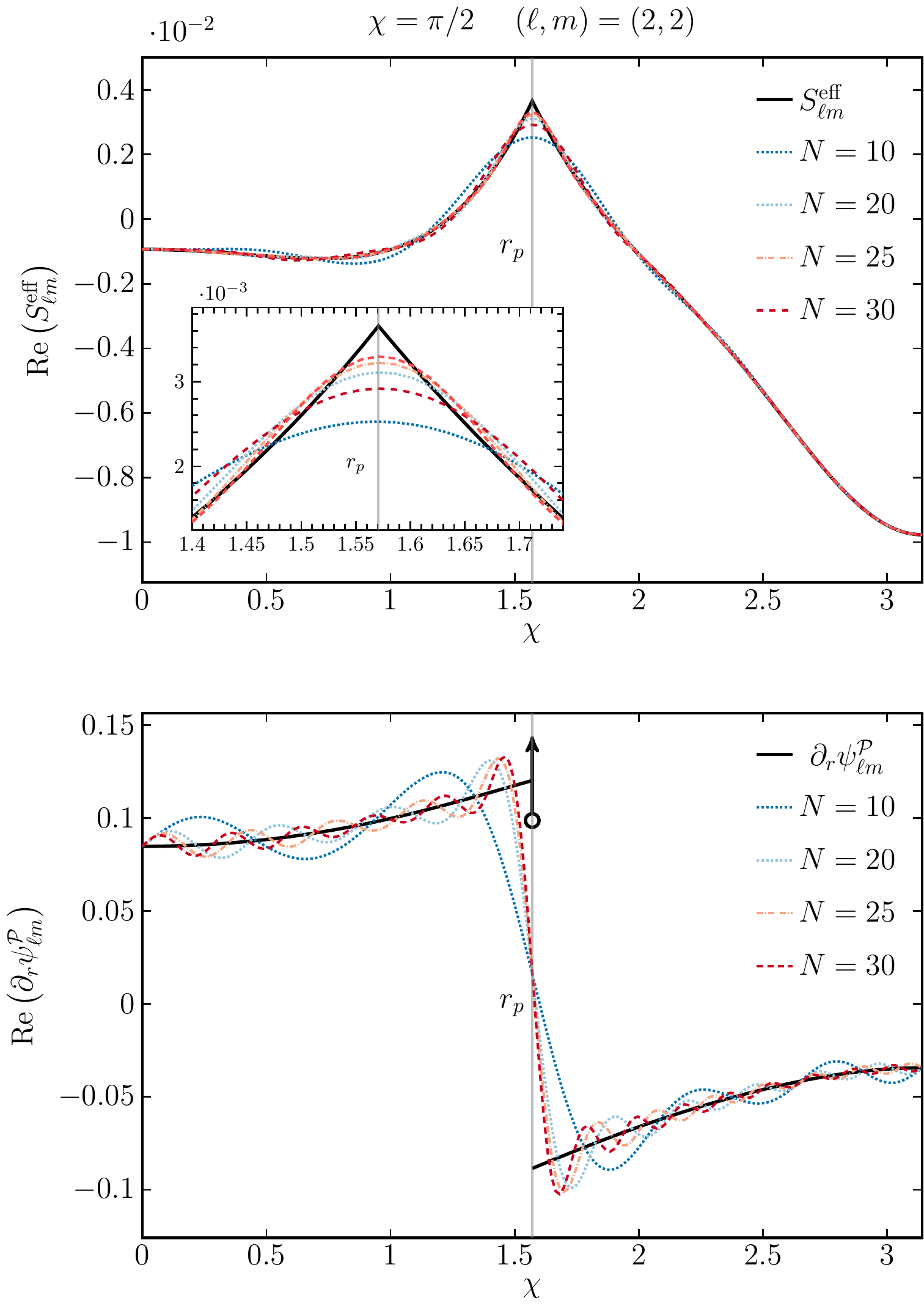}
\caption{
The standard FD approach to reconstructing the TD effective-source and	the derivative of the puncture field with respect to $r$.  
The top panel shows $S^{\text{eff}}_{\lm}$ and the bottom shows $\partial_{r}\psi^{\mathcal{P}}_{\lm}$ for the $(\ell, m) = (2, 2)$ mode at $\chi = \pi/2$ for a particle orbiting with $p = 10M$ and $e = 0.2$.  
Partial sums are computed with \eqn{seff_fouriersum} and shown for different $N$, where $N$ is the maximum of the partial sum.  
For comparison purposes we also display (black) the TD effective-source (top panel), $S^{\text{eff}}_{22}$, and the derivative of the TD puncture field with respect to $r$ (bottom panel), $\partial_{r}\psi^{\mathcal{P}}_{22}$, which we have obtained using the TD puncture, $\psi^{\mathcal{P}}_{\lm}$.
The Gibbs phenomenon is clear in both cases.
}
\label{fig:effective_source_gibbs}
\end{figure}

\section{Construction of the standard effective-source in the frequency domain}
\label{sec:standard_effective_source}

We now wish to calculate the modes of the effective-source in the frequency domain.
One approach would be to use Eqs.~\eqref{eq:time_domain_seff} and \eqref{eq:fourier_transform_field} to write
\begin{align}
	\psi_{\lmn}^{\res}(r) &= \psi_{\lmn}(r) -  \frac{1}{T_r}\int_{0}^{T_r}\psi^{\P}_{\lm}(t,r) e^{-i\omega_{mn}t} \label{eq:psiR_lmn}\\
	                      &:= \psi_{\lmn}(r) - \psi^{\P}_{\lmn}(r)
\end{align}
Acting on this equation with the radial domain operator, $\square_{\lmn}$
defined in \eqn{retarded_field_eqn} we get
\begin{align}\label{eq:FDwave_Seff}
	\square_{\lmn}\psi^{\res}(r) = J_{\lmn}(r) - \square_{\lmn}\psi^{\P}_{\lmn}(r) := \fdSeff(r)
\end{align}
The challenge with this approach is that, although we know $\psi^{\P}_{lm}(t,r)$ analytically \cite{Heffernan:2017cad}, due to the eccentric orbital motion we do not analytically know its Fourier transform.
We could still numerically evaluate the Fourier integral in Eq.~\eqref{eq:psiR_lmn} on a dense grid of radial values and interpolate the result.
This has to be done to very high precision in order get an accurate result
after applying the $\square_{\lmn}$ operator.
Furthermore, the result must then cancel the Fourier transform of the delta
function, $J_{\lmn}(r)$ and from Eq.~\eqref{eq:ft_delta_source} we see this
diverges as $1/u^r$ at the orbital turning points making numerical cancellation
very challenging.

We find the above approach of computing the Fourier transform of the puncture and then constructing the FD effective-source to be unworkable.
Instead, we can first compute the time-domain effective-source, as in Eq.~\eqref{eq:time_domain_seff}, and then take the Fourier transform.
That is, we define
\begin{align}
	\label{eq:fourier_domain_seff}
	\fdSeff(r) = \frac{1}{T_r}\int_0^{T_r}  \Seff(t,r) e^{-i\omega_{mn} t}\, dt
\end{align}
where $\Seff(t,r)$ is defined in Eq.~\eqref{eq:time_domain_seff}.
This has the distinct advantage that before carrying out the Fourier
decomposition (i) the operator $\Box_{\lm}$ can be applied analytically during
the construction of $\Seff(t,r)$ and (ii) the cancellation of the distributional
term in the source can thus be done analytically.

\begin{figure}[t!]
         \centering
         \includegraphics[width=0.48\textwidth]{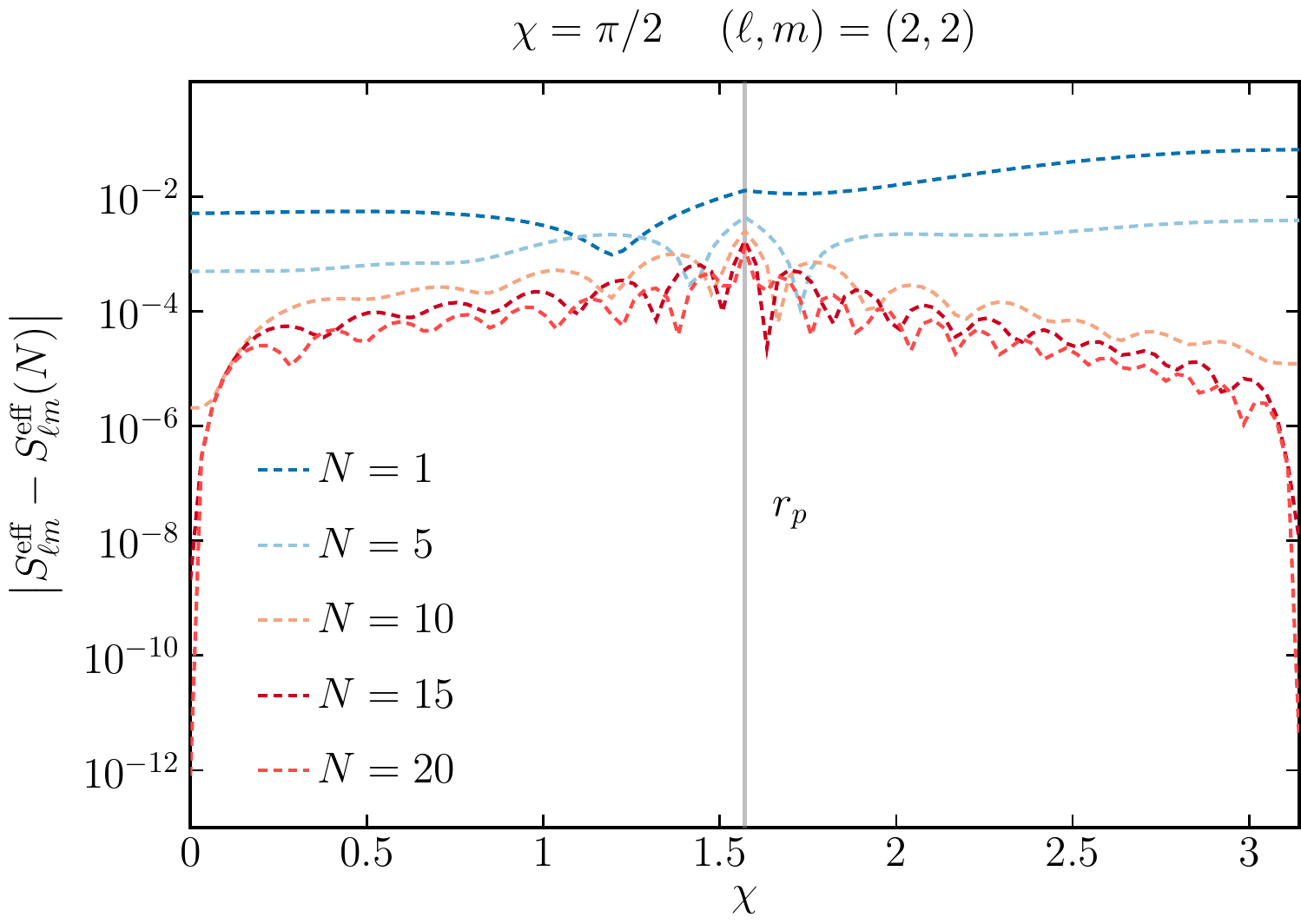}
		 \caption{
	 	The absolute error of the TD effective-source computed using the standard method.  
	 	Here the orbital parameters are $p = 10M$ and $e = 0.2$ and we consider the $(l,m) = (2,2)$ mode.
		We show the difference between the TD effective-source and the result computed using the partial sum  in Eq.~\eqref{eq:seff_fouriersum} at $\chi = \pi/2$.
	 	For the standard approach we see the convergence is algebraic as it is blighted by Gibbs phenomenon --- see Fig.~\ref{fig:effective_source_gibbs}.
		 }
		 \label{fig:effective_source_gibbs_diff}
\end{figure}

We still have to numerically evaluate the Fourier integral in Eq.~\eqref{eq:fourier_domain_seff} on a grid of radial values.
This is still challenging as the time-domain effective-source is piecewise continuous ($C^{0}$) at the instantaneous particle location $\rp$.
This lack of smoothness results in the well-known Gibbs phenomenon and the Fourier sum
\beq
	\tdSeff(t, r) = \lim_{N\rightarrow\infty}\sum_{n=-N}^N \fdSeff(r)\etminus,
	\label{eq:seff_fouriersum}
\eeq
converges very slowly as $1/N$. 
This slow convergence is shown with numerical results in Figs.~\ref{fig:effective_source_gibbs} and \ref{fig:effective_source_gibbs_diff}.
The non-smoothness of the effective-source also hampers the efficient calculation of $\fdSeff(r)$ as we cannot directly apply the efficient Fast Fourier Transform (FFT) algorithm. 
One option is to directly numerically evaluate the integral in Eq.~\eqref{eq:fourier_domain_seff} for each frequency mode, though we find this to be quite inefficient.
Fortunately, we find an alternative convolution method that allows the FFT to be employed -- see Sec.~\ref{sec:convolution} below for details.

\section{Calculation of the residual field using a worldtube method}
\label{sec:standard_res_field}

We now present the standard calculation of the residual field using a worldtube method.
As we shall see, the convergence rate of the partial sum over Fourier modes of the residual field is very slow when using the standard effective-source approach.

\subsection{Frequency domain worldtube method}
\label{sec:eccentric_worldtube_method}
Our goal is compute the residual field inside a worldtube of finite size around the particle's worldline.
To achieve this we need a scheme for working with a worldtube in the frequency domain.  
A version of such a scheme was outlined in Ref.~\cite{Warburton:2013lea} but here we present a simpler form based on notes originally made by Barack \cite{Barack:2019}.

The frequency domain field equation we want to solve takes the form of Eq.~\eqref{eq:FDwave_Seff} where the effective-source is non-zero inside a worldtube with boundaries coinciding with the edges of the libration region, $\rmin$ and $\rmax$. 
Outside of the worldtube we will solve $\fdbox \retfield = 0$.
Let $\fdh$ be two independent homogeneous solutions satisfying retarded boundary conditions in the respective domains $r \longrightarrow 2M$ and $r \longrightarrow \infty$.  
\begin{widetext}
We seek a solution of the following form,
\beq
	\fdregfield(r) = \bec
	a_{\lmn}^{h}\fdhminus(r), & r \leq  \rmin, \\
    b_{\lmn}^{\infty}\fdhplus(r) + b_{\lmn}^{h}\fdhminus(r) + \fdinhfield(r), & \rmin < r < \rmax,\\
	a_{\lmn}^{\infty}\fdhplus(r), & r \geq  \rmax.
\eec
\label{eq:worldtube_ansatz}
\eeq
where $a_{\lmn}^{\infty/h}$ and $b_{\lmn}^{\infty/h}$ are constants.
\end{widetext}
Here $\fdinhfield(r)$ is the particular inhomogeneous solution found from the standard variations of parameters approach
\begin{align}
	\fdinhfield(r) = C_{\lmn}^\infty(r) \fdhplus(r) + C_{\lmn}^h(r) \fdhminus(r) 	
\end{align}
with
\begin{align}\label{eq:eccentric_source_integrals}
C^{\infty}_{\lmn}(r) &=
\int_{\rmin}^{r}\frac{\psi_{\lmn}^h(r^{\prime})\fdSeff(r^{\prime})}{W
[\fdhminus(r^ {\prime}), \fdhplus(r^{\prime})]}dr^{\prime} \\
C^{h}_{\lmn}(r)    &=
\int^{\rmax}_{r}\frac{\psi_{\lmn}^\infty(r^{\prime})\fdSeff(r^{\prime
})}{W [\fdhminus(r^ {\prime}), \fdhplus(r^{\prime})]}dr^{\prime}
\end{align}
where $W[\psi_{1}, \psi_{2}] := \psi_{1}\diff{\psi_{2}}{r} - \psi_{2}\diff{\psi_{1}}{r} $ is the Wronskian.
The unknown coefficients $a_{\lmn}^{\pm}$ and $b_{\lmn}^{\pm}$ are determined from the conditions that $\fdretfield = \fdregfield + \fdsingfield$ and $d\fdretfield/dr$ are continuous at the worldtube boundaries, $r = \rmin$ and $r = \rmax$.
This gives
\begin{align}
	a_{\lmn}^{\infty} =& \frac{1}{\fdhplus(\rmax)}\big\{\fdhplus(\rmax)[b_{\lmn}^{\infty} + C_{\lmn}^\infty(\rmax) ]\nonumber \\
					  & + b_{\lmn}^{h}\fdhminus(\rmax)+ \fdsingfield(\rmax)\big\}, 
\end{align}
\begin{align}
	a_{\lmn}^{h}      =& \frac{1}{\fdhminus(\rmin)}\big\{b_{\lmn}^{\infty}\fdhplus(\rmin)  \nonumber \\
	  				  & + \fdhminus(\rmin)[b_{\lmn}^{h} + C_{\lmn}^h(\rmin)] + \fdsingfield(\rmin)\big\}.
	\label{eq:a_coefficients}
\end{align}
Similarly, the coefficients $b^{h/\infty}_{\lmn}$ are given by
\begin{align}
	\begin{split}
b^{\infty} 	&= \left. \frac{W[\fdsingfield(r),
\fdhminus(r)]}{W[\fdhminus(r),\fdhplus(r)]}\right|_{r=\rmin} , \\
b^{h} 		&= \left. \frac{W[\fdsingfield(r),
\fdhplus(r)]}{W[\fdhplus(r),\fdhminus(r)]}\right|_{r=\rmax}.
	\label{eq:b_coefficients}
	\end{split}	
\end{align}

\subsection{The residual field computed from the standard effective-source}
\label{sec:regular_field_from_standard_source}

\begin{figure}[!tp]
	\centering
    \includegraphics[width=0.48\textwidth]{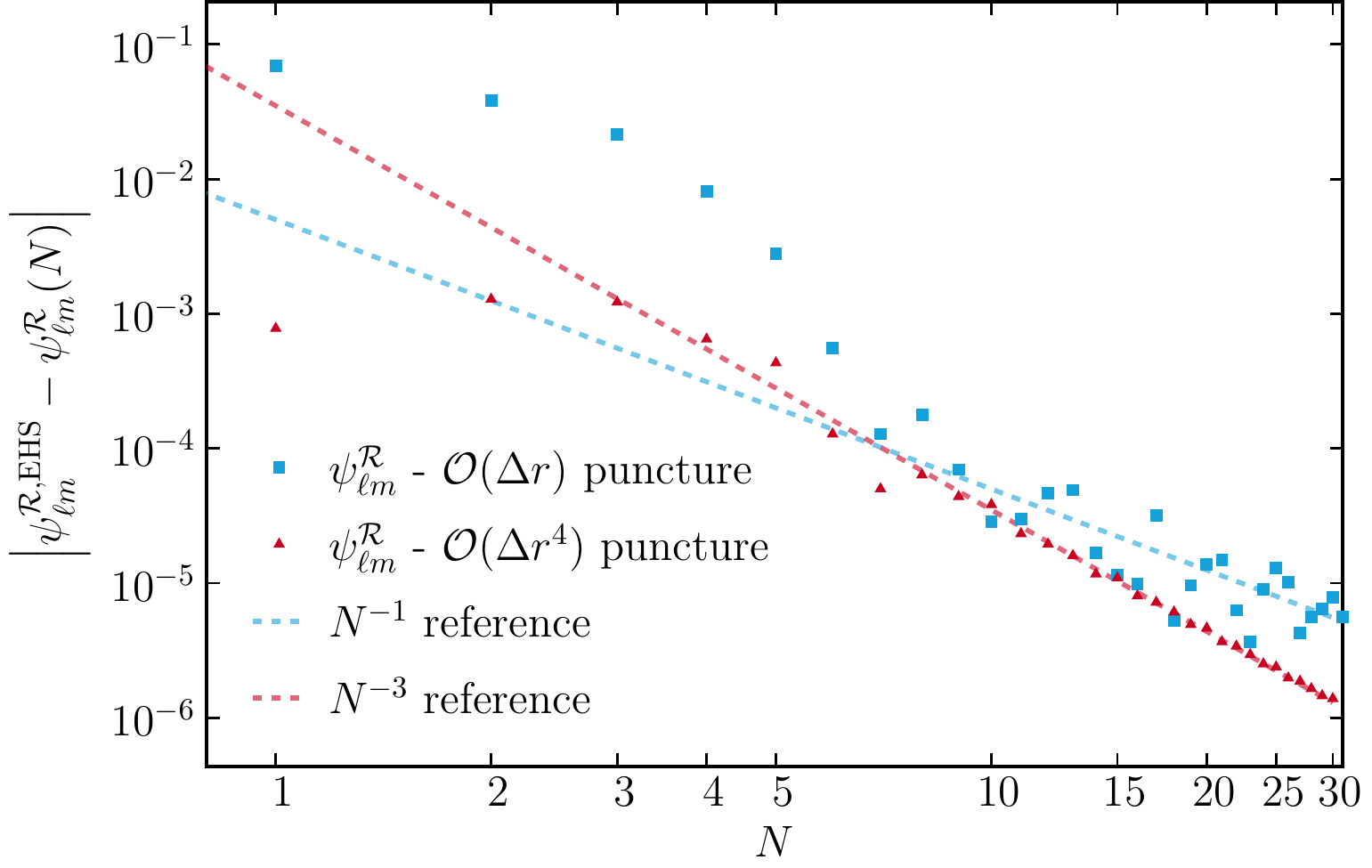}
	\caption{
	The convergence of the residual field computed with the standard method frequency domain approach for different orders of the puncture for a particle with orbital parameters $p = 10M$ and $e = 0.2$ and for $(\ell, m) = (2, 2)$.
	We compute the reference residual field, $\tdresfieldEHS = \psi^{\rm EHS}_{\lm} - \tdsingfield$, using the method of EHS. 
	For the residual field computed with the ${\cal O}(\Delta r)$ puncture (blue squares) we see that the series appears to converges very slowly as $\sim N^{-1}$. 
	If we increase the order of the puncture to ${\cal O}(\Delta r^{4})$ then we see the rate of convergence of the residual field (red triangles) increases to $\sim N^{-3}$.  
	The convergence is still algebraic and very slow in comparison to the exponential convergence of, e.g., the method of EHS. 
	We note that the noisy behaviour observed when using the ${\cal O}(\Delta r)$ puncture is reminiscent of the  behaviour in the partial Fourier sum of the retarded field for the monopole mode using the standard Fourier approach -- see Fig.~2 of Ref.~\cite{Barack:2008ms}.
	}
    \label{fig:residual_field_slow_convergence}
\end{figure}

Given an effective-source, the Fourier modes of the regular field can be calculated using the variation of parameters with a worldtube approach outlined above.
The $\lm$-mode of the residual field are then constructed via
\begin{align}\label{eq:phiR_lm_sum}
	\psi^\res_{\lm}(t,r) = \lim_{N\rightarrow\infty} \sum_{n=-N}^N
	\psi_{\lmn}^\res(r)e^{-i\omega_{mn}t}
\end{align}
The smoothness of the resulting regular field, $\psi_{\lm}^\res(t,r)$, at the
location of the particle depends upon the order of the puncture used to
construct the effective-source.
This in turn effects the rate of convergence of the Fourier sum in
Eq.~\eqref{eq:phiR_lm_sum}.
In Appendix~\ref{apdx:scalar_field_puncture} we present the puncture through
$\mathcal{O}(\Delta r)$ where $\Delta r = r-r_p(t)$.
This puncture gives a regular field that is $C^1$ in the radial direction at
the location of the particle and is thus sufficient to calculate the
self-force.
We find the partial sum when including up to $n=\pm N$ terms in Eq.~\eqref{eq:phiR_lm_sum} converges very slowly as $1/N$.
This rate of convergence can be improved by using a high-order puncture, though the convergence remains a power law.
Using a puncture through $\mathcal{O}(\Delta r^4)$ the convergence improves to $1/N^3$ -- see Fig.~\ref{fig:residual_field_slow_convergence}. 
The explicit form of the higher-order puncture can be found in the supplmental material accompanying this work \cite{supplemental_material}.

In principle higher-order punctures could be derived to further accelerate the
convergence of the Fourier sum.
Deriving higher-order punctures becomes increasingly more challenging as the
order in $\Delta r$ increases \cite{Heffernan:2012su,Heffernan:2017cad} and
even with higher-order punctures the convergence would still be algebraic.
Instead we now seek a method to restore exponential convergence to the Fourier
sum inside the worldtube.  
Note in the worldtube method with a source that is zero outside the libration region means that the convergence outside the worldtube is exponential.

\section{Construction of the Extended Effective-Sources}
\label{sec:ees_method}

The slow convergence of the partial sum over Fourier modes of the residual field in Eq.~\eqref{eq:phiR_lm_sum} is reminiscent of the slow convergence of the retarded field calculation when using the standard delta function source \cite{Barack:2008ms}.
In that case the method of \textit{extended homogeneous solutions} was developed in order to restore exponential convergence \cite{Barack:2008ms}.
For a review of the EHS method, see Appendix \ref{apdx:extended_homogeneous_solutions}.
That approach relied crucially on the perturbation away from the particle's worldline being a solution to the vacuum field equations.
As such this approach cannot be applied with an effective-source as it is non-zero in a finite region around the wordline.
Later, in the context of gauge transformations, Ref.~\cite{Hopper:2012ty}, devised the method of \textit{extended particular solutions} as a way restoring exponential convergence when summing over inhomogeneous frequency modes from an extended source.
In this section we present an equivalent approach that also incorporates the puncture scheme through a worldtube method.
We call our approach the method of \emph{extended effective-sources} (EES).

\begin{figure}[htb]
 	\centering
          \includegraphics[width=0.48\textwidth]{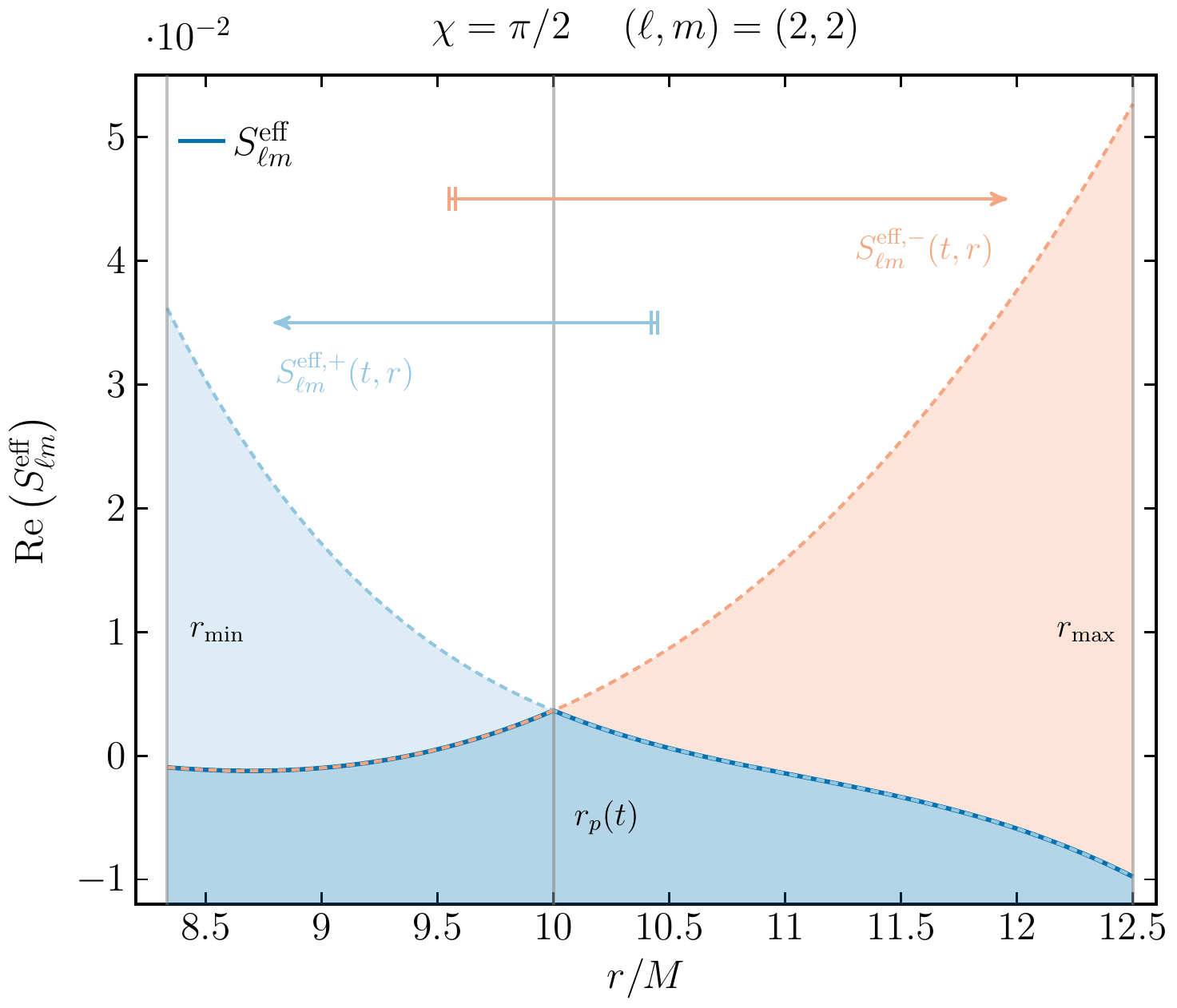}
          \label{fig:ExtendedEffectiveSourceDiagram}
 \caption{A plot illustrating the construction of extended effective-source terms in the
 TD.  Here we present $S^{\rm{eff}}_{\lm}(r)$, $S^{\rm{eff}, +}_{\lm}(r)$ and $S^{\rm{eff}, -}_{\lm}(r)$,
 with $(\ell, m) = (2, 2)$, for a particle at $\chi = \pi/2$ with orbital parameters $p = 10M$ and $e = 0.2$.
 The position of the particle is indicated by $\rp$.
 The function $\tdSeffminus(r)$ is an extension of the effective-source from $\rmin
 \leq r < \rp$ to $\rmax$ whilst $\tdSeffplus(r)$ is an extension of the effective
 source from $\rp < r \leq \rmax$ to $\rmin$.}
     \label{fig:EESConstruction}
\end{figure}

\begin{figure}[htb]
	\vspace{1em}
	\centering
    \includegraphics[width=0.48\textwidth]{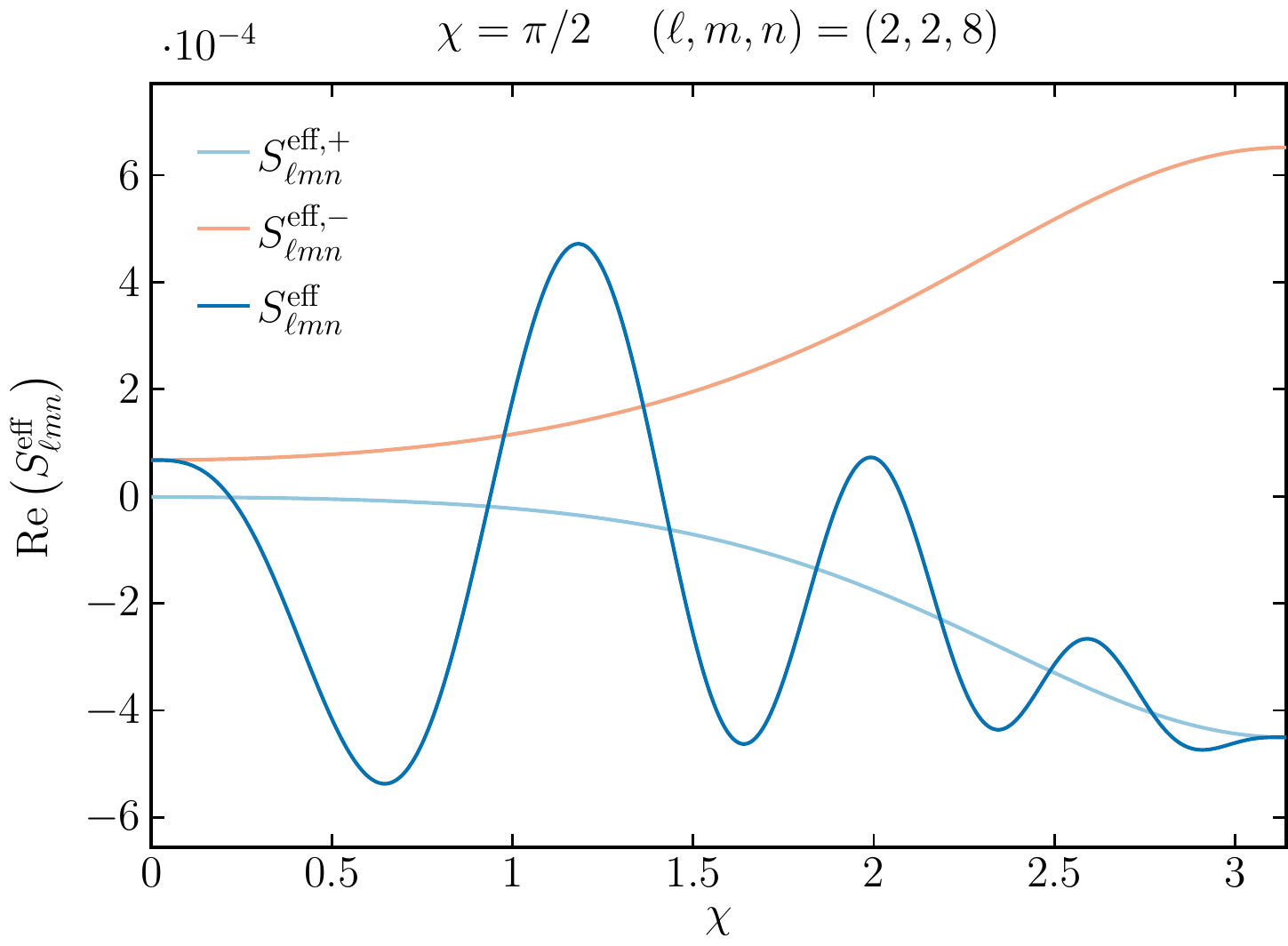}
	\caption{A plot of the Fourier modes of the effective-source $\fdSeff(r)$, and the extended effective-sources$, \fdSeffpm(r)$ for $(\ell, m, n) = (2,2,8)$.
	The effective-source is highly oscillatory with the number of oscillations growing as $|n|$ increases.
	Note the extended effective-source $\fdSeffplus(r)$ coincides with the effective-source at $r=\rmax (\chi=\pi)$ and similarly for $\fdSeffminus(r)$ at 
	$r=\rmin (\chi = 0)$.}
    \label{fig:effective_source_comparison}
\end{figure}

\begin{figure*}[ht!]
	 \centering
     \includegraphics[width=0.98\textwidth]{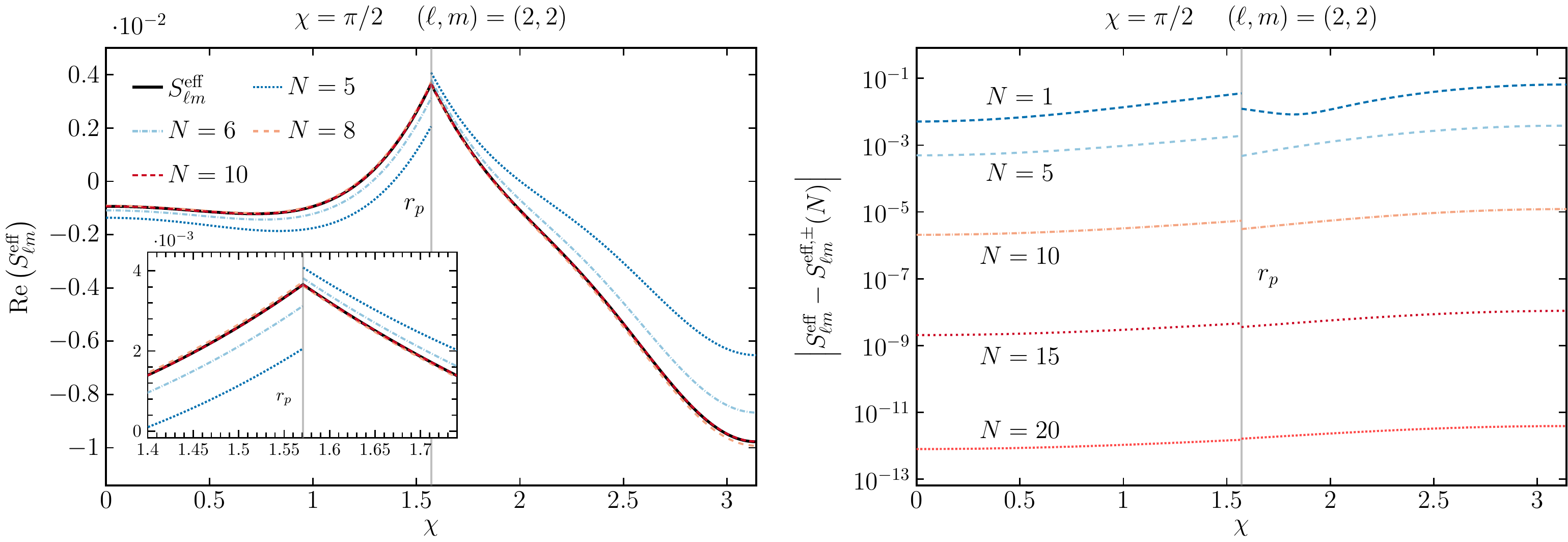}
     \label{fig:seff_ees_22}
	\caption{
	The reconstruction of the TD effective-source using EES for a particle orbiting with $p = 10$ and $e = 0.2$. 
	The left panel shows $S^{\text{eff}}_{\lm}$ for the $(\ell, m) = (2,2)$ mode with $\chi = \pi/2$.  
	Partial sums are computed with \eqn{seff_plusminus} and shown for different $N$, where $N$ is the maximum term included in the partial sum.  
	For comparison purposes we also display (black) the TD effective-source, $S^{\text{eff}}_{22}$ computed from \eqn{time_domain_seff}.
	Exponentially fast converge is manifest with the extended effective-source (bottom pannel).
	The Gibbs phenomenon that previously disrupted convergence --- see Figs.~\ref{fig:effective_source_gibbs} and \ref{fig:effective_source_gibbs_diff} --- is circumvented completely.
	We observe similar results for, e.g.,  puncture field and its radial derivatives.
	}
    \label{fig:effective_source_ees}
\end{figure*}

We begin by analytically extending the effective-source either side of the particle's location to form two smooth functions across the libration region: $\tdSeffplus(t, r)$ and $\tdSeffminus(t, r)$.
For any $t$ and $r$, the true effective-source in the TD is given by
\beq
\tdSeff(t, r) = \tdSeffplus(t, r) \Theta^+(t,r) + \tdSeffminus(t,
r)\Theta^-(t,r).
	\label{eq:actual_fd_seff2}
\eeq
where the Heaviside functions are given by $\Theta^\pm(t,r) =
\Theta[\pm(r-r_p(t))]$.
 The construction of the extended effective-sources is illustrated in \fig{EESConstruction}.
We plot the an example of the Fourier modes of the effective-source and the extended effective-sources in Fig.~\ref{fig:effective_source_comparison}.
In the FD these extended sources transform as
\beq
	\fdSeffpm(r) = \frac{1}{\T}\int^{\T}_{0}\tdSeffpm(t, r)\etplus dt
	\label{eq:seff_plusminus_ft},
\eeq
and the respective series representations of these extended sources in the TD
is then
\beq
	\tdSeffpm(t, r) = \nsum \fdSeffpm(r) \etminus.
	\label{eq:seff_plusminus}
\eeq
We will also need extended puncture fields which we defined in an analogous
way:
\beq
	\tdsingfield(t,r) = \tdsingfieldplus(t,r)\Theta^+(t,r) +
	\tdsingfieldminus(t,r)\Theta^-(t,r)
\label{eq:eq:tdsing_heaviside}
\eeq
The extended punctures can be be expanded as Fourier series via
\beq
	\tdsingfieldpm(t,r) = \sum_{n=-\infty}^\infty \fdsingfieldpm(r) \etminus,
	\label{eq:tdsing_pm}
\eeq
where
\beq
	\fdsingfieldpm(r) = \frac{1}{T_r}\int_0^{T_r} \tdsingfieldpm(t,r)\etplus\,dt.
	\label{eq:fdsing_plusminus}
\eeq
It is useful to note that $\fdsingfieldplus(\rmax) = \fdsingfield(\rmax)$ and $\fdsingfieldminus(\rmin) = \fdsingfield(\rmin)$.
Since $\tdSeffpm(t, r)$ and $\tdsingfieldpm(t,r)$ are smooth functions the Fourier sums in Eqs.~\eqref{eq:seff_plusminus} and \eqref{eq:tdsing_pm} converge exponentially -- see Fig.~\ref{fig:effective_source_ees}.
Furthermore, the FFT algorithm can be employed to evaluate the integrals in \eqref{eq:seff_plusminus_ft} and \eqref{eq:fdsing_plusminus}.
We now proceed using a modified version of the worldtube method presented in
Sec.~\ref{sec:eccentric_worldtube_method} whereby we integrate over the
extended effective-sources and ensure the resulting solution smoothly attaches
to the retarded solution at the worldtube boundaries.

\begin{widetext}	
To calculate the residual field we begin by defining extended regular solutions
$\psi^{\mathcal{R},\pm}_{\lmn}$ via
\beq
	\psi^{\mathcal{R},-}_{\lmn}(r) = \bec
	a_{\lmn}^{h}\psi_{\lmn}^h(r), & r \leq  \rmin \\
	b_{\lmn}^{\infty,-}\psi_{\lmn}^\infty(r) + b_{\lmn}^{h,-}\psi_{\lmn}^{h}(r) + \psi_{\lmn}^{{\rm inh},-}(r), & \rmin < r < \rmax.\\
\eec
	\label{eq:residual_field_ees_minus}
\eeq
and
\beq
	\fdregfieldplus(r) = \bec
	b_{\lmn}^{\infty,+}\psi_{\lmn}^\infty(r) + b_{\lmn}^{h,+}\psi_{\lmn}^{h}(r) + \fdinhfieldplus(r), & \rmin < r < \rmax,\\
	a_{\lmn}^{\infty}\psi_{\lmn}^\infty(r), & r \geq  \rmax.
\eec
	\label{eq:residual_field_ees_plus}
\eeq
where $a_{\lmn}^h$ and $a_{\lmn}^\infty$ are the constants computed by integrating over the true effective-source as given by Eq.~\eqref{eq:a_coefficients}.
\end{widetext}

The extended inhomogeneous solutions are computed via the usual variation of parameter approach such that
\begin{align}
	\psi_{\lmn}^{{\rm inh},\pm}(r) = C^{\infty,\pm}_{\lmn}(r)\psi_{\lmn}^\infty(r) + C^{h,\pm}_{\lmn}(r)\psi_{\lmn}^h(r)
\end{align}
with
\begin{align}
	C_{\lmn}^{\infty,\pm}(r) &=  \int_{\rmin}^r \frac{\psi^h_{\lmn}(r')\fdSeffpm(r')}{W[\psi^h_{\lmn}(r'),\psi^\infty_{\lmn} (r')]}\,dr' 
	\label{eq:extended_eccentric_source_integral_inf} \\
	C_{\lmn}^{h,\pm}(r)      &=  \int_r^{\rmax} \frac{\psi^\infty_{\lmn}(r')\fdSeffpm(r')}{W[\psi^h_{\lmn}(r'),\psi^\infty_{\lmn}(r')]}\,dr'.
	\label{eq:extended_eccentric_source_integral_h}
\end{align}

We find the values of $b^{h,\pm}_{\lmn}$ and $b^{\infty,\pm}_{\lmn}$ by
requiring that $\fdretfieldplus := \fdregfieldplus + \fdsingfieldplus =
\psi_{\lmn}$ at $r=\rmax$ and $\fdretfieldminus := \fdregfieldminus +
\fdsingfieldminus = \psi_{\lmn}$ at $r=\rmin$. This gives
\begin{align}
	b^{\infty,+}_{\lmn} &= \left. \frac{W[\kappa^{\infty,+}_{\lmn}(r)\psi_{\lmn}^\infty(r) - \fdsingfieldplus(r),\psi_{\lmn}^h(r)]}{W[\psi_{\lmn}^\infty(r),\psi_{\lmn}^h(r) ]}\right|_{r=\rmax}, \\
	b^{\infty,-}_{\lmn} &= \left. \frac{W[\kappa^{h,-}_{\lmn}(r)\psi^h_{\lmn}(r) - \fdsingfieldminus(r),\psi^\infty_{\lmn}(r)]}{W[\psi_{\lmn}^h(r),\psi_{\lmn}^{\infty}(r)]} \right|_{r=\rmin},
\end{align}
where 
\begin{align}
	\kappa^{\infty,+}_{\lmn}(r) &:= a^{\infty}_{\lmn} - C^{\infty,+}_{\lmn}(r), \\
	\kappa^{h,-}_{\lmn}(r) &:= a^h_{\lmn}-C^{h,-}_{\lmn}(r).
\end{align}
and 
\begin{align}
	b^{h,+}_{\lmn} 		&= \left. \frac{W[\fdsingfieldplus(r),\psi_{\lmn}^\infty(r)]}{W[\psi_{\lmn}^\infty(r),\psi_{\lmn}^h(r)]}\right|_{r=\rmax}, \\
	b^{h,-}_{\lmn} 		&= \left.\frac{W[\fdsingfieldminus(r),\psi^h_{\lmn}(r)]}{W[\psi_{\lmn}^h(r),\psi_{\lmn}^{\infty}(r)]}\right|_{r=\rmin}.
\end{align}

The time domain residual solution is then given by
\begin{align}
	\tdresfield(t,r) = \tdresfieldplus(t,r)\Theta^+(t,r) + \tdresfieldminus(t,r)\Theta^-(t,r),
	\label{eq:tdresfield}
\end{align}
where
\begin{align}
	\tdresfieldpm(t,r) = \sum_{n=-\infty}^\infty \fdregfieldpm(r) \etminus.
	\label{eq:tdpsipm}
\end{align}
The key result of our method is that Eq.~\eqref{eq:tdpsipm} convergences exponentially to the true residual field.
We empirically demonstrate this in the results section \ref{sec:effective_source_eccentric_results} below.

\section{Implementation and numerical results}
\label{sec:numerical_scheme}

In this section we describe our numerical scheme and present some sample
results.

\subsection{Fourier expansion of the effective-source by convolution}
\label{sec:convolution}

One challenge that arises with our method is the practical calculation of
Fourier transform of the effective-source, as defined in
Eq.~\eqref{eq:fourier_domain_seff}.
For any given radius within the libration region the effective-source is
non-smooth at the time the particle crosses that radius.
Let us define this time to be $t_p(r)$ with $0 \le t_p(r) \le T_r/2$.
This non-smoothness hampers the  efficient calculation of the Fourier modes of
the
effective-source as the Fast Fourier Transform (FFT) will converge very slowly.

An alternative approach that will allow us to employ the FFT is to make use of
the
form of the effective-source given in Eq.~\eqref{eq:actual_fd_seff2}.
We proceed by noting that the Fourier transform of a product of functions is
given by the convolution of the Fourier transforms of the individual functions.
The functions $\tdSeffpm(r)$ are smooth and so the coefficients of their Fourier
expansions, given by Eq.~\eqref{eq:seff_plusminus}, can be computed efficiently
using the FFT algorithm.
The Fourier series of the Heaviside step function is given analytically by:
\begin{align}
	\Theta^\pm(t,r) = \sum_{n = -\infty}^\infty  b^\pm_n(r) e^{-i n \Omega_r t}
\end{align}
where
\begin{align}
	b^{+}_{0}(r)  &= \frac{2 t_p(r)}{T_r}, 
	\label{eq:fft_b0_plus} \\
	b^{-}_{0}(r) &= 1 - \frac{2 t_p(r)}{T_r},
	\label{eq:fft_b0_minus} \\
	b^\pm_n(r)  &= \pm \frac{1}{n\pi}\sin\left(\frac{2 n\pi\, t_p(r)}{T_r}\right)
	\label{eq:fft_b}
\end{align}
We can now calculate the Fourier transform of $\tdSeff$ using its form in
Eq.~\eqref{eq:actual_fd_seff2}, via convolution
\begin{align}
\fdSeff(r) =& \sum_{n'=-\infty}^\infty\left[ b^+_{n' - n}(r)\fdSeffplus(r) +
b^-_{n' - n}(r)\fdSeffminus(r)\right]
	\label{eq:effective_source_convolution}
\end{align}
We find this approach allows us to efficiently calculate $\fdSeff(r)$.

\subsection{Interpolating the Fourier modes of the effective-source using Chebyshev polynomials}
\label{sec:chebyshev_interpolation}

The above technique allows us to efficiently compute the Fourier transform of the modes of the effective-source at a given radius.
In order to compute the associated residual field we need to integrate this source across the radial libration region as in Eqs.~(\ref{eq:eccentric_source_integrals}) and Eqs.~(\ref{eq:extended_eccentric_source_integral_inf})-(\ref{eq:extended_eccentric_source_integral_h}).
We achieve this using an efficient Chebyshev interpolation scheme which we
describe now.

One can express a smooth function $g(x)$ in terms of Chebyshev series,
\beq
	g(x) = \sum^{\cal N}_{k = 1} c^{({\cal N},\, g)}_{k} T_{k}(x),
	\label{eq:chebyshev_expansion}
\eeq
where $\cal N \in \mathbb{Z}$ is the Chebyshev expansion order, $c^{(\cal
N)}_{k}$ are the (spectral) Chebyshev coefficients which implicitly depend on
the order ${\cal N}$, and $T_{k}(x) = \cos [k \arccos (x) ]$ are the Chebyshev
polynomials of the first kind.  The Chebyshev polynomials form an orthonormal
basis on the interval $x \in [-1,1]$ such that
\beq
	\int^{1}_{-1} \frac{T_{n}(x)T_{m}(x)}{\sqrt{1 - x^{2}}}dx = \delta_{nm}.
	\label{eq:chebyshev_orthogonality}
\eeq
While it is not immediately obvious from their definition, the Chebyshev
functions $T_{k}(x)$, are in fact simple polynomials in $x$ of degree $k$.  As
such, these polynomials have $k$ real and distinct zeros within the interval $x
\in [-1,1]$:
\beq
	x_{k} = \cos \left(\frac{\pi(2k + 1)}{2j + 2}\right),
	\quad k = 0,1,\dots,j.
	\label{eq:chebyshev_gauss_points}
\eeq
One can leverage this property to fix the Chebyshev coefficients $c^{(\cal
N)}_{k}$ in order to obtain a global polynomial interpolant for the function
$g(x)$.  We introduce a discrete grid that coinicide with the roots of the
Chebyshev polynomials and require at these points \eqn{chebyshev_expansion} is
exactly equal to the function $g(x)$.  There are several possible choices one
can make for such a grid, and thereby different resultant expressions for the
Chebyshev coefficients.  Here we choose a Chebyshev-Lobatto grid, whereby
\beq
	x_{k} = \cos \left(\frac{\pi k}{{\cal N} - 1}\right),
	\quad k = 0,1,\,\dots,{\cal N} - 1.
	\label{eq:chebyshev_lobatto_points}
\eeq
The reason for this choice is that our worldtube method requires the FT of the  puncture and, by virtue, the resultant effective-source to be evaluated at the boundaries of the worldtube. 
Hence, we require an accurate interpolant at these extrema.  
In this case, the Chebyshev coefficients are given by
\begin{align}
	c^{({\cal N},\, g)}_{k} &= (-1)^{k} 
	\frac{2 - \delta_{k,0} - \delta_{k,\,{\cal N} - 1}}{{\cal N} - 1}
	\nn \\
	&\times \left( \frac{1}{2}[g(x_{0}) + (-1)^{k}g(x_{{\cal N} - 1})] 
	+ \sum^{{\cal N} - 2}_{j = 1} g(x_{j}) T_{j}(x_{k}) \right).
	\label{eq:chebyshev_coefficients_original_form}
\end{align}
Inserting \eqn{chebyshev_lobatto_points} into
\eqn{chebyshev_coefficients_original_form}, one finds 
\beq
	c^{({\cal N},\, g)}_{k} = (-1)^{k} 
	\sqrt{\frac{2 - \delta_{k,0} - \delta_{k,\,{\cal N} - 1}}
	{{\cal N} - 1}}\,
	{\cal G}_{k},
	\label{eq:chebyshev_coefficients_dct}
\eeq
where ${\cal G}_{k}$ are the real Fourier coefficients given by the discrete
cosine transform of type I (DCT-1), 
\begin{align}
	{\cal G}_{k} = \sqrt{\frac{2}{{\cal N} - 1}}
 	\times \Bigg( &\frac{1}{2}[g(x_{0}) + (-1)^{k}g(x_{{\cal N} - 1})] 
 	\nn \\
	&+ \sum^{{\cal N} - 2}_{j = 1} g(x_{j}) 
	\cos \left(\frac{\pi k}{{\cal N} - 1} j \right)
	\Bigg).
	\label{eq:fourier_coefficients_dct}
\end{align}
This is not a surprising result as a Chebyshev series is simply a Fourier
cosine series under a change of variable; but the implication allows one to use
FFT methods in order to compute the nodes for the radial interpolants. 
Furthermore, as the extended effective-sources are $C^{\infty}$-differentiable
then we will expect the Chebyshev coefficients to converge exponentially as
opposed to algebraically as in interpolation with splines.

Thus far, we have described Chebyshev expansion for a generic function defined
on the interval $x \in [-1,1]$.  We now, as an example, specialise to the case of the
interpolation of the effective-source.  The effective-sources are defined only
within the worldtube region that we prescribe to be exactly the libration
region
of the particle, i.e. $r \in [\rmin, \rmax]$.  We map this radial interval onto
the domain of the Chebyshev polynomials via the affine transformation, 
\beq
	x = \frac{2 r - (\rmax + \rmin)}{\rmax - \rmin}.
	\label{eq:radial_coordinate_rescaled}
\eeq
Inverting this relation yields the Chebyshev-Lobatto grid in terms of the
radial coordinate,
\begin{align}
	r_{k} = &\frac{1}{2}(\rmin + \rmax) + \frac{1}{2}(\rmax - \rmin) x_{k},\nn \\
	&k = 0,1,\dots,\,{\cal N} - 1,
	\label{eq:chebyshev_lobatto_rescaled}
\end{align}
allowing us to compute the relevant Chebyshev nodes of our effective-source.
We can then expand the effective-source in \eqn{effective_source_convolution}
in terms of Chebyshev polynomials in the form of \eqn{chebyshev_expansion},
\beq
	\fdSeff(r) = \sum^{\cal N}_{k = 1} c^{({\cal N}, S^{\rm eff})}_{k} T_{k}(x),
	\label{eq:chebyshev_expansion_seff}
\eeq
where the Chebshev coefficients are given by \eqn{chebyshev_coefficients_dct}
with 
\begin{align}
	{\cal G}_{k} = \sqrt{\frac{2}{{\cal N} - 1}}
 	\times \Bigg( &\frac{1}{2}[\fdSeff(\rmin) + (-1)^{k}
 	\fdSeff(\rmax)] 
 	\nn \\
	&+ \sum^{{\cal N} - 2}_{j = 1} \fdSeff(r_{j}) 
	\cos \left(\frac{\pi k}{{\cal N} - 1} j \right)
	\Bigg).
	\label{eq:chebyshev_coefficients_dct_seff}
\end{align}
The same expressions are also applicable to the extended effective-sources
$\fdSeffpm(r)$.

There is one additional subtlety that must be considered when considering the interpolation of the convolved source $\fdSeff(r)$.  
To achieve the desired spectral (exponential) convergence of the interpolant the sampled function needs to be analytic throughout the domain.  
The extended sources, $\fdSeffpm(r)$, are indeed $C^{\infty}$-differentiable and therefore one finds the exponential decay of the Chebyshev coefficients.  
In the form written in \eqn{effective_source_convolution}, however, $\fdSeff(r)$ is only finitely differentiable at $\rmin$ and $\rmax$.  
This is because we have introduced the function $t_{p}(r)$, in Eqs.~(\ref{eq:fft_b0_plus})-(\ref{eq:fft_b}), which as we mentioned previously is non-smooth.
If left in this form, the convergence of the interpolation of $\fdSeff(r)$ would be merely algebraic.  
One can avoid this problem by changing variables from $r \longrightarrow r_{p}(\chi)$, which leads to  $t_{p}(r) \longrightarrow t_{p}(\chi)$, which is entirely smooth throughout the domain.  Note that our expressions 
in \secref{ees_method} will remain in the same as previously, except with a change of variables from 
$r \longrightarrow r_{p}(\chi)$.  As such the integrals in Eqs.~(\ref{eq:fft_b0_plus})-(\ref{eq:fft_b})
become
\begin{align}
	C_{\lmn}^{\infty,\pm}(\chi) &=  \int_{0}^{\chi} \frac{\psi^h_{\lmn}(r_{p}(\chi'))S^\pm_{\lmn}(\chi')}{W[\psi^h_{\lmn}(r_{p}(\chi')),\psi^\infty_{\lmn}	(r_{p}(\chi'))]}\diff{r_{p}}{\chi'}\,d\chi' \label{eq:ees_integral_inf_chi} \\
	C_{\lmn}^{h,\pm}(\chi)      &=  \int_{\chi}^{\pi} \frac{\psi^\infty_{\lmn}(r_{p}(\chi'))S^\pm_{\lmn}(\chi')}{W[\psi^h_{\lmn}(r_{p}(\chi')),\psi^\infty_{\lmn}(r_{p}(\chi'))]}\diff{r_{p}}{\chi'}\,d\chi'.
\end{align}
One can then transform the resultant functions back to radial functions by a simple inversion of \eqn{rp_chi} to yield,
\beq
	\chi(r) = \arccos\left(\frac{p - r}{e\,r}\right).
\eeq

\begin{figure}[t]
	\centering
    \includegraphics[width=0.48\textwidth]{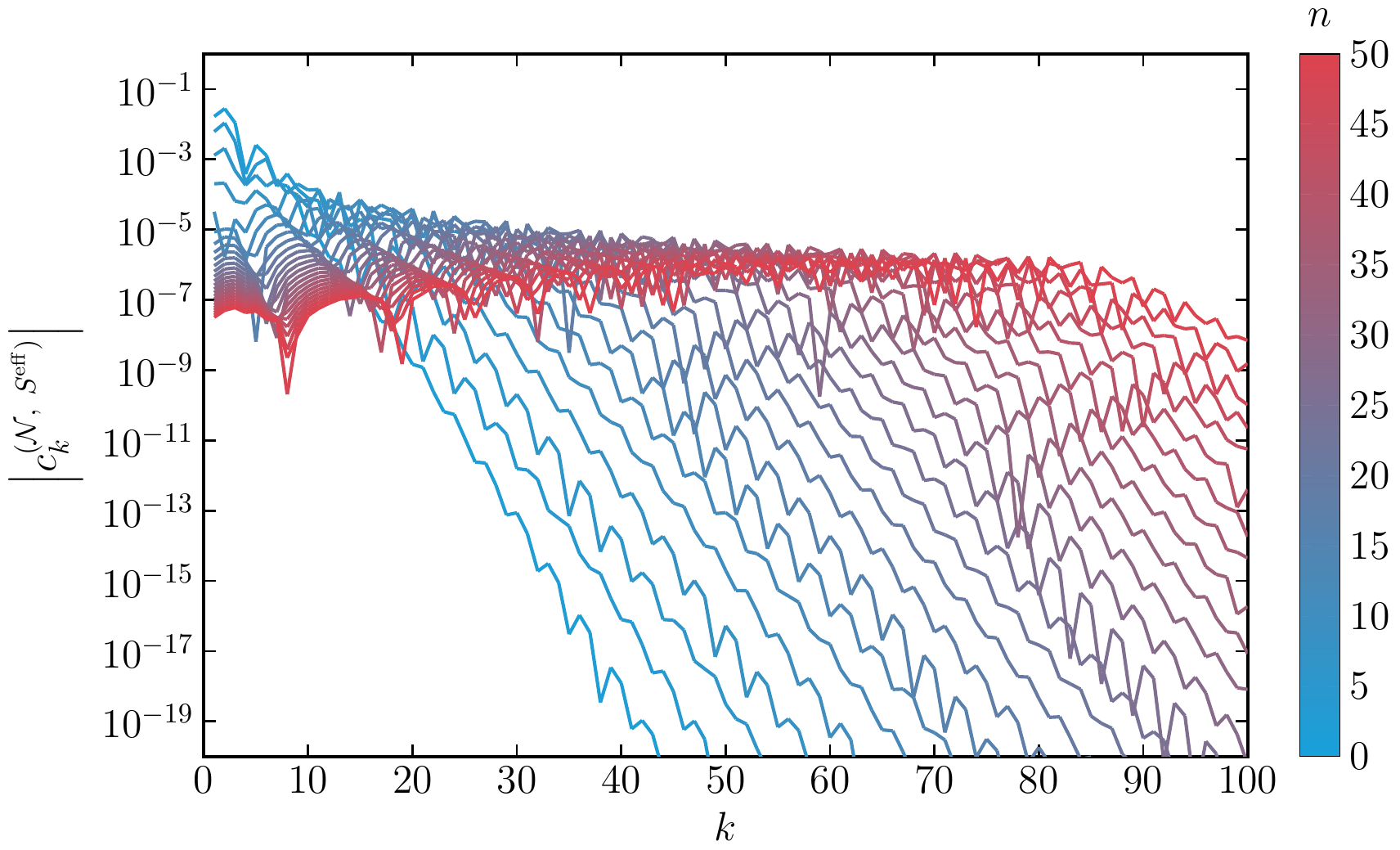}
	\caption{The Chebyshev coefficients, $c_k$, [see Eq.~\eqref{eq:chebyshev_expansion_seff}] for 
	$S^{\rm eff}_{22n}(r)$.
	As the effective-source becomes more oscillatory as $|n|$ grows we find the number, $\mathcal{N}$, 
	of Chebyshev nodes needed to reach a given accuracy increases.}
    \label{fig:chebyshev_coefficients}
\end{figure}

The number of grid points, $\mathcal{N}$, we use in practice depends on the function being  interpolated.
We find the Fourier modes of the effective-source, $\fdSeff$ are very oscillatory  --- see \fig{effective_source_comparison} --- and thus we need a high resolution grid Chebyshev grid  to reach a good accuracy in the interpolation --- see \fig{chebyshev_coefficients}.
On the other hand, the extended functions, e.g., $\fdSeffpm$ do not have oscillations  --- again see \fig{effective_source_comparison} --- and so only low  resolution grid is needed.

\subsection{Numerical Algorithm}

The following steps describe how we compute the regular field in practice.  
Our code is implemented in \textit{Mathematica} and often makes use of the Black Hole Perturbation Toolkit (BHPToolkit) \cite{BHPToolkit}.
\begin{enumerate}
	\item
		Pick a $p$ and $e$ and calculate the orbital frequencies $\Omega_{r}$ and $\Omega_{\varphi}$ using in \eqn{omega_r} and \eqn{omega_phi}. 
		In practice we compute these using the \texttt{KerrGeodesics} package from the BHPToolkit.
		Then, for each $lm$-mode complete the following steps.
	\item 
		Construct the extended effective-sources, $\tdSeffpm(t, r)$ defined by Eq.~\eqref{eq:actual_fd_seff2} and construct the extended punctures $\fdsingfieldpm(r)$ defined by Eq.~\eqref{eq:eq:tdsing_heaviside}
		As these are smooth functions, for a given radius we can use the FFT to calculate all Fourier harmonics with $|n|<50$.
		We calculate this Fourier transform at each radius of Chebyshev-Lobatto grid given by Eq.~\eqref{eq:chebyshev_lobatto_rescaled}.
		The number of points, $\mathcal{N}$, in the radial grid depends on the particular mode --- see Fig.~\ref{fig:chebyshev_coefficients}.
		For each $\lmn$-mode we interpolate the functions using the Chebyshev polynomials as outlined in Sec.~\ref{sec:chebyshev_interpolation}. 
	\item
		For each $\lmn$-mode we now compute the standard effective-source using the convolution formula given in Eq.~\eqref{eq:effective_source_convolution}.
	\item
		 We use \textit{Mathematica}'s \texttt{NDSolve} function to compute the homogeneous solutions, $\fdhminus$ and $\fdhplus$, of the wave equation \eqref{eq:retarded_field_eqn}.
		 The boundary conditions at $\rout = 1000M$ and $\rin = (2+10^{-12})M$ are computed as outlined in Appendix A of \cite{Warburton:2013lea}.
		 For some low-frequency modes the location of the infinity boundary is moved out to ensure convergence of the asymptotic boundary condition series.
	\item
		We compute the $a^\infty_{\lmn}$ and $a^h_{\lmn}$ weighting coefficients using Eqs.~\eqref{eq:a_coefficients}. 
		In practice we evaluate the integrals in Eq.~\eqref{eq:eccentric_source_integrals} using \textit{Mathematica}'s \texttt{NDSolve}.
	\item
		We now compute the $b^{\infty,\pm}_{\lmn}$ and $b^{h,\pm}_{\lmn}$ weighting coefficients in Eqs.~\eqref{eq:b_coefficients}.
		Again, we evaluate the integrals in Eq.~\eqref{eq:extended_eccentric_source_integral_h} using \textit{Mathematica}'s \texttt{NDSolve}.
	\item
		The extended regular fields, $\fdregfieldpm(r)$, are then given by Eqs.~\eqref{eq:residual_field_ees_minus} and \eqref{eq:residual_field_ees_plus}.
		We can now obtain the TD residual field, $\tdresfield(t,r)$, from Eqs.~\eqref{eq:tdresfield} and \eqref{eq:tdpsipm}.
\end{enumerate}

\subsection{Numerical Results}
\label{sec:effective_source_eccentric_results}

\begin{figure}[t!]
	\centering
    \includegraphics[width=0.48\textwidth]{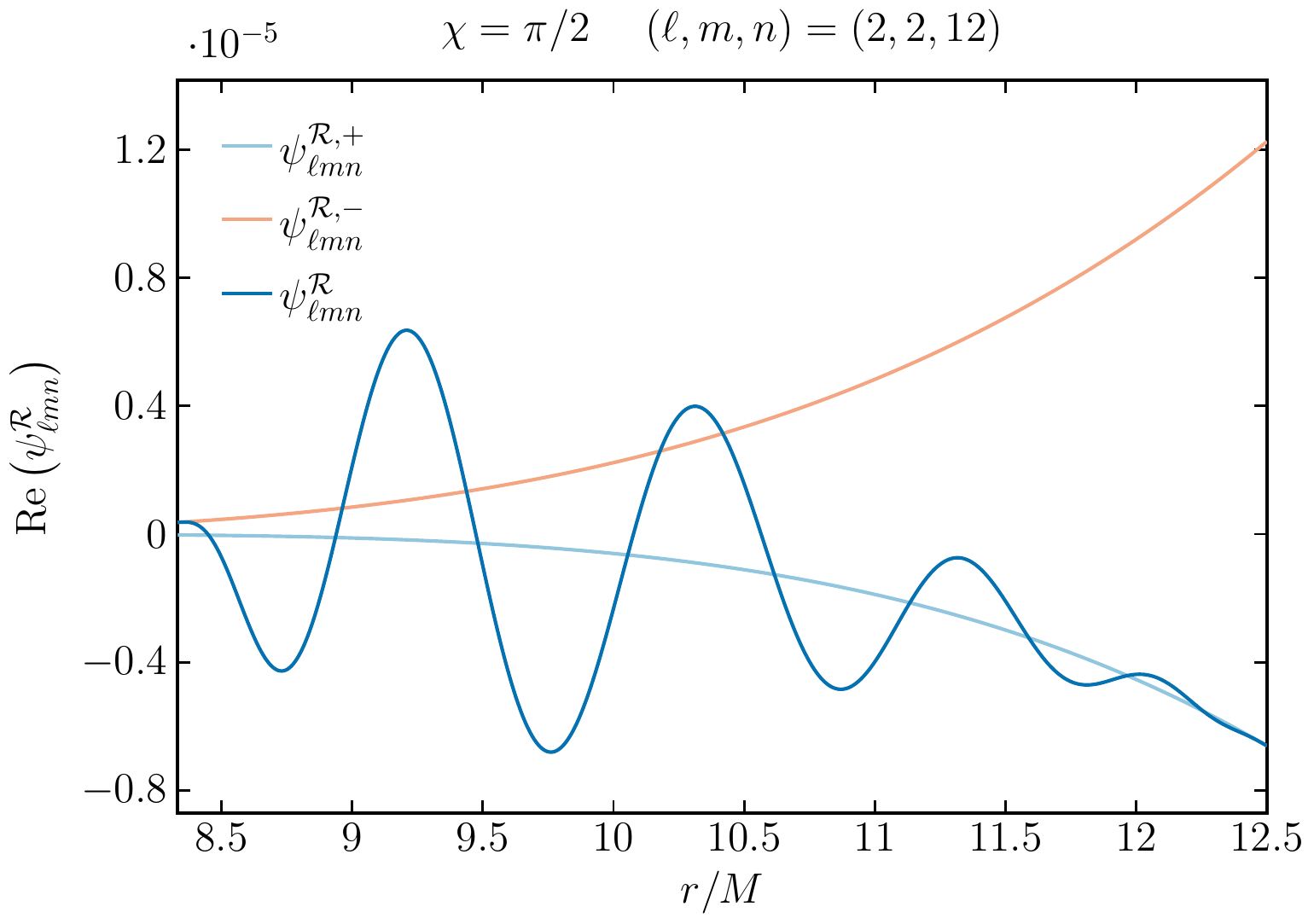}
	\caption{The FD residual field(s), $\fdregfield$, for $(\ell, m, n) = (2,2,12)$.  
	We observe oscillations in the residual field computed with the standard method which result from the behaviour of the source - see \fig{effective_source_comparison}.  
	However, no such oscillations can be seen in the extended regular fields, $\eesplus$ and $\eesminus$.  
	Note how $\eesplus$ and $\eesminus$ coincide exactly with $\fdregfield$ at $\rmax$ and $\rmin$ respectively, the boundaries where the respective effective-sources have been extended from.  
	This is expected since, in formulating our method, we demanded that  $\fdretfieldplus := \fdregfieldplus + \fdsingfieldplus = \psi_{\lmn}$ at $r=\rmax$ and $\fdretfieldminus := \fdregfieldminus + \fdsingfieldminus  = \psi_{\lmn}$ at $r=\rmin$.}
    \label{fig:regular_field_comparison}
\end{figure}

\begin{figure*}[ht!]
	\centering
    \includegraphics[width=0.85\textwidth]{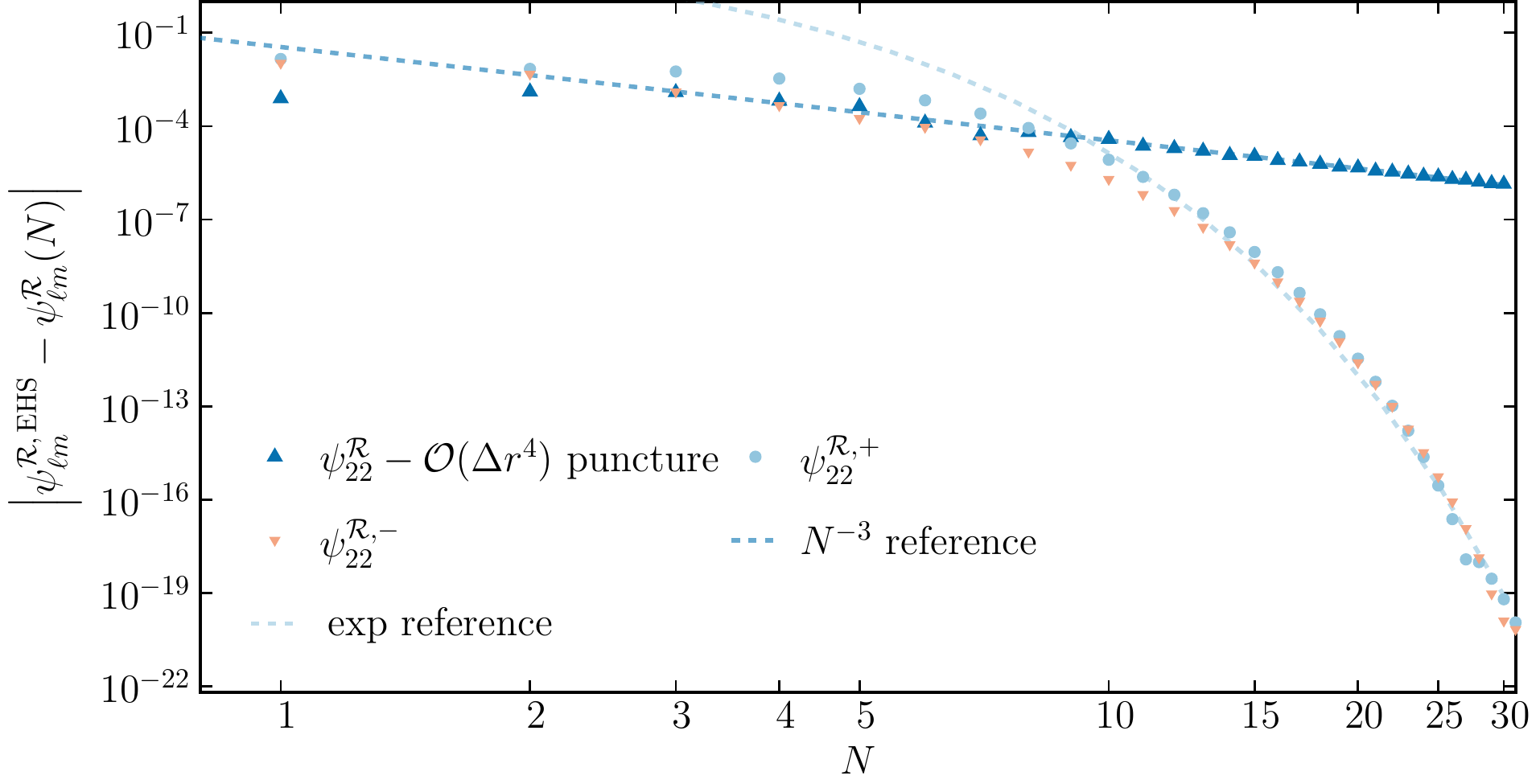}
	\caption
	{
	The convergence of the partial sum of the Fourier modes of the residual field using the standard method and our new extended effective-sources approach.
	Here we present results for $p = 10M$ and $e = 0.2$,  $(l,m) = (2,2)$ mode and the field is computed when $\chi=\pi/2$.
	For each partial sum over the Fourier modes of the residual field we plot the absolute difference compared to the result computed using the method of EHS, where $\tdresfieldEHS(t,r) = \psi^{\rm EHS}_{\lm}(t, r) - \tdsingfield(t,r)$.
	The (dark blue) triangles show the result using the standard method expanded through $\mathcal{O}(\Delta r^4)$ which converges algebraically as $N^{-3}$.
	This data is the same as presented in Fig.~\ref{fig:residual_field_slow_convergence}.
	Using our new extended effective-sources method, the partial sum over the extended regular fields, $\psi^{\mathcal{R},\pm}_{lm}$, converges exponentially to the time-domain result.
	} 
    \label{fig:convergence_summary}
\end{figure*}

The methods outlined above allow us calculate the residual scalar field, $\tdresfield$, using an effective-source method in the frequency domain.
For reference values to compare against we use the method of EHS \cite{Barack:2008ms} to solve Eq.~\eqref{eq:td_retarded_field_eqn} to construct the time-domain retarded field, $\psi_{\lm}(t, r)$, for a given $(l,m)$-mode.
We briefly review the method of EHS in Appendix \ref{apdx:extended_homogeneous_solutions}.
In using the method we compute the homogeneous solutions to frequency domain wave field equation \eqref{eq:retarded_field_eqn} using the \texttt{Teukolsky} package in the BHPToolkit.
This makes the reference EHS calculation of the retarded field completely independent of our effective-source calculation.
To construct the reference residual field in the time-domain we subtract the time-domain puncture from the retarded solution, i.e., $\tdresfieldEHS(t,r) = \psi^{\rm EHS}_{\lm}(t, r) - \tdsingfield(t,r)$.
In computing the retarded field using the method of EHS we use $N_{\rm max} = 50$ Fourier modes.

Before discussing the main results of the EES method we note that some numerical results have been presented in earliest sections.
In Fig.~\ref{fig:effective_source_gibbs} we showed the poor convergence Fourier sum for the effective-source and the radial derivative of the puncture.
We then showed in Fig.~\ref{fig:residual_field_slow_convergence} how using a higher-order puncture can accelerated the convergence of the Fourier sums for the residual field but the convergence remains algebraic.
We gave an example of the Fourier modes of the effective-source, $\fdSeff(r)$, and the extended effective-sources, $\fdSeffpm(r)$, in Fig.~\ref{fig:effective_source_comparison}.
In Fig.~\ref{fig:effective_source_ees} we showed the rapid convergence of the extended effective-sources is exponential whereas the convergence of the standard effective-source is extremely slow.
In Fig.~\ref{fig:chebyshev_coefficients} we discuss the number of Chebyshev nodes needed to interpolate the highly oscillatory $\fdSeff$ and note that more modes are needed for higher $|n|$.

Using the method of EES we can efficiently compute the time-domain residual field by calculating extended residual fields, $\fdregfieldpm(r)$.
We give an example of the Fourier domain residual field computed using the standard method outlined in Sec.~\ref{sec:eccentric_worldtube_method} and the extended residual fields computed using the EES method in Fig.~\ref{fig:regular_field_comparison}.
The standard residual field is found to be highly oscillatory whereas the extended fields are slowly varying.

The time-domain TD residual field, $\tdresfield(t,r)$, can then be constructed using from  Eqs.~\eqref{eq:tdresfield} and \eqref{eq:tdpsipm}.
Our main result is that the Fourier sum in \eqn{tdpsipm} now converges exponentially to the correct value -- see Fig.~\ref{fig:convergence_summary} where we give an example for the $(l,m)=(2,2)$ mode for an orbit with $(p,e) = (10,0.2)$.
We find similar results for other modes and orbital configurations.

\begin{figure}[!t]
	 \centering
         \includegraphics[width=0.48\textwidth]{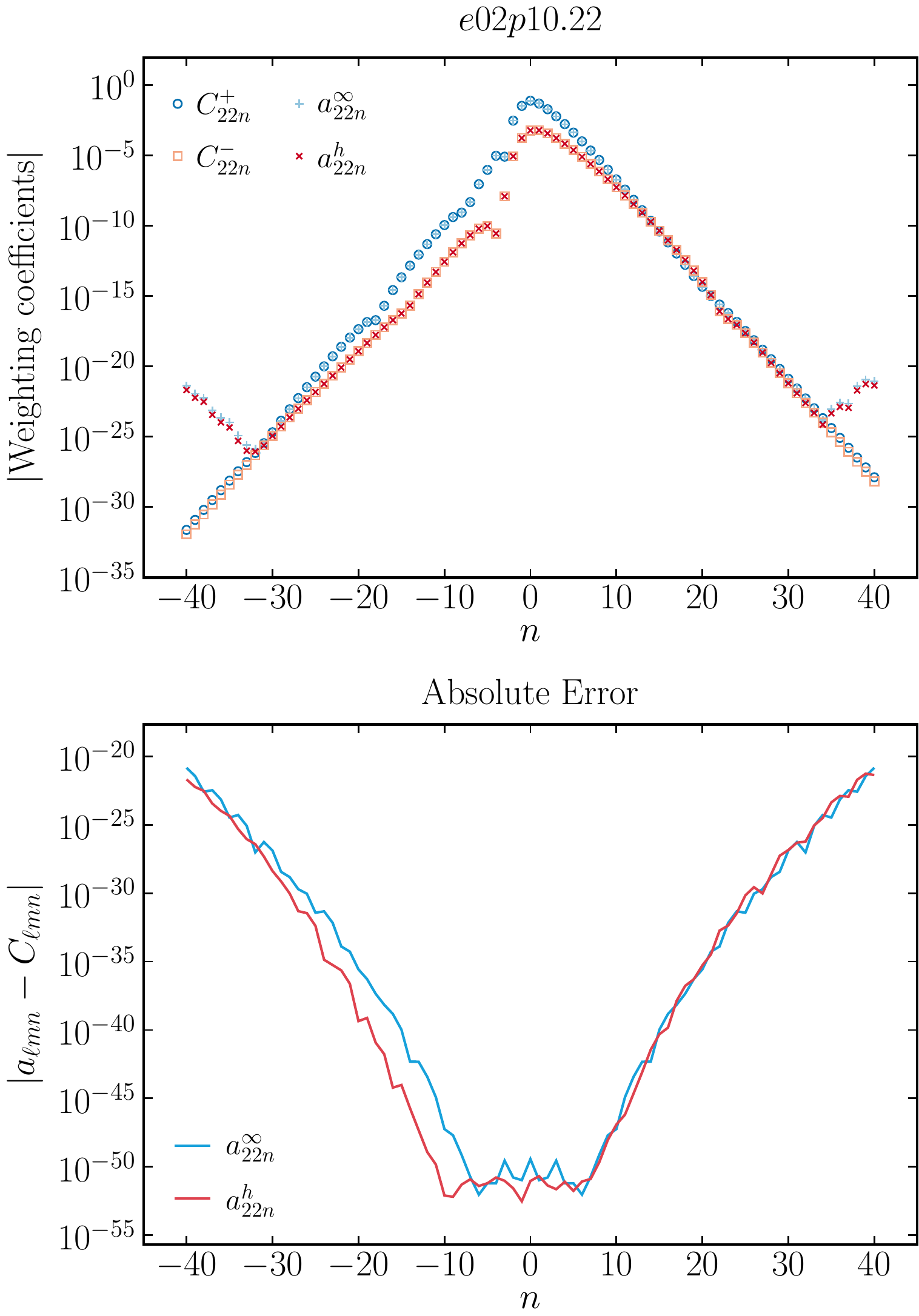}
\caption{
A comparison of the weighting coefficients, $a^{\infty/h}_{\lmn}$ and $C^{+/-}_{\lmn}$, for a particle with orbital parameters $p = 10M$ and $e = 0.2$ with $(\ell, m) = (2, 2)$, computed with EHS and EES respectively.  
The weighting coefficients for successive $n$-modes falls off exponentially when computed away from the peak harmonic. 
For higher $|n|$, the $a^{\infty/h}_{\lmn}$ coefficients reach Nyquist-like notches, beyond which the coefficients increase in magnitude due to alaising of the convolved source term.
}
\label{fig:ehs_coefficients_comparison_22}
\end{figure}

As a further check our results we verify that our code recovers the correct field outside of the worldtube region for a range of orbital configurations and modes.  
In this section we shall only present results for a particle with the same orbital configuration as \fig{convergence_summary}.  In \fig{ehs_coefficients_comparison_22}, we compute the weighting coefficients, $a^{\infty/h}_{\lmn}$, appearing in Eqs.~(\ref{eq:residual_field_ees_plus}) and (\ref{eq:residual_field_ees_minus}) that scale the homogeneous outside the worldtube to recover the retarded field.  
These coefficients are then compared to the equivalent weighting coefficients, $C^{\infty/h}_{\lmn}$, that are computed through EHS as described in Appendix~\ref{apdx:extended_homogeneous_solutions}.  
One observes the weighting coefficients have an exponentially decaying spectrum centered around a peak harmonic until reaching some Nyquist point around $\sim {\cal N}/2$, where ${\cal N}$ is the number of Chebyshev nodes used to interpolate the standard effective-source.  
This aliasing phenomenon can be explained by the nature of the convolved effective-source.  
Specifically, if we recall \fig{effective_source_comparison}, one observes that for increasing $|n|$, the standard effective-source becomes more oscillatory. 
Hence for source harmonics with high $|n|$, one therefore requires more Chebyshev coefficients to ensure the source's oscillatory behaviour is not under-sampled.  
This is seen in \fig{chebyshev_coefficients} as for higher $|n|$, there is an initial accuracy floor is reached until the number of Chebyshev nodes becomes sufficient to suitably capture the oscillatory behaviour of the source.  
From a practical perspective, the Nyquist-like notches only begin once the weighting coefficients are far below machine precision.  
In fact, for the models considered in this work, these points of inflection would only be observable thanks to \textit{Mathematica}’s arbitrary precision arithmetic.
If one was to extend the implementation beyond this work, for a given precision, one could choose to halt the calculation at a given $|n|$ when this limit is reached, with the weighting coefficients presented here being a good measure.

\begin{figure}[!t]
	 \centering
         \includegraphics[width=0.48\textwidth]{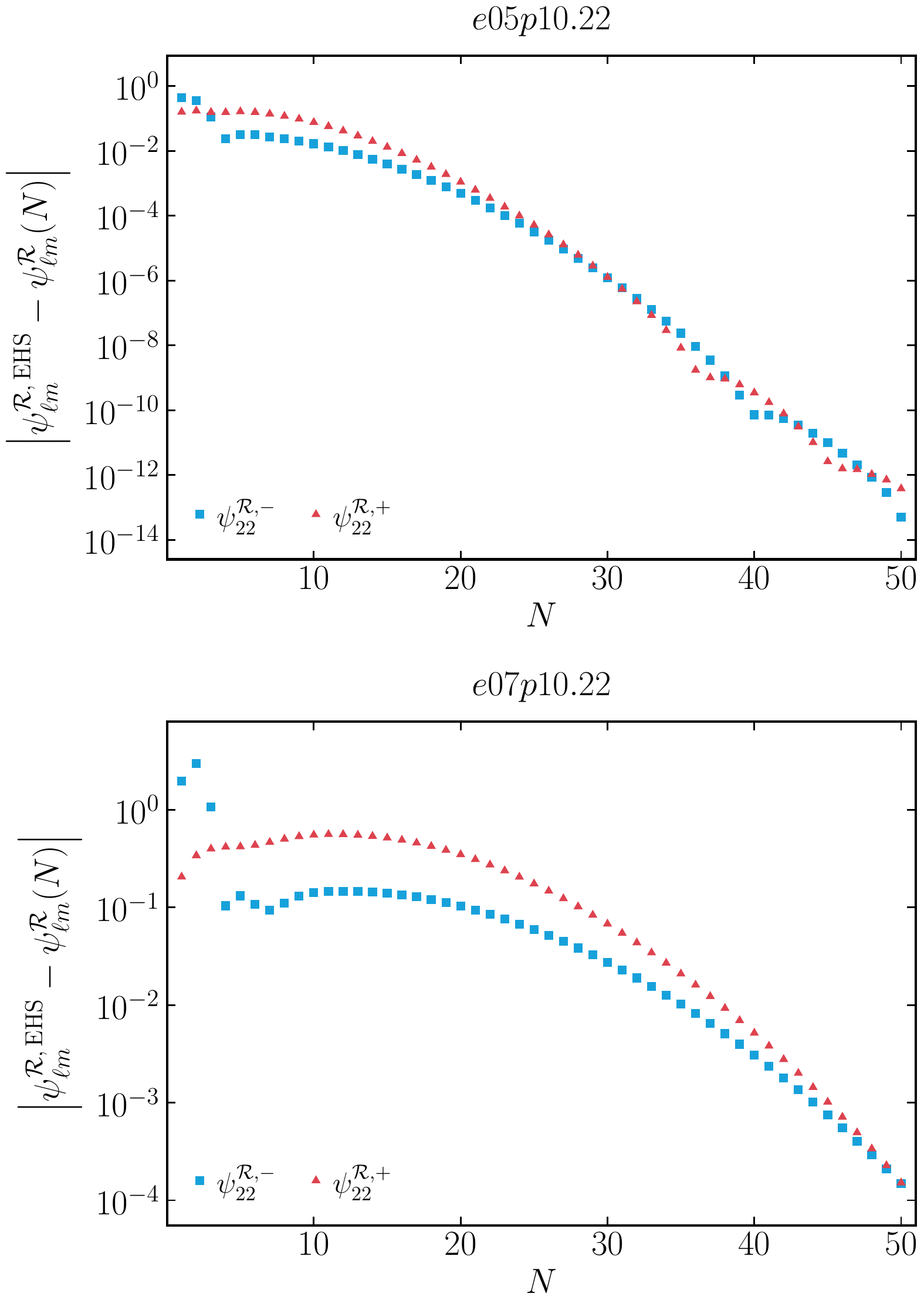}
\caption{
The convergence of the partial sum of the Fourier modes of the residual field calculated through EES for two models with high eccentricities: $e05p10.22$ (top panel) and $e07p10.22$ (bottom panel).
We find the partial sums still converge exponentially to the reference values computed through EES, but find for increasing eccentricity the initial value of the partial sum is more disparate when compared to the EHS reference value and the decay rate slows.  
Therefore for higher eccentricity one would require a higher number of $n$-modes to achieve the same level of accuracy to that of \fig{convergence_summary}.
}
\label{fig:EES_convergences_high_e}
\end{figure}

For all of the models, we find exponential convergence to an absolute error of at least $\sim 10^{-15}$.
We find, as with most FD methods including EHS, that increasing eccentricity leads to a slower exponential decay of the partial sum and hence for larger eccentricities approaching $e = 1$ that our method becomes less practical.
Nonetheless, as we show in Fig.~\ref{fig:EES_convergences_high_e}, the EES method can still handle up to eccentricities of $e \lesssim 0.7$.
In \fig{effective_source_gibbs}, we show the exponential convergence of the partial sum for models with high eccentricities of $e = 0.5$ and $e = 0.7$. 
We give further details of these comparisons in Appendix~\ref{apdx:num_results}.

    Formally our frequency domain approach is valid for all eccentricities.
	Practically the calculation becomes increasingly more difficult as the number of $n$-modes needed to reach a given precision diverges in the $e \longrightarrow 1$ limit.
	For $e \ge 1$, which describes parabolic and hyperbolic orbits, the spectrum of radial harmonics becomes continuous.
	The frequency domain approach can still be used for these orbits but with additional techniques needed -- see \cite{Hopper:2017iyq, Whittall:2023xjp}.
	In more recent works, however, many authors have had no issues reaching eccentricities of $e = 0.8$ \cite{Osburn:2014hoa}.
	
	For our own calculation it is difficult to predetermine the number of $n$-modes from certain set of parameters, but we can say for larger eccentricity for a given semi-latus rectum, the spectrum of the modes broadens around the peak mode which occurs at low $|n|$.
    As one can see from Fig.~\ref{fig:ehs_coefficients_comparison_eccentricity} and Fig.~\ref{fig:EES_convergences_high_e} in Appendix~\ref{apdx:num_results}, the spectrums are not always centered around $n = 0$ and can be asymmetric around the peak harmonic.
    We also observe similar behaviour to \cite{Hopper:2010uv, Warburton:2011hp}, that the negative $n$-modes decay more rapidly than positive $n$-modes, especially for orbits with high eccentricites. See $e07p10.22$ in Fig.~\ref{fig:ehs_coefficients_comparison_eccentricity}.
    
    It is therefore difficult to come up with a precise scaling requirement for $n$-modes in order to produce the prerequisite accuracy. 
	In reality, other similar eccentric orbit codes, see \cite{Hopper:2010uv, Warburton:2011hp}, set some numerical threshold and truncate the $n$-mode calculation when the mode contribution either side of $n = 0$ drops below this threshold.  For example, in \cite{Hopper:2010uv}, the authors calculate energy and angular momentum fluxes, which are related to the weighting coefficients $a^{\infty/h}_{\ell m n}$ and $C^{\infty/h}_{\ell m n}$.
	They find to reach a similar prescribed accuracy with $p = 8.75455, e = 0.764124$, and $(\ell, m) = (2, 2)$, they sum $n$-modes from $n_{\rm min} = 47$ to $n_{\rm max} = 82$\footnote{This is shown in Table III of \cite{Hopper:2010uv}.}.
	
	For $p < 10M$ and a given eccentricity, as we reach further into the strong-field the mode spectrum broadens in a similar manner
	to increasing $e$ for a given $p$, as stated in the previous item.  
	This broadening is especially apparent for ``Zoom-Whirl" type orbits close to the seperatrix \cite{Glampedakis:2002ya}, and as similarly observed in \cite{Warburton:2011hp}.  
	For $p > 10M$, the Fourier spectrum will narrow but if we increase $p \gg 10M$, far away from the strong field regime then other we will encounter difficulties with other aspects of the calculation.  
	The difficulties mainly lie in calculating the homogeneous solutions but, as pointed out in \cite{PanossoMacedo:2022fdi}, this could be avoided by considering a novel hyperboloidal approach with compactification that utilises spectral methods to solve our wave equation.
	This circumvents the difficulty in calculating homogeneous solutions for large-$p$ orbits and the associated issues with variation of parameters.  We would, however, need to reformulate the method of extended effective sources to work with this numerical approach.

\section{Conclusion}
\label{sec:conclusion}
In this paper we have formulated an effective-source approach for eccentric orbits in the frequency domain.
The method allows one to overcome the Gibbs phenomenon and associated slow convergence experienced with naive frequency domain computations that use an effective-source.  
As an example we show how the method can be used to compute the residual scalar field for a compact source moving on an eccentric orbit in a Schwarzschild background.  
Our results were validated against those obtained using an independent implementation of the method of extended homogeneous solutions. 
Crucially, with our new method we find the Fourier modes of the residual scalar field converge exponentially.

We find our method to be reasonably computational efficient with the main bottleneck being the need to interpolate the highly oscillatory standard effective-source.
It would be interesting to explore if the oscillatory nature of the effective-source could be understood analytically and thus removed to leave behind a more slowly varying numerical residual to be interpolated.

The main motivation for the development of our extended effective-sources approach was for application to second-order gravitational self-force calculations.
Although in this paper we only considered a source with support in a finite region around the libration region our method should extend to sources with unbounded support such as appear in the second-order field equations \cite{Miller:2020bft}.
With our new method there is now no remaining obstacle to computing Lorenz-gauge second-order results for eccentric orbits on a Schwarzschild background, although in practice this will be a very significant undertaking.

Looking further to the future, it will be important push second-order GSF calculations to Kerr spacetime.
This will likely first be attempted for circular, equatorial orbits in either the Lorenz gauge \cite{Dolan:2021ijg} or a emerging second-order Teukolsky frameworks \cite{Spiers:2023cip}.
As the majority of EMRIs are expected to be quite eccentric whilst emitting gravitational waves in the LISA band it will be important to apply the results of this paper to second-order perturbations for a body on an eccentric orbit around a Kerr black hole.

\section*{Acknowledgements}
We thank Barry Wardell for informative discussions about the puncture field.
We also thank Leor Barack and Adam Pound for discussions about worldtube
methods.
This work makes use of the Black Hole Perturbation Toolkit \cite{BHPToolkit}.
NW acknowledges support from a Royal Society - Science Foundation Ireland
University Research Fellowship.
This publication has emanated from research conducted with the financial
support of Science Foundation Ireland under Grant numbers 16/RS-URF/3428 and
17/RS-URF-RG/3490.
For the purpose of Open Access, the author has applied a CC BY public copyright
licence to any Author Accepted Manuscript version arising from this submission.

\appendix

\section{Scalar-field puncture fields for eccentric orbits in Schwarzschild spacetime}
\label{apdx:scalar_field_puncture}
In this section, we give our explicit expression for the scalar-field puncture
field, $\psi^{\mathcal{P}}_{\lm}(t, r)$, for the case of an eccentric orbit in
Schwarzschild spacetime.  This puncture is decomposed into spherical
harmonic $\lm$-mode contains all the necessary pieces of
the Detweiler-Whiting singular field required to compute the regular scalar
field and the extended effective-sources $\tdSeffpm(t, r)$.  In the usual
Schwarzschild coordinates, $\tensor*{x}{^{\mu}} = (t, r, \theta, \varphi)$, the
puncture can be found from the expressions in \cite{Heffernan:2017cad} (by
setting the four-acceleration terms $\tensor*{a}{^{t}} = \tensor*{a}{^{r}} =
\tensor*{a}{^{\varphi}} = 0$) and is given through $\Delta r(t)$ by
 \begin{widetext}
\begin{align}
	\psi^{\mathcal{P}}_{\lm}(t, r) 
	=\, &2r\, e^{i m \left[ c(t) -\varphi_{p}(t) \right]}
	D_{m,0}^\ell \sqrt{\frac{\pi}{2 \ell+1}}
	\Bigg\{ \frac{2 K}{\pi
	\sqrt{\rp^2+\mathcal{L}^2}} - \Delta r(t) \Bigg[ \frac{ (2 \ell+1)
	E\, \rp\, \text{sgn}(\Delta r(t)) }{ 2 (\rp-2 M)
    \left( \rp^2+\mathcal{L}^2 \right) } \nn \\
    &- 
    \frac{E \left[\mathcal{L}\, \rp^2 \left(4 M-\left(2-3 E^2\right)
    \rp \right)+\mathcal{L}^3 (4 M-2\rp) \right]- 2E^2 \mathcal{L}\, 
    \rp^3 K}{\pi \mathcal{L}\, \rp (\rp- 2M)
	\left(\rp^2+\mathcal{L}^2\right)^{3/2}} \Bigg] \Bigg\},
\end{align}
\end{widetext}
where
\beq
	c(t) := \Delta r(t)\frac{\tensor*{u}{^{r}}\mathcal{L}\, \rp}{(\rp -
	2M)(\mathcal{L}^{2} + \rp^{2})},
\eeq
$D_{m,0}^{\ell} := D_{m,0}^\ell\left(\pi ,\frac{\pi
}{2},\frac{\pi}{2}\right)$ is the Wigner-D matrix and
$K := \int^{\pi/2}_{0}\left( 1 - w \sin^{2}\theta \right)^{-1/2}d\theta$ and $E
:= \int^{\pi/2}_{0}\left( 1 - w \sin^{2}\theta \right)^{1/2}d\theta$
are the complete elliptic integrals of the first and second kind, respectively,
where $w = \mathcal{L}^2/(\mathcal{L}^2+\rp^2)$.

For our calculations in Sec.~\ref{fig:residual_field_slow_convergence} we also used a higher-order puncture in $\Delta r$ which was original computed in Ref.~\cite{Heffernan:2017cad}.  
We thank Barry Wardell for providing us with the full expression which is given explicitly in the supplemental material accompanying this work \cite{supplemental_material}.

\section{Extended Homogeneous Solutions}
\label{apdx:extended_homogeneous_solutions}

\begin{figure}[htb]
	\centering
    \includegraphics[width=0.48\textwidth]{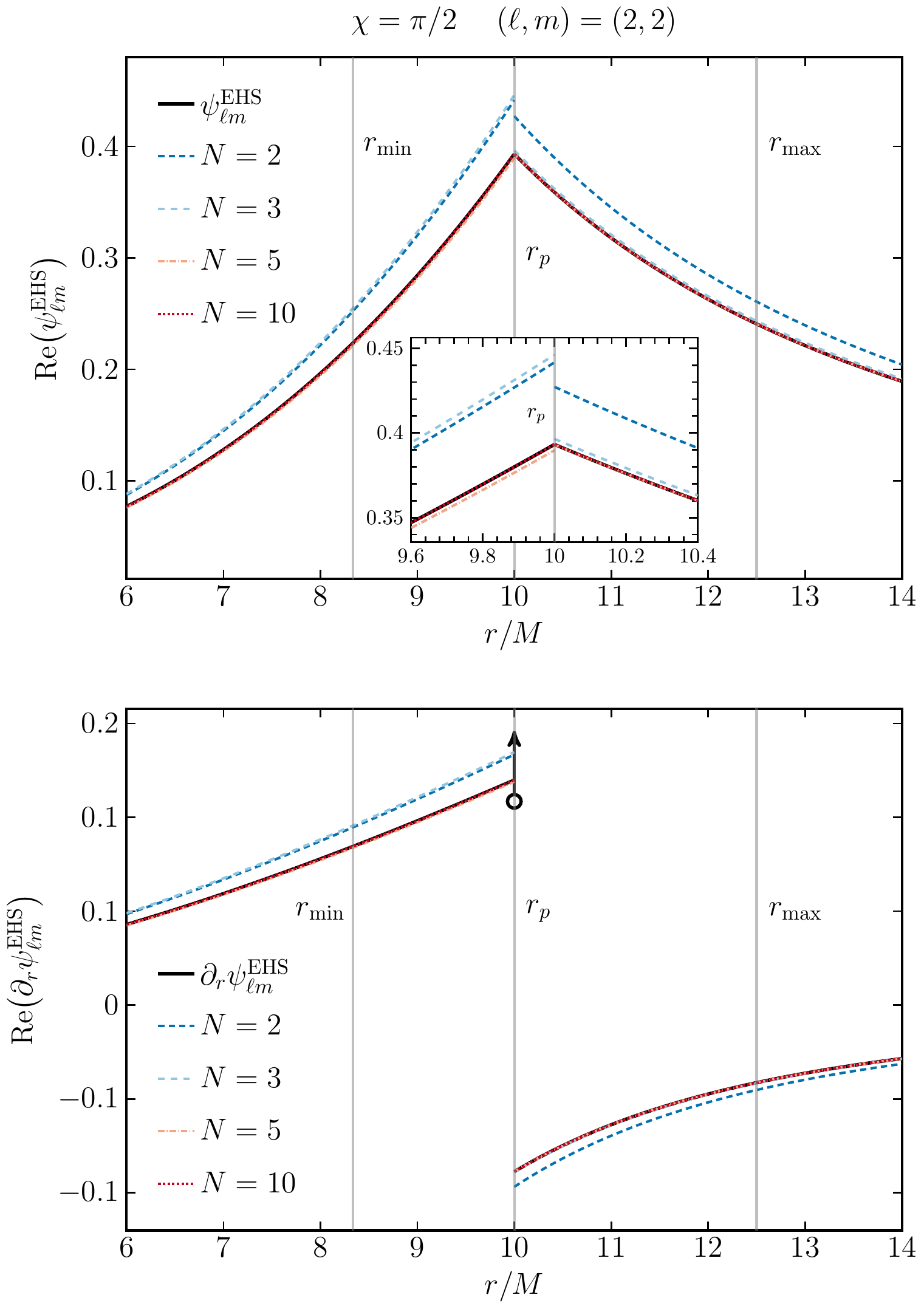}
 	\caption[Fourier reconstruction of the TD retarded monopole field and its
 	derivative using extended homogeneous solutions.]{The reconstruction of the
 	retarded $(2,2$-mode (top panel) and its derivative with respect to
 	$r$ (bottom panel) using EHS. The
 	orbital	parameters are the same as in \fig{convergence_summary}, $p = 10M$ and
 	$e = 0.2$, with $\rmin = 8.3M$ and $\rmax = 12.5M$.  Partial sums are
 	computed with \eqn{ehs_td_sum} and shown for different $N$, 
 	where $N_{\rm max}$ is the maximum of the partial sum.  For comparison
 	purposes we also display (black) the ``full" $(2,2)$-mode solution,
 	$\psi^{\text{EHS}}_{22}$, which is the EHS solution for $N_{\rm max} = 30$ and for our
 	purposes, indistinguishable from the true solution.  Exponentially fast
 	converge is manifest in both the retarded field and its derivative with the
 	Gibbs phenomenon that previously disrupted convergence circumvented completely
 	with EHS.  This figure is inspired by Fig.~3 in
 	\cite{Barack:2008ms}}
         \label{fig:22_mode_ehs}
 \end{figure}
 \begin{figure}[htb]
       \centering
       \includegraphics[width=0.48\textwidth]{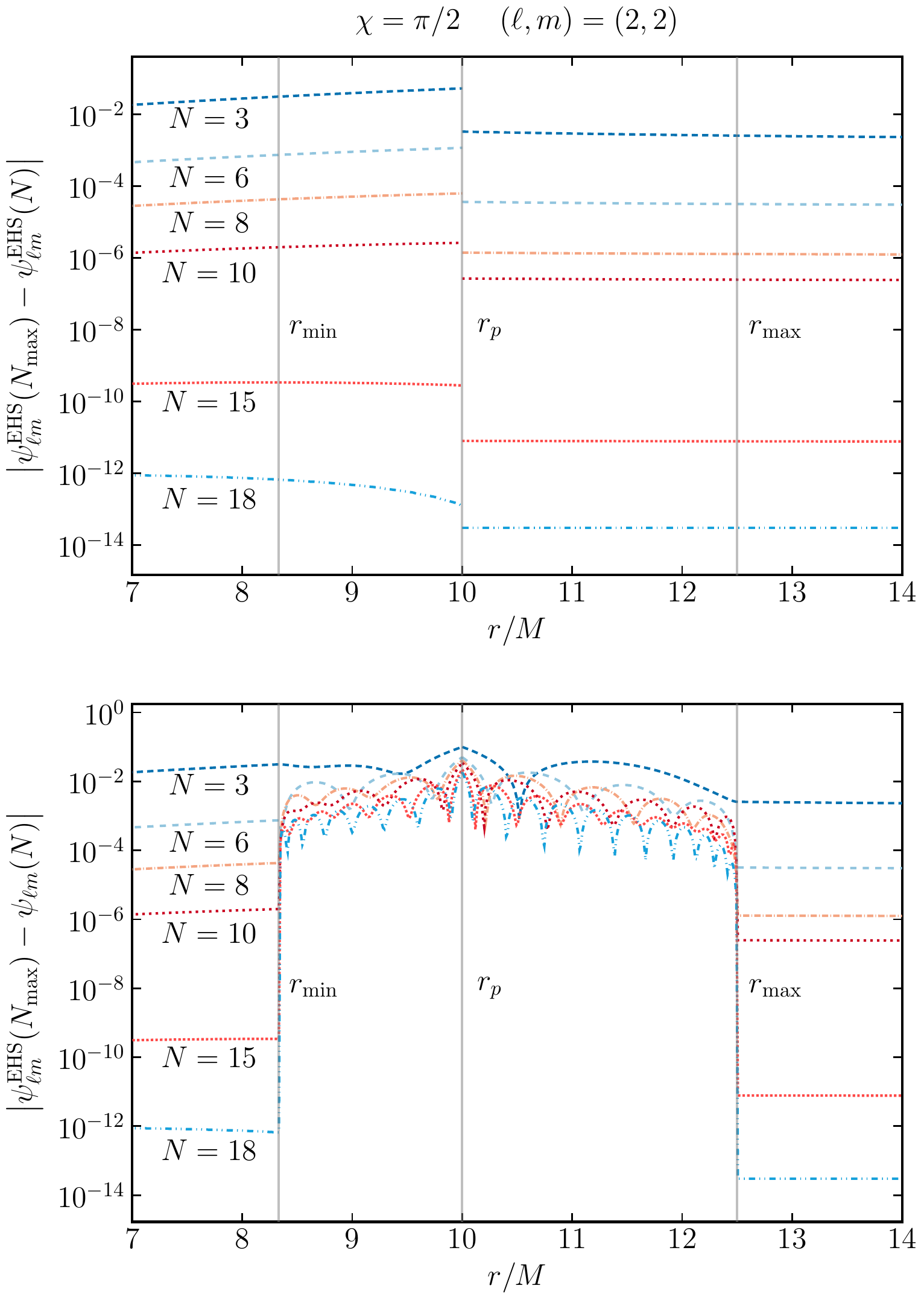}
 	\caption[A comparison of the relative error of Fourier reconstruction of the
 	monopole mode of retarded field using the standard method and EHS.] {A plot of
 	the relative error of the TD reconstruction of the retarded field using the
 	standard method and EHS.  For a particle orbiting with the same parameters as
 	\fig{22_mode_ehs}, i.e. $p = 10M$ and $e = 0.2$ with $\rmin = 8.3M$ and
 	$\rmax = 12.5M$, we compute the retarded
 	field, $\psi_{\lm}$, by summing over modes ranging from $-N_{\text{max}} \leq
 	N \leq N_{\text{max}}$, where $N_{\text{max}} = 30$.  We plot the log of the
 	absolute error between $\psi_{22} (N)$ and $\psi_{22} (N_{\text{max}})$ for a
 	range of $N < N_{\text{max}}$.  For the standard approach (top panel) we see
 	that outside the libration we have exponential convergence.  Inside the
 	libration region the convergence becomes algebraic.  Whereas the EHS method
 	(bottom panel) obtains exponentially converging results throughout the entire
 	domain.  EHS also yields exponentially convergent solutions for
 	$\partial_{r}\psi_{\lm}$ for all points outside and inside the libration
 	region.}
     \label{fig:22_self_ehs}
 \end{figure}
In this appendix we give a brief overview of the method of \emph{extended homogeneous solutions} to computing the retarded field for a point particle moving on an eccentric orbit \cite{Barack:2008ms}.
Firstly, as the name suggests, we consider an extension of the homogeneous solutions $\fdh$ to the entire domain, defined through
\beq
	\tilde{\psi}^{\pm}_{\lmn}(r) := C^{\pm}_{\lmn} \fdh(r),
\eeq
where the coefficients $C^{\pm}_{\lmn}$ are the same as those given in \eqn{eccentric_source_integrals} with $\fdSeffpm(r)$ replaced by $J_{\lmn}(r)$ from \eqn{ft_delta_source}.
One then defines two time-domain extended homogeneous solutions $\tilde{\psi}^{+}_{\lm}$ and $\tilde{\psi}^{-}_{\lm}$ by
\beq
	\tilde{\psi}^{\pm}_{\lm}(t, r) := \sum^{\infty}_{n = -\infty}
	\tilde{\psi}^{\pm}_{\lmn}(r) \etminus.
	\label{eq:ehs_td_sum}
\eeq
While these solutions exist within the entire domain, we emphasise that $\tilde{\psi}^{-}_{\lm}(t, r)$ and $\tilde{\psi}^{+}_{\lm}(t, r)$ are not solutions to the inhomogeneous $n$-mode equation given in \eqn{retarded_field_eqn} in the sourced domain $\rmin \leq r \leq \rmax$ and ordinarily only coincide with $\fdinhfield$ in their respective domains $r < \rmin$ and $r > \rmax$.  
In \cite{Barack:2008ms}, it was shown that the $n$-mode sum given in \eqn{ehs_td_sum} converges exponentially fast in $|n|$ and uniformly in $t$ and $r$ throughout the entire domain.  Furthermore, as $N
\longrightarrow \infty$ in the partial sums,
\beq
	\lim_{r \rightarrow \rp} \tilde{\psi}^{-}_{\lm}(t, r) = \lim_{r \rightarrow
	\rp} \tilde{\psi}^{+}_{\lm}(t, r).
\eeq
Ref.~\cite{Barack:2008ms} argued that the EHS can be used to construct the actual solution to the inhomogeneous wave-equation given in \eqn{retarded_field_eqn} such that $\psi^{inh}_{\lm}(t, r) = \psi^{\text{EHS}}_{\lm}(t, r)$ where
\begin{align}
	\psi^{\text{EHS}}_{\lm}(t, r) :=\, \tilde{\psi}^{+}_{\lm}(t, r)& \Theta[r
	- \rp] \nonumber\\
	&+ \tilde{\psi}^{-}_{\lm}(t, r) \Theta[\rp - r].
	\label{eq:actual_fd_seff}
\end{align}
Their argument is based on analytic continuation. 
Ref.~\cite{Barack:2008ms} also demonstrated this numerically for the example of the scalar-field monopole.
Similar results to \cite{Barack:2008ms} are presented from our own calculations for the $(\ell, m) = (2,2)$ in Figs.~\ref{fig:22_mode_ehs} and \ref{fig:22_self_ehs}.
The EHS treatment has been used extensively in calculations of the first-order self-force in conjunction with the mode-sum method \cite{Akcay:2013wfa,Osburn:2014hoa,vandeMeent:2015lxa,vandeMeent:2017bcc}.  

\section{Further numerical results}
\label{apdx:num_results}

\begin{figure*}[htb]
	\centering
    \includegraphics[width=\textwidth]{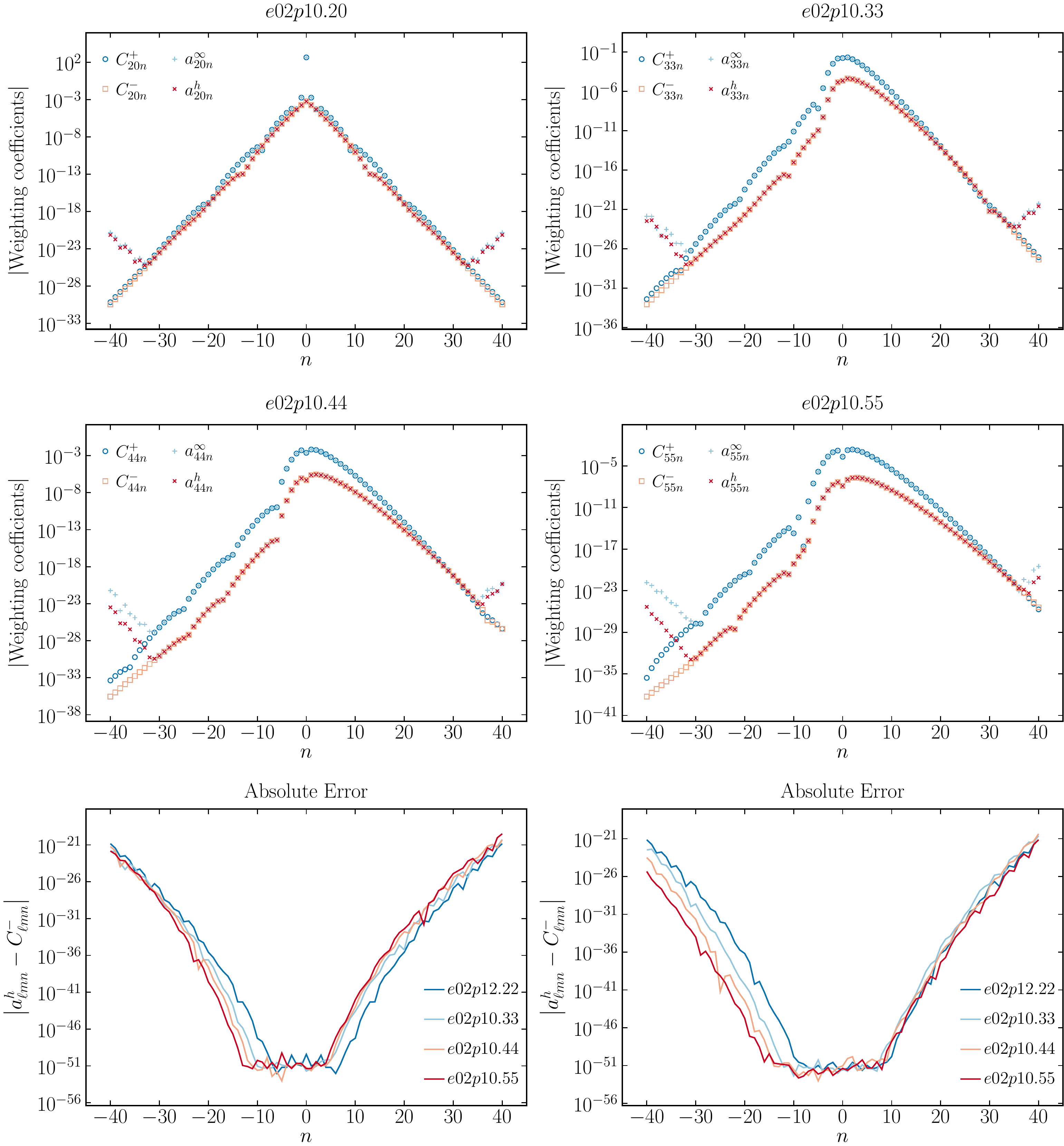}
	\caption{A comparison of the weighting coefficients, $a^{\infty/h}_{\lmn}$ and 
	$C^{+/-}_{\lmn}$, for different $\lm$-modes listed in Table \ref{tbl:labels}
	computed with EHS and EES respectively.  The weighting coefficients for successive 
	$n$-modes falls off exponentially when computed away from the peak harmonic. For
	higher $|n|$, the $a^{\infty/h}_{\lmn}$ coefficients reach Nyquist-like notches,
	beyond which the coefficients increase in magnitude due to alaising of the convolved
	source term.  The minima scale as $\sim {\cal N}/2$, where ${\cal N}$ is the number of
	Chebyshev nodes used to interpolate the effective-source.}
    \label{fig:ehs_coefficients_comparison_modes}
\end{figure*}

\begin{figure*}[htb]
	\centering
    \includegraphics[width=\textwidth]{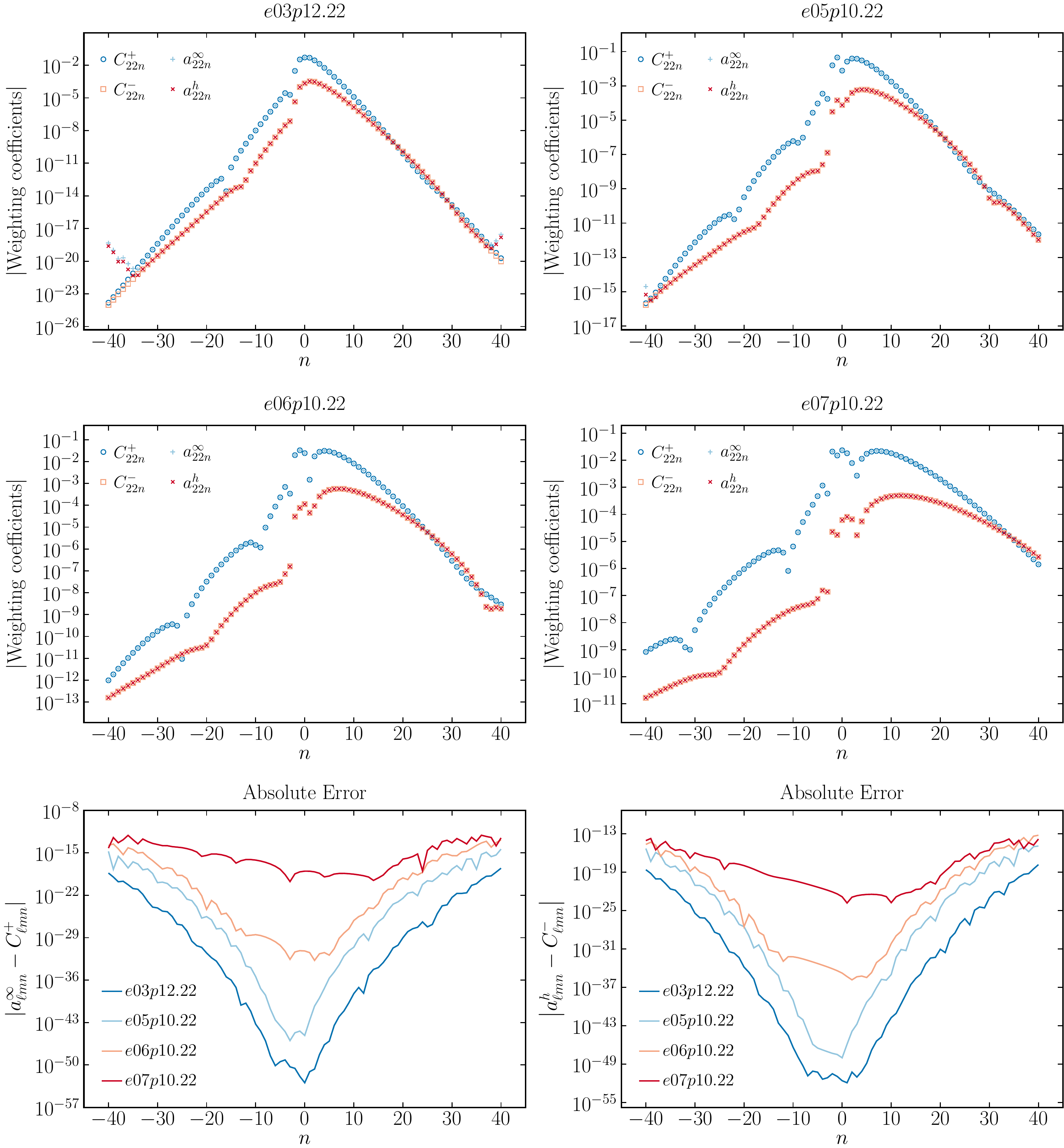}
	\caption{A comparison of the weighting coefficients, $a^{\infty/h}_{\lmn}$ and 
	$C^{\infty/h}_{\lmn}$, for the $(2,2)$-mode orbtial configurations with 
	different eccentricities listed in Table \ref{tbl:labels} computed with EHS and EES respectively.  
	As in Fig.~\ref{fig:ehs_coefficients_comparison_modes}, the weighting coefficients for successive 
	$n$-modes falls off exponentially when computed away from the peak harmonic.  We observe
	that for increasing eccentricity, the decay-rate is slower.}
    \label{fig:ehs_coefficients_comparison_eccentricity}
\end{figure*}

{\renewcommand{\arraystretch}{1.4}
\begin{table}[htb]
\begin{tabular*}{\columnwidth}{c @{\extracolsep{\fill}} c c c}
\hline
\hline
Label & $e$ & $p$ & $(\ell, m)$ \\
\hline
$e02p10.22$ & 0.2 & 10 & (2, 2) \\
$e02p10.20$ & 0.2 & 10 & (2, 0) \\
$e02p10.33$ & 0.2 & 10 & (3, 3) \\
$e02p10.44$ & 0.2 & 10 & (4, 4) \\
$e02p10.55$ & 0.2 & 10 & (5, 5) \\
$e03p12.22$ & 0.3 & 12 & (2, 2) \\
$e05p10.22$ & 0.5 & 10 & (2, 2) \\
$e06p10.22$ & 0.6 & 10 & (2, 2) \\
$e07p10.22$ & 0.7 & 10 & (2, 2) \\
\hline

\hline
\hline
\end{tabular*}
\caption{A table summarising the orbital parameters used in the results in this appendix.} 
\label{tbl:labels}
\end{table}
}
As a further check our results we verify that our code recovers the correct field outside of the worldtube region.
We do this by computing the weighting coefficients of the homogeneous solutions independently using EES and EHS for a variety of orbital configurations, which are listed in Table \ref{tbl:labels}.
In Fig.~\ref{fig:ehs_coefficients_comparison_modes}, we see good agreement between the EES weighting coefficients, $a^{\infty/h}_{\lmn}$ and the EHS weighting coefficients, $C^{\infty/h}_{\lmn}$ for different $\lm$-mode with $e = 0.2$ and $p = 10$.  
Furthermore, in Fig.~\ref{fig:ehs_coefficients_comparison_eccentricity}, we present further comparisons between the weighting coefficients for $(\ell, m) = (2,2)$ mode with different eccentricities.
Similarly to Fig.~\ref{fig:EES_convergences_high_e}, we find that for higher eccentricities the exponential fall-off the weighting coefficients is slower but this is inline with the coefficients calculated using EHS.
As a consequence, we see the Nyquist-like minima appear deeper into the spectra of weighting coefficients.
For example, for $e07p10.22$, these minima do not occur at all for $|n| \leq 40$.  
We also see that that increasing eccentricity decreases the absolute error between the EHS and EES weighting coefficients and whilst this error is still at a tolerable level, further work should investigate how this discrepancy could be improved upon.

\begin{figure*}[htb]
	\centering
    \includegraphics[width=\textwidth]{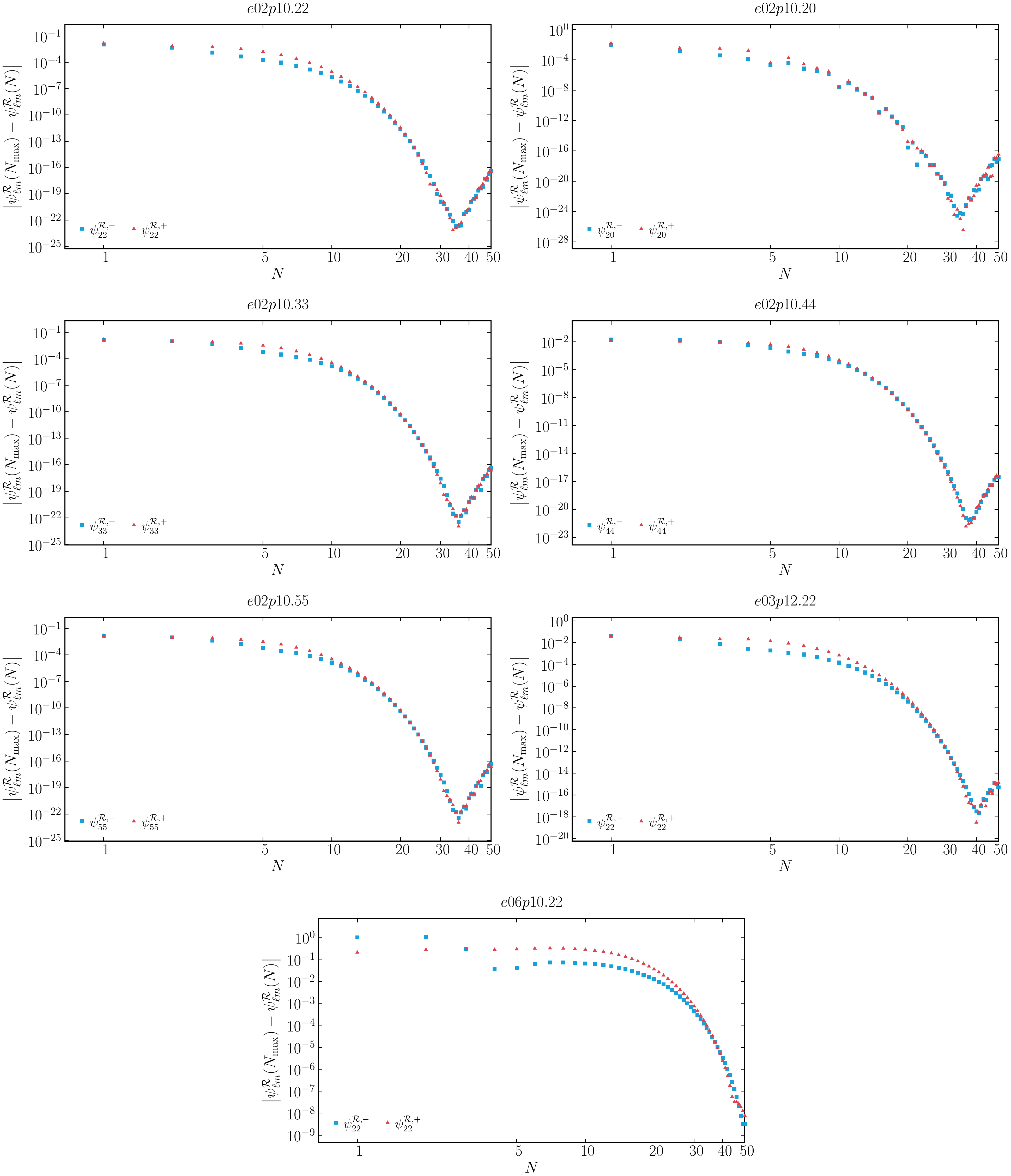}
	\caption{Convergence of the residual fields, $\tdresfieldpm$, computed with EES to a reference 
	value computed with EHS for orbtial configurations listed in Table \ref{tbl:labels}.
	We compute the residual field, $\tdresfieldpm$, by summing over modes ranging from 
	$-N_{\text{max}} \leq N \leq N_{\text{max}}$, where $N_{\text{max}} = 50$.
	In all of our cases we observe exponential convergence of the partial sum of the 
	residual fields constructed with EES to the reference value computed independantly with
	EHS.
	For large $N$, we begin to encounter the same aliasing effect as observed in 
	\fig{ehs_coefficients_comparison_22} and \fig{ehs_coefficients_comparison_modes} for 
	corresponding large $|n|$.}
    \label{fig:ees_convergence_comparison}
\end{figure*}

Finally, in Fig.~\ref{fig:ees_convergence_comparison}, we present convergence of the extended residual fields, $\tdresfieldpm(r)$, for a rest of the orbital configurations in Table \ref{tbl:labels} not yet shown in this paper.  
For ease of comparison, we also again show the orbital configuration, $e02p10.22$, but for the full range of $N$ up to $N_{\rm max} = 50$.  
We find exponential convergence of the partial sum for all of the models listed in Table \ref{tbl:labels} and we find the aliasing effect observed in Figs.~\ref{fig:ehs_coefficients_comparison_22} and \ref{fig:ehs_coefficients_comparison_modes} manifests itself for high $N_{\rm max}$ as notches where the residual fields begin to diverge away from the reference value computed with EHS.  
We have verified the self-convergence of the EHS retarded field and therefore the residual field $\psi^{\mathcal{R}}_{\lm}(N)$ is exponential and thus what we are seeing here is simply limitations of the calculation with the set numerical parameters.  
As we said previously, this could be utilised as a stopping point as the absolute error here is far below machine precision or could be improved even further by increasing the number of Chebyshev nodes used in the interpolation of the effective-source.

\bibliographystyle{apsrev4-2}
\bibliography{Bibliography}

\begin{thebibliography}{81}%
\makeatletter
\providecommand \@ifxundefined [1]{%
 \@ifx{#1\undefined}
}%
\providecommand \@ifnum [1]{%
 \ifnum #1\expandafter \@firstoftwo
 \else \expandafter \@secondoftwo
 \fi
}%
\providecommand \@ifx [1]{%
 \ifx #1\expandafter \@firstoftwo
 \else \expandafter \@secondoftwo
 \fi
}%
\providecommand \natexlab [1]{#1}%
\providecommand \enquote  [1]{``#1''}%
\providecommand \bibnamefont  [1]{#1}%
\providecommand \bibfnamefont [1]{#1}%
\providecommand \citenamefont [1]{#1}%
\providecommand \href@noop [0]{\@secondoftwo}%
\providecommand \href [0]{\begingroup \@sanitize@url \@href}%
\providecommand \@href[1]{\@@startlink{#1}\@@href}%
\providecommand \@@href[1]{\endgroup#1\@@endlink}%
\providecommand \@sanitize@url [0]{\catcode `\\12\catcode `\$12\catcode
  `\&12\catcode `\#12\catcode `\^12\catcode `\_12\catcode `\%12\relax}%
\providecommand \@@startlink[1]{}%
\providecommand \@@endlink[0]{}%
\providecommand \url  [0]{\begingroup\@sanitize@url \@url }%
\providecommand \@url [1]{\endgroup\@href {#1}{\urlprefix }}%
\providecommand \urlprefix  [0]{URL }%
\providecommand \Eprint [0]{\href }%
\providecommand \doibase [0]{https://doi.org/}%
\providecommand \selectlanguage [0]{\@gobble}%
\providecommand \bibinfo  [0]{\@secondoftwo}%
\providecommand \bibfield  [0]{\@secondoftwo}%
\providecommand \translation [1]{[#1]}%
\providecommand \BibitemOpen [0]{}%
\providecommand \bibitemStop [0]{}%
\providecommand \bibitemNoStop [0]{.\EOS\space}%
\providecommand \EOS [0]{\spacefactor3000\relax}%
\providecommand \BibitemShut  [1]{\csname bibitem#1\endcsname}%
\let\auto@bib@innerbib\@empty
\bibitem [{\citenamefont {Abbott}\ \emph
  {et~al.}(2016{\natexlab{a}})\citenamefont {Abbott} \emph
  {et~al.}}]{LIGOScientific:2016vbw}%
  \BibitemOpen
  \bibfield  {author} {\bibinfo {author} {\bibfnamefont {B.~P.}\ \bibnamefont
  {Abbott}} \emph {et~al.} (\bibinfo {collaboration} {LIGO Scientific,
  Virgo}),\ }\href {https://doi.org/10.1103/PhysRevD.93.122003} {\bibfield
  {journal} {\bibinfo  {journal} {Phys. Rev. D}\ }\textbf {\bibinfo {volume}
  {93}},\ \bibinfo {pages} {122003} (\bibinfo {year} {2016}{\natexlab{a}})},\
  \Eprint {https://arxiv.org/abs/1602.03839} {arXiv:1602.03839 [gr-qc]}
  \BibitemShut {NoStop}%
\bibitem [{\citenamefont {Abbott}\ \emph
  {et~al.}(2016{\natexlab{b}})\citenamefont {Abbott} \emph
  {et~al.}}]{LIGOScientific:2016sjg}%
  \BibitemOpen
  \bibfield  {author} {\bibinfo {author} {\bibfnamefont {B.~P.}\ \bibnamefont
  {Abbott}} \emph {et~al.} (\bibinfo {collaboration} {LIGO Scientific,
  Virgo}),\ }\href {https://doi.org/10.1103/PhysRevLett.116.241103} {\bibfield
  {journal} {\bibinfo  {journal} {Phys. Rev. Lett.}\ }\textbf {\bibinfo
  {volume} {116}},\ \bibinfo {pages} {241103} (\bibinfo {year}
  {2016}{\natexlab{b}})},\ \Eprint {https://arxiv.org/abs/1606.04855}
  {arXiv:1606.04855 [gr-qc]} \BibitemShut {NoStop}%
\bibitem [{\citenamefont {Abbott}\ \emph
  {et~al.}(2017{\natexlab{a}})\citenamefont {Abbott} \emph
  {et~al.}}]{LIGOScientific:2017bnn}%
  \BibitemOpen
  \bibfield  {author} {\bibinfo {author} {\bibfnamefont {B.~P.}\ \bibnamefont
  {Abbott}} \emph {et~al.} (\bibinfo {collaboration} {LIGO Scientific,
  VIRGO}),\ }\href {https://doi.org/10.1103/PhysRevLett.118.221101} {\bibfield
  {journal} {\bibinfo  {journal} {Phys. Rev. Lett.}\ }\textbf {\bibinfo
  {volume} {118}},\ \bibinfo {pages} {221101} (\bibinfo {year}
  {2017}{\natexlab{a}})},\ \bibinfo {note} {[Erratum: Phys.Rev.Lett. 121,
  129901 (2018)]},\ \Eprint {https://arxiv.org/abs/1706.01812}
  {arXiv:1706.01812 [gr-qc]} \BibitemShut {NoStop}%
\bibitem [{\citenamefont {Abbott}\ \emph
  {et~al.}(2017{\natexlab{b}})\citenamefont {Abbott} \emph
  {et~al.}}]{LIGOScientific:2017vox}%
  \BibitemOpen
  \bibfield  {author} {\bibinfo {author} {\bibfnamefont {B.~. P.~.}\
  \bibnamefont {Abbott}} \emph {et~al.} (\bibinfo {collaboration} {LIGO
  Scientific, Virgo}),\ }\href {https://doi.org/10.3847/2041-8213/aa9f0c}
  {\bibfield  {journal} {\bibinfo  {journal} {Astrophys. J. Lett.}\ }\textbf
  {\bibinfo {volume} {851}},\ \bibinfo {pages} {L35} (\bibinfo {year}
  {2017}{\natexlab{b}})},\ \Eprint {https://arxiv.org/abs/1711.05578}
  {arXiv:1711.05578 [astro-ph.HE]} \BibitemShut {NoStop}%
\bibitem [{\citenamefont {Abbott}\ \emph
  {et~al.}(2017{\natexlab{c}})\citenamefont {Abbott} \emph
  {et~al.}}]{LIGOScientific:2017ycc}%
  \BibitemOpen
  \bibfield  {author} {\bibinfo {author} {\bibfnamefont {B.~P.}\ \bibnamefont
  {Abbott}} \emph {et~al.} (\bibinfo {collaboration} {LIGO Scientific,
  Virgo}),\ }\href {https://doi.org/10.1103/PhysRevLett.119.141101} {\bibfield
  {journal} {\bibinfo  {journal} {Phys. Rev. Lett.}\ }\textbf {\bibinfo
  {volume} {119}},\ \bibinfo {pages} {141101} (\bibinfo {year}
  {2017}{\natexlab{c}})},\ \Eprint {https://arxiv.org/abs/1709.09660}
  {arXiv:1709.09660 [gr-qc]} \BibitemShut {NoStop}%
\bibitem [{\citenamefont {Abbott}\ \emph
  {et~al.}(2017{\natexlab{d}})\citenamefont {Abbott} \emph
  {et~al.}}]{LIGOScientific:2017vwq}%
  \BibitemOpen
  \bibfield  {author} {\bibinfo {author} {\bibfnamefont {B.~P.}\ \bibnamefont
  {Abbott}} \emph {et~al.} (\bibinfo {collaboration} {LIGO Scientific,
  Virgo}),\ }\href {https://doi.org/10.1103/PhysRevLett.119.161101} {\bibfield
  {journal} {\bibinfo  {journal} {Phys. Rev. Lett.}\ }\textbf {\bibinfo
  {volume} {119}},\ \bibinfo {pages} {161101} (\bibinfo {year}
  {2017}{\natexlab{d}})},\ \Eprint {https://arxiv.org/abs/1710.05832}
  {arXiv:1710.05832 [gr-qc]} \BibitemShut {NoStop}%
\bibitem [{\citenamefont {Abbott}\ \emph
  {et~al.}(2020{\natexlab{a}})\citenamefont {Abbott} \emph
  {et~al.}}]{LIGOScientific:2020aai}%
  \BibitemOpen
  \bibfield  {author} {\bibinfo {author} {\bibfnamefont {B.~P.}\ \bibnamefont
  {Abbott}} \emph {et~al.} (\bibinfo {collaboration} {LIGO Scientific,
  Virgo}),\ }\href {https://doi.org/10.3847/2041-8213/ab75f5} {\bibfield
  {journal} {\bibinfo  {journal} {Astrophys. J. Lett.}\ }\textbf {\bibinfo
  {volume} {892}},\ \bibinfo {pages} {L3} (\bibinfo {year}
  {2020}{\natexlab{a}})},\ \Eprint {https://arxiv.org/abs/2001.01761}
  {arXiv:2001.01761 [astro-ph.HE]} \BibitemShut {NoStop}%
\bibitem [{\citenamefont {Abbott}\ \emph {et~al.}(2021)\citenamefont {Abbott}
  \emph {et~al.}}]{LIGOScientific:2021qlt}%
  \BibitemOpen
  \bibfield  {author} {\bibinfo {author} {\bibfnamefont {R.}~\bibnamefont
  {Abbott}} \emph {et~al.} (\bibinfo {collaboration} {LIGO Scientific, KAGRA,
  VIRGO}),\ }\href {https://doi.org/10.3847/2041-8213/ac082e} {\bibfield
  {journal} {\bibinfo  {journal} {Astrophys. J. Lett.}\ }\textbf {\bibinfo
  {volume} {915}},\ \bibinfo {pages} {L5} (\bibinfo {year} {2021})},\ \Eprint
  {https://arxiv.org/abs/2106.15163} {arXiv:2106.15163 [astro-ph.HE]}
  \BibitemShut {NoStop}%
\bibitem [{\citenamefont {Amaro-Seoane}\ \emph {et~al.}(2015)\citenamefont
  {Amaro-Seoane}, \citenamefont {Gair}, \citenamefont {Pound}, \citenamefont
  {Hughes},\ and\ \citenamefont {Sopuerta}}]{Amaro-Seoane:2014ela}%
  \BibitemOpen
  \bibfield  {author} {\bibinfo {author} {\bibfnamefont {P.}~\bibnamefont
  {Amaro-Seoane}}, \bibinfo {author} {\bibfnamefont {J.~R.}\ \bibnamefont
  {Gair}}, \bibinfo {author} {\bibfnamefont {A.}~\bibnamefont {Pound}},
  \bibinfo {author} {\bibfnamefont {S.~A.}\ \bibnamefont {Hughes}},\ and\
  \bibinfo {author} {\bibfnamefont {C.~F.}\ \bibnamefont {Sopuerta}},\ }\href
  {https://doi.org/10.1088/1742-6596/610/1/012002} {\bibfield  {journal}
  {\bibinfo  {journal} {J. Phys. Conf. Ser.}\ }\textbf {\bibinfo {volume}
  {610}},\ \bibinfo {pages} {012002} (\bibinfo {year} {2015})},\ \Eprint
  {https://arxiv.org/abs/1410.0958} {arXiv:1410.0958 [astro-ph.CO]}
  \BibitemShut {NoStop}%
\bibitem [{\citenamefont {Barack}\ \emph {et~al.}(2019)\citenamefont {Barack}
  \emph {et~al.}}]{Barack:2018yly}%
  \BibitemOpen
  \bibfield  {author} {\bibinfo {author} {\bibfnamefont {L.}~\bibnamefont
  {Barack}} \emph {et~al.},\ }\href {https://doi.org/10.1088/1361-6382/ab0587}
  {\bibfield  {journal} {\bibinfo  {journal} {Class. Quant. Grav.}\ }\textbf
  {\bibinfo {volume} {36}},\ \bibinfo {pages} {143001} (\bibinfo {year}
  {2019})},\ \Eprint {https://arxiv.org/abs/1806.05195} {arXiv:1806.05195
  [gr-qc]} \BibitemShut {NoStop}%
\bibitem [{\citenamefont {Barack}\ and\ \citenamefont
  {Cutler}(2007)}]{Barack:2006pq}%
  \BibitemOpen
  \bibfield  {author} {\bibinfo {author} {\bibfnamefont {L.}~\bibnamefont
  {Barack}}\ and\ \bibinfo {author} {\bibfnamefont {C.}~\bibnamefont
  {Cutler}},\ }\href {https://doi.org/10.1103/PhysRevD.75.042003} {\bibfield
  {journal} {\bibinfo  {journal} {Phys. Rev. D}\ }\textbf {\bibinfo {volume}
  {75}},\ \bibinfo {pages} {042003} (\bibinfo {year} {2007})},\ \Eprint
  {https://arxiv.org/abs/gr-qc/0612029} {arXiv:gr-qc/0612029} \BibitemShut
  {NoStop}%
\bibitem [{\citenamefont {Gair}\ \emph {et~al.}(2004)\citenamefont {Gair},
  \citenamefont {Barack}, \citenamefont {Creighton}, \citenamefont {Cutler},
  \citenamefont {Larson}, \citenamefont {Phinney},\ and\ \citenamefont
  {Vallisneri}}]{Gair:2004iv}%
  \BibitemOpen
  \bibfield  {author} {\bibinfo {author} {\bibfnamefont {J.~R.}\ \bibnamefont
  {Gair}}, \bibinfo {author} {\bibfnamefont {L.}~\bibnamefont {Barack}},
  \bibinfo {author} {\bibfnamefont {T.}~\bibnamefont {Creighton}}, \bibinfo
  {author} {\bibfnamefont {C.}~\bibnamefont {Cutler}}, \bibinfo {author}
  {\bibfnamefont {S.~L.}\ \bibnamefont {Larson}}, \bibinfo {author}
  {\bibfnamefont {E.~S.}\ \bibnamefont {Phinney}},\ and\ \bibinfo {author}
  {\bibfnamefont {M.}~\bibnamefont {Vallisneri}},\ }\href
  {https://doi.org/10.1088/0264-9381/21/20/003} {\bibfield  {journal} {\bibinfo
   {journal} {Class. Quant. Grav.}\ }\textbf {\bibinfo {volume} {21}},\
  \bibinfo {pages} {S1595} (\bibinfo {year} {2004})},\ \Eprint
  {https://arxiv.org/abs/gr-qc/0405137} {arXiv:gr-qc/0405137} \BibitemShut
  {NoStop}%
\bibitem [{\citenamefont {Hinderer}\ and\ \citenamefont
  {Flanagan}(2008)}]{Hinderer:2008dm}%
  \BibitemOpen
  \bibfield  {author} {\bibinfo {author} {\bibfnamefont {T.}~\bibnamefont
  {Hinderer}}\ and\ \bibinfo {author} {\bibfnamefont {E.~E.}\ \bibnamefont
  {Flanagan}},\ }\href {https://doi.org/10.1103/PhysRevD.78.064028} {\bibfield
  {journal} {\bibinfo  {journal} {Phys. Rev. D}\ }\textbf {\bibinfo {volume}
  {78}},\ \bibinfo {pages} {064028} (\bibinfo {year} {2008})},\ \Eprint
  {https://arxiv.org/abs/0805.3337} {arXiv:0805.3337 [gr-qc]} \BibitemShut
  {NoStop}%
\bibitem [{\citenamefont {Seoane}\ \emph {et~al.}(2023)\citenamefont {Seoane}
  \emph {et~al.}}]{LISA:2022yao}%
  \BibitemOpen
  \bibfield  {author} {\bibinfo {author} {\bibfnamefont {P.~A.}\ \bibnamefont
  {Seoane}} \emph {et~al.} (\bibinfo {collaboration} {LISA}),\ }\href
  {https://doi.org/10.1007/s41114-022-00041-y} {\bibfield  {journal} {\bibinfo
  {journal} {Living Rev. Rel.}\ }\textbf {\bibinfo {volume} {26}},\ \bibinfo
  {pages} {2} (\bibinfo {year} {2023})},\ \Eprint
  {https://arxiv.org/abs/2203.06016} {arXiv:2203.06016 [gr-qc]} \BibitemShut
  {NoStop}%
\bibitem [{\citenamefont {van~de Meent}(2018)}]{vandeMeent:2017bcc}%
  \BibitemOpen
  \bibfield  {author} {\bibinfo {author} {\bibfnamefont {M.}~\bibnamefont
  {van~de Meent}},\ }\href {https://doi.org/10.1103/PhysRevD.97.104033}
  {\bibfield  {journal} {\bibinfo  {journal} {Phys. Rev. D}\ }\textbf {\bibinfo
  {volume} {97}},\ \bibinfo {pages} {104033} (\bibinfo {year} {2018})},\
  \Eprint {https://arxiv.org/abs/1711.09607} {arXiv:1711.09607 [gr-qc]}
  \BibitemShut {NoStop}%
\bibitem [{\citenamefont {Skoupy}\ \emph {et~al.}(2023)\citenamefont {Skoupy},
  \citenamefont {Lukes-Gerakopoulos}, \citenamefont {Drummond},\ and\
  \citenamefont {Hughes}}]{Skoupy:2023lih}%
  \BibitemOpen
  \bibfield  {author} {\bibinfo {author} {\bibfnamefont {V.}~\bibnamefont
  {Skoupy}}, \bibinfo {author} {\bibfnamefont {G.}~\bibnamefont
  {Lukes-Gerakopoulos}}, \bibinfo {author} {\bibfnamefont {L.~V.}\ \bibnamefont
  {Drummond}},\ and\ \bibinfo {author} {\bibfnamefont {S.~A.}\ \bibnamefont
  {Hughes}},\ }\href@noop {} {\  (\bibinfo {year} {2023})},\ \Eprint
  {https://arxiv.org/abs/2303.16798} {arXiv:2303.16798 [gr-qc]} \BibitemShut
  {NoStop}%
\bibitem [{\citenamefont {Warburton}\ \emph {et~al.}(2012)\citenamefont
  {Warburton}, \citenamefont {Akcay}, \citenamefont {Barack}, \citenamefont
  {Gair},\ and\ \citenamefont {Sago}}]{Warburton:2011fk}%
  \BibitemOpen
  \bibfield  {author} {\bibinfo {author} {\bibfnamefont {N.}~\bibnamefont
  {Warburton}}, \bibinfo {author} {\bibfnamefont {S.}~\bibnamefont {Akcay}},
  \bibinfo {author} {\bibfnamefont {L.}~\bibnamefont {Barack}}, \bibinfo
  {author} {\bibfnamefont {J.~R.}\ \bibnamefont {Gair}},\ and\ \bibinfo
  {author} {\bibfnamefont {N.}~\bibnamefont {Sago}},\ }\href
  {https://doi.org/10.1103/PhysRevD.85.061501} {\bibfield  {journal} {\bibinfo
  {journal} {Phys. Rev. D}\ }\textbf {\bibinfo {volume} {85}},\ \bibinfo
  {pages} {061501} (\bibinfo {year} {2012})},\ \Eprint
  {https://arxiv.org/abs/1111.6908} {arXiv:1111.6908 [gr-qc]} \BibitemShut
  {NoStop}%
\bibitem [{\citenamefont {Osburn}\ \emph {et~al.}(2016)\citenamefont {Osburn},
  \citenamefont {Warburton},\ and\ \citenamefont {Evans}}]{Osburn:2015duj}%
  \BibitemOpen
  \bibfield  {author} {\bibinfo {author} {\bibfnamefont {T.}~\bibnamefont
  {Osburn}}, \bibinfo {author} {\bibfnamefont {N.}~\bibnamefont {Warburton}},\
  and\ \bibinfo {author} {\bibfnamefont {C.~R.}\ \bibnamefont {Evans}},\ }\href
  {https://doi.org/10.1103/PhysRevD.93.064024} {\bibfield  {journal} {\bibinfo
  {journal} {Phys. Rev. D}\ }\textbf {\bibinfo {volume} {93}},\ \bibinfo
  {pages} {064024} (\bibinfo {year} {2016})},\ \Eprint
  {https://arxiv.org/abs/1511.01498} {arXiv:1511.01498 [gr-qc]} \BibitemShut
  {NoStop}%
\bibitem [{\citenamefont {Warburton}\ \emph {et~al.}(2017)\citenamefont
  {Warburton}, \citenamefont {Osburn},\ and\ \citenamefont
  {Evans}}]{Warburton:2017sxk}%
  \BibitemOpen
  \bibfield  {author} {\bibinfo {author} {\bibfnamefont {N.}~\bibnamefont
  {Warburton}}, \bibinfo {author} {\bibfnamefont {T.}~\bibnamefont {Osburn}},\
  and\ \bibinfo {author} {\bibfnamefont {C.~R.}\ \bibnamefont {Evans}},\ }\href
  {https://doi.org/10.1103/PhysRevD.96.084057} {\bibfield  {journal} {\bibinfo
  {journal} {Phys. Rev. D}\ }\textbf {\bibinfo {volume} {96}},\ \bibinfo
  {pages} {084057} (\bibinfo {year} {2017})},\ \Eprint
  {https://arxiv.org/abs/1708.03720} {arXiv:1708.03720 [gr-qc]} \BibitemShut
  {NoStop}%
\bibitem [{\citenamefont {Van De~Meent}\ and\ \citenamefont
  {Warburton}(2018)}]{VanDeMeent:2018cgn}%
  \BibitemOpen
  \bibfield  {author} {\bibinfo {author} {\bibfnamefont {M.}~\bibnamefont {Van
  De~Meent}}\ and\ \bibinfo {author} {\bibfnamefont {N.}~\bibnamefont
  {Warburton}},\ }\href {https://doi.org/10.1088/1361-6382/aac8ce} {\bibfield
  {journal} {\bibinfo  {journal} {Class. Quant. Grav.}\ }\textbf {\bibinfo
  {volume} {35}},\ \bibinfo {pages} {144003} (\bibinfo {year} {2018})},\
  \Eprint {https://arxiv.org/abs/1802.05281} {arXiv:1802.05281 [gr-qc]}
  \BibitemShut {NoStop}%
\bibitem [{\citenamefont {Fujita}\ and\ \citenamefont
  {Shibata}(2020)}]{Fujita:2020zxe}%
  \BibitemOpen
  \bibfield  {author} {\bibinfo {author} {\bibfnamefont {R.}~\bibnamefont
  {Fujita}}\ and\ \bibinfo {author} {\bibfnamefont {M.}~\bibnamefont
  {Shibata}},\ }\href {https://doi.org/10.1103/PhysRevD.102.064005} {\bibfield
  {journal} {\bibinfo  {journal} {Phys. Rev. D}\ }\textbf {\bibinfo {volume}
  {102}},\ \bibinfo {pages} {064005} (\bibinfo {year} {2020})},\ \Eprint
  {https://arxiv.org/abs/2008.13554} {arXiv:2008.13554 [gr-qc]} \BibitemShut
  {NoStop}%
\bibitem [{\citenamefont {Lynch}\ \emph {et~al.}(2022)\citenamefont {Lynch},
  \citenamefont {van~de Meent},\ and\ \citenamefont
  {Warburton}}]{Lynch:2021ogr}%
  \BibitemOpen
  \bibfield  {author} {\bibinfo {author} {\bibfnamefont {P.}~\bibnamefont
  {Lynch}}, \bibinfo {author} {\bibfnamefont {M.}~\bibnamefont {van~de
  Meent}},\ and\ \bibinfo {author} {\bibfnamefont {N.}~\bibnamefont
  {Warburton}},\ }\href {https://doi.org/10.1088/1361-6382/ac7507} {\bibfield
  {journal} {\bibinfo  {journal} {Class. Quant. Grav.}\ }\textbf {\bibinfo
  {volume} {39}},\ \bibinfo {pages} {145004} (\bibinfo {year} {2022})},\
  \Eprint {https://arxiv.org/abs/2112.05651} {arXiv:2112.05651 [gr-qc]}
  \BibitemShut {NoStop}%
\bibitem [{\citenamefont {Hughes}\ \emph {et~al.}(2021)\citenamefont {Hughes},
  \citenamefont {Warburton}, \citenamefont {Khanna}, \citenamefont {Chua},\
  and\ \citenamefont {Katz}}]{Hughes:2021exa}%
  \BibitemOpen
  \bibfield  {author} {\bibinfo {author} {\bibfnamefont {S.~A.}\ \bibnamefont
  {Hughes}}, \bibinfo {author} {\bibfnamefont {N.}~\bibnamefont {Warburton}},
  \bibinfo {author} {\bibfnamefont {G.}~\bibnamefont {Khanna}}, \bibinfo
  {author} {\bibfnamefont {A.~J.~K.}\ \bibnamefont {Chua}},\ and\ \bibinfo
  {author} {\bibfnamefont {M.~L.}\ \bibnamefont {Katz}},\ }\href
  {https://doi.org/10.1103/PhysRevD.103.104014} {\bibfield  {journal} {\bibinfo
   {journal} {Phys. Rev. D}\ }\textbf {\bibinfo {volume} {103}},\ \bibinfo
  {pages} {104014} (\bibinfo {year} {2021})},\ \Eprint
  {https://arxiv.org/abs/2102.02713} {arXiv:2102.02713 [gr-qc]} \BibitemShut
  {NoStop}%
\bibitem [{\citenamefont {Lynch}\ \emph {et~al.}(2023)\citenamefont {Lynch},
  \citenamefont {van~de Meent},\ and\ \citenamefont
  {Warburton}}]{Lynch:2023gpu}%
  \BibitemOpen
  \bibfield  {author} {\bibinfo {author} {\bibfnamefont {P.}~\bibnamefont
  {Lynch}}, \bibinfo {author} {\bibfnamefont {M.}~\bibnamefont {van~de
  Meent}},\ and\ \bibinfo {author} {\bibfnamefont {N.}~\bibnamefont
  {Warburton}},\ }\href@noop {} {\  (\bibinfo {year} {2023})},\ \Eprint
  {https://arxiv.org/abs/2305.10533} {arXiv:2305.10533 [gr-qc]} \BibitemShut
  {NoStop}%
\bibitem [{\citenamefont {Pound}\ \emph {et~al.}(2020)\citenamefont {Pound},
  \citenamefont {Wardell}, \citenamefont {Warburton},\ and\ \citenamefont
  {Miller}}]{Pound:2019lzj}%
  \BibitemOpen
  \bibfield  {author} {\bibinfo {author} {\bibfnamefont {A.}~\bibnamefont
  {Pound}}, \bibinfo {author} {\bibfnamefont {B.}~\bibnamefont {Wardell}},
  \bibinfo {author} {\bibfnamefont {N.}~\bibnamefont {Warburton}},\ and\
  \bibinfo {author} {\bibfnamefont {J.}~\bibnamefont {Miller}},\ }\href
  {https://doi.org/10.1103/PhysRevLett.124.021101} {\bibfield  {journal}
  {\bibinfo  {journal} {Phys. Rev. Lett.}\ }\textbf {\bibinfo {volume} {124}},\
  \bibinfo {pages} {021101} (\bibinfo {year} {2020})},\ \Eprint
  {https://arxiv.org/abs/1908.07419} {arXiv:1908.07419 [gr-qc]} \BibitemShut
  {NoStop}%
\bibitem [{\citenamefont {Warburton}\ \emph {et~al.}(2021)\citenamefont
  {Warburton}, \citenamefont {Pound}, \citenamefont {Wardell}, \citenamefont
  {Miller},\ and\ \citenamefont {Durkan}}]{Warburton:2021kwk}%
  \BibitemOpen
  \bibfield  {author} {\bibinfo {author} {\bibfnamefont {N.}~\bibnamefont
  {Warburton}}, \bibinfo {author} {\bibfnamefont {A.}~\bibnamefont {Pound}},
  \bibinfo {author} {\bibfnamefont {B.}~\bibnamefont {Wardell}}, \bibinfo
  {author} {\bibfnamefont {J.}~\bibnamefont {Miller}},\ and\ \bibinfo {author}
  {\bibfnamefont {L.}~\bibnamefont {Durkan}},\ }\href@noop {} {\  (\bibinfo
  {year} {2021})},\ \Eprint {https://arxiv.org/abs/2107.01298}
  {arXiv:2107.01298 [gr-qc]} \BibitemShut {NoStop}%
\bibitem [{\citenamefont {Wardell}\ \emph {et~al.}(2023)\citenamefont
  {Wardell}, \citenamefont {Pound}, \citenamefont {Warburton}, \citenamefont
  {Miller}, \citenamefont {Durkan},\ and\ \citenamefont
  {Le~Tiec}}]{Wardell:2021fyy}%
  \BibitemOpen
  \bibfield  {author} {\bibinfo {author} {\bibfnamefont {B.}~\bibnamefont
  {Wardell}}, \bibinfo {author} {\bibfnamefont {A.}~\bibnamefont {Pound}},
  \bibinfo {author} {\bibfnamefont {N.}~\bibnamefont {Warburton}}, \bibinfo
  {author} {\bibfnamefont {J.}~\bibnamefont {Miller}}, \bibinfo {author}
  {\bibfnamefont {L.}~\bibnamefont {Durkan}},\ and\ \bibinfo {author}
  {\bibfnamefont {A.}~\bibnamefont {Le~Tiec}},\ }\href
  {https://doi.org/10.1103/PhysRevLett.130.241402} {\bibfield  {journal}
  {\bibinfo  {journal} {Phys. Rev. Lett.}\ }\textbf {\bibinfo {volume} {130}},\
  \bibinfo {pages} {241402} (\bibinfo {year} {2023})},\ \Eprint
  {https://arxiv.org/abs/2112.12265} {arXiv:2112.12265 [gr-qc]} \BibitemShut
  {NoStop}%
\bibitem [{\citenamefont {Abbott}\ \emph
  {et~al.}(2020{\natexlab{b}})\citenamefont {Abbott} \emph
  {et~al.}}]{LIGOScientific:2020stg}%
  \BibitemOpen
  \bibfield  {author} {\bibinfo {author} {\bibfnamefont {R.}~\bibnamefont
  {Abbott}} \emph {et~al.} (\bibinfo {collaboration} {LIGO Scientific,
  Virgo}),\ }\href {https://doi.org/10.1103/PhysRevD.102.043015} {\bibfield
  {journal} {\bibinfo  {journal} {Phys. Rev. D}\ }\textbf {\bibinfo {volume}
  {102}},\ \bibinfo {pages} {043015} (\bibinfo {year} {2020}{\natexlab{b}})},\
  \Eprint {https://arxiv.org/abs/2004.08342} {arXiv:2004.08342 [astro-ph.HE]}
  \BibitemShut {NoStop}%
\bibitem [{\citenamefont {Abbott}\ \emph
  {et~al.}(2020{\natexlab{c}})\citenamefont {Abbott} \emph
  {et~al.}}]{LIGOScientific:2020zkf}%
  \BibitemOpen
  \bibfield  {author} {\bibinfo {author} {\bibfnamefont {R.}~\bibnamefont
  {Abbott}} \emph {et~al.} (\bibinfo {collaboration} {LIGO Scientific,
  Virgo}),\ }\href {https://doi.org/10.3847/2041-8213/ab960f} {\bibfield
  {journal} {\bibinfo  {journal} {Astrophys. J. Lett.}\ }\textbf {\bibinfo
  {volume} {896}},\ \bibinfo {pages} {L44} (\bibinfo {year}
  {2020}{\natexlab{c}})},\ \Eprint {https://arxiv.org/abs/2006.12611}
  {arXiv:2006.12611 [astro-ph.HE]} \BibitemShut {NoStop}%
\bibitem [{\citenamefont {Hopman}\ and\ \citenamefont
  {Alexander}(2005)}]{Hopman:2005vr}%
  \BibitemOpen
  \bibfield  {author} {\bibinfo {author} {\bibfnamefont {C.}~\bibnamefont
  {Hopman}}\ and\ \bibinfo {author} {\bibfnamefont {T.}~\bibnamefont
  {Alexander}},\ }\href {https://doi.org/10.1086/431475} {\bibfield  {journal}
  {\bibinfo  {journal} {Astrophys. J.}\ }\textbf {\bibinfo {volume} {629}},\
  \bibinfo {pages} {362} (\bibinfo {year} {2005})},\ \Eprint
  {https://arxiv.org/abs/astro-ph/0503672} {arXiv:astro-ph/0503672}
  \BibitemShut {NoStop}%
\bibitem [{\citenamefont {Mino}\ \emph {et~al.}(1997)\citenamefont {Mino},
  \citenamefont {Sasaki},\ and\ \citenamefont {Tanaka}}]{Mino:1996nk}%
  \BibitemOpen
  \bibfield  {author} {\bibinfo {author} {\bibfnamefont {Y.}~\bibnamefont
  {Mino}}, \bibinfo {author} {\bibfnamefont {M.}~\bibnamefont {Sasaki}},\ and\
  \bibinfo {author} {\bibfnamefont {T.}~\bibnamefont {Tanaka}},\ }\href
  {https://doi.org/10.1103/PhysRevD.55.3457} {\bibfield  {journal} {\bibinfo
  {journal} {Phys. Rev.}\ }\textbf {\bibinfo {volume} {D55}},\ \bibinfo {pages}
  {3457} (\bibinfo {year} {1997})},\ \Eprint
  {https://arxiv.org/abs/gr-qc/9606018} {arXiv:gr-qc/9606018 [gr-qc]}
  \BibitemShut {NoStop}%
\bibitem [{\citenamefont {Quinn}\ and\ \citenamefont
  {Wald}(1997)}]{Quinn:1996am}%
  \BibitemOpen
  \bibfield  {author} {\bibinfo {author} {\bibfnamefont {T.~C.}\ \bibnamefont
  {Quinn}}\ and\ \bibinfo {author} {\bibfnamefont {R.~M.}\ \bibnamefont
  {Wald}},\ }\href {https://doi.org/10.1103/PhysRevD.56.3381} {\bibfield
  {journal} {\bibinfo  {journal} {Phys.\ Rev.\ D}\ }\textbf {\bibinfo {volume}
  {56}},\ \bibinfo {pages} {3381} (\bibinfo {year} {1997})},\ \Eprint
  {https://arxiv.org/abs/gr-qc/9610053} {arXiv:gr-qc/9610053} \BibitemShut
  {NoStop}%
\bibitem [{\citenamefont {Gralla}\ and\ \citenamefont
  {Wald}(2008)}]{Gralla:2008fg}%
  \BibitemOpen
  \bibfield  {author} {\bibinfo {author} {\bibfnamefont {S.~E.}\ \bibnamefont
  {Gralla}}\ and\ \bibinfo {author} {\bibfnamefont {R.~M.}\ \bibnamefont
  {Wald}},\ }\href {https://doi.org/10.1088/0264-9381/25/20/205009} {\bibfield
  {journal} {\bibinfo  {journal} {Class. Quant. Grav.}\ }\textbf {\bibinfo
  {volume} {25}},\ \bibinfo {pages} {205009} (\bibinfo {year} {2008})},\
  \bibinfo {note} {[Erratum: Class.Quant.Grav. 28, 159501 (2011)]},\ \Eprint
  {https://arxiv.org/abs/0806.3293} {arXiv:0806.3293 [gr-qc]} \BibitemShut
  {NoStop}%
\bibitem [{\citenamefont {Pound}(2010)}]{Pound:2009sm}%
  \BibitemOpen
  \bibfield  {author} {\bibinfo {author} {\bibfnamefont {A.}~\bibnamefont
  {Pound}},\ }\href {https://doi.org/10.1103/PhysRevD.81.024023} {\bibfield
  {journal} {\bibinfo  {journal} {Phys. Rev. D}\ }\textbf {\bibinfo {volume}
  {81}},\ \bibinfo {pages} {024023} (\bibinfo {year} {2010})},\ \Eprint
  {https://arxiv.org/abs/0907.5197} {arXiv:0907.5197 [gr-qc]} \BibitemShut
  {NoStop}%
\bibitem [{\citenamefont {Barack}\ and\ \citenamefont
  {Ori}(2000)}]{Barack:1999wf}%
  \BibitemOpen
  \bibfield  {author} {\bibinfo {author} {\bibfnamefont {L.}~\bibnamefont
  {Barack}}\ and\ \bibinfo {author} {\bibfnamefont {A.}~\bibnamefont {Ori}},\
  }\href {https://doi.org/10.1103/PhysRevD.61.061502} {\bibfield  {journal}
  {\bibinfo  {journal} {Phys. Rev.}\ }\textbf {\bibinfo {volume} {D61}},\
  \bibinfo {pages} {061502} (\bibinfo {year} {2000})},\ \Eprint
  {https://arxiv.org/abs/gr-qc/9912010} {arXiv:gr-qc/9912010 [gr-qc]}
  \BibitemShut {NoStop}%
\bibitem [{\citenamefont {Miller}\ and\ \citenamefont
  {Pound}(2021)}]{Miller:2020bft}%
  \BibitemOpen
  \bibfield  {author} {\bibinfo {author} {\bibfnamefont {J.}~\bibnamefont
  {Miller}}\ and\ \bibinfo {author} {\bibfnamefont {A.}~\bibnamefont {Pound}},\
  }\href {https://doi.org/10.1103/PhysRevD.103.064048} {\bibfield  {journal}
  {\bibinfo  {journal} {Phys. Rev. D}\ }\textbf {\bibinfo {volume} {103}},\
  \bibinfo {pages} {064048} (\bibinfo {year} {2021})},\ \Eprint
  {https://arxiv.org/abs/2006.11263} {arXiv:2006.11263 [gr-qc]} \BibitemShut
  {NoStop}%
\bibitem [{\citenamefont {Pound}\ and\ \citenamefont
  {Miller}(2014)}]{Pound:2014xva}%
  \BibitemOpen
  \bibfield  {author} {\bibinfo {author} {\bibfnamefont {A.}~\bibnamefont
  {Pound}}\ and\ \bibinfo {author} {\bibfnamefont {J.}~\bibnamefont {Miller}},\
  }\href {https://doi.org/10.1103/PhysRevD.89.104020} {\bibfield  {journal}
  {\bibinfo  {journal} {Phys. Rev. D}\ }\textbf {\bibinfo {volume} {89}},\
  \bibinfo {pages} {104020} (\bibinfo {year} {2014})},\ \Eprint
  {https://arxiv.org/abs/1403.1843} {arXiv:1403.1843 [gr-qc]} \BibitemShut
  {NoStop}%
\bibitem [{\citenamefont {Upton}\ and\ \citenamefont
  {Pound}(2021)}]{Upton:2021oxf}%
  \BibitemOpen
  \bibfield  {author} {\bibinfo {author} {\bibfnamefont {S.~D.}\ \bibnamefont
  {Upton}}\ and\ \bibinfo {author} {\bibfnamefont {A.}~\bibnamefont {Pound}},\
  }\href {https://doi.org/10.1103/PhysRevD.103.124016} {\bibfield  {journal}
  {\bibinfo  {journal} {Phys. Rev. D}\ }\textbf {\bibinfo {volume} {103}},\
  \bibinfo {pages} {124016} (\bibinfo {year} {2021})},\ \Eprint
  {https://arxiv.org/abs/2101.11409} {arXiv:2101.11409 [gr-qc]} \BibitemShut
  {NoStop}%
\bibitem [{\citenamefont {Barack}\ and\ \citenamefont
  {Golbourn}(2007)}]{Barack:2007jh}%
  \BibitemOpen
  \bibfield  {author} {\bibinfo {author} {\bibfnamefont {L.}~\bibnamefont
  {Barack}}\ and\ \bibinfo {author} {\bibfnamefont {D.~A.}\ \bibnamefont
  {Golbourn}},\ }\href {https://doi.org/10.1103/PhysRevD.76.044020} {\bibfield
  {journal} {\bibinfo  {journal} {Phys. Rev. D}\ }\textbf {\bibinfo {volume}
  {76}},\ \bibinfo {pages} {044020} (\bibinfo {year} {2007})},\ \Eprint
  {https://arxiv.org/abs/0705.3620} {arXiv:0705.3620 [gr-qc]} \BibitemShut
  {NoStop}%
\bibitem [{\citenamefont {Vega}\ and\ \citenamefont
  {Detweiler}(2008)}]{Vega:2007mc}%
  \BibitemOpen
  \bibfield  {author} {\bibinfo {author} {\bibfnamefont {I.}~\bibnamefont
  {Vega}}\ and\ \bibinfo {author} {\bibfnamefont {S.~L.}\ \bibnamefont
  {Detweiler}},\ }\href {https://doi.org/10.1103/PhysRevD.77.084008} {\bibfield
   {journal} {\bibinfo  {journal} {Phys. Rev. D}\ }\textbf {\bibinfo {volume}
  {77}},\ \bibinfo {pages} {084008} (\bibinfo {year} {2008})},\ \Eprint
  {https://arxiv.org/abs/0712.4405} {arXiv:0712.4405 [gr-qc]} \BibitemShut
  {NoStop}%
\bibitem [{\citenamefont {Dolan}\ \emph {et~al.}(2011)\citenamefont {Dolan},
  \citenamefont {Barack},\ and\ \citenamefont {Wardell}}]{Dolan:2011dx}%
  \BibitemOpen
  \bibfield  {author} {\bibinfo {author} {\bibfnamefont {S.~R.}\ \bibnamefont
  {Dolan}}, \bibinfo {author} {\bibfnamefont {L.}~\bibnamefont {Barack}},\ and\
  \bibinfo {author} {\bibfnamefont {B.}~\bibnamefont {Wardell}},\ }\href
  {https://doi.org/10.1103/PhysRevD.84.084001} {\bibfield  {journal} {\bibinfo
  {journal} {Phys. Rev. D}\ }\textbf {\bibinfo {volume} {84}},\ \bibinfo
  {pages} {084001} (\bibinfo {year} {2011})},\ \Eprint
  {https://arxiv.org/abs/1107.0012} {arXiv:1107.0012 [gr-qc]} \BibitemShut
  {NoStop}%
\bibitem [{\citenamefont {Dolan}\ and\ \citenamefont
  {Barack}(2013)}]{Dolan:2012jg}%
  \BibitemOpen
  \bibfield  {author} {\bibinfo {author} {\bibfnamefont {S.~R.}\ \bibnamefont
  {Dolan}}\ and\ \bibinfo {author} {\bibfnamefont {L.}~\bibnamefont {Barack}},\
  }\href {https://doi.org/10.1103/PhysRevD.87.084066} {\bibfield  {journal}
  {\bibinfo  {journal} {Phys. Rev. D}\ }\textbf {\bibinfo {volume} {87}},\
  \bibinfo {pages} {084066} (\bibinfo {year} {2013})},\ \Eprint
  {https://arxiv.org/abs/1211.4586} {arXiv:1211.4586 [gr-qc]} \BibitemShut
  {NoStop}%
\bibitem [{\citenamefont {Diener}\ \emph {et~al.}(2012)\citenamefont {Diener},
  \citenamefont {Vega}, \citenamefont {Wardell},\ and\ \citenamefont
  {Detweiler}}]{Diener:2011cc}%
  \BibitemOpen
  \bibfield  {author} {\bibinfo {author} {\bibfnamefont {P.}~\bibnamefont
  {Diener}}, \bibinfo {author} {\bibfnamefont {I.}~\bibnamefont {Vega}},
  \bibinfo {author} {\bibfnamefont {B.}~\bibnamefont {Wardell}},\ and\ \bibinfo
  {author} {\bibfnamefont {S.}~\bibnamefont {Detweiler}},\ }\href
  {https://doi.org/10.1103/PhysRevLett.108.191102} {\bibfield  {journal}
  {\bibinfo  {journal} {Phys. Rev. Lett.}\ }\textbf {\bibinfo {volume} {108}},\
  \bibinfo {pages} {191102} (\bibinfo {year} {2012})},\ \Eprint
  {https://arxiv.org/abs/1112.4821} {arXiv:1112.4821 [gr-qc]} \BibitemShut
  {NoStop}%
\bibitem [{\citenamefont {Warburton}\ and\ \citenamefont
  {Wardell}(2014)}]{Warburton:2013lea}%
  \BibitemOpen
  \bibfield  {author} {\bibinfo {author} {\bibfnamefont {N.}~\bibnamefont
  {Warburton}}\ and\ \bibinfo {author} {\bibfnamefont {B.}~\bibnamefont
  {Wardell}},\ }\href {https://doi.org/10.1103/PhysRevD.89.044046} {\bibfield
  {journal} {\bibinfo  {journal} {Phys. Rev. D}\ }\textbf {\bibinfo {volume}
  {89}},\ \bibinfo {pages} {044046} (\bibinfo {year} {2014})},\ \Eprint
  {https://arxiv.org/abs/1311.3104} {arXiv:1311.3104 [gr-qc]} \BibitemShut
  {NoStop}%
\bibitem [{\citenamefont {Wardell}\ and\ \citenamefont
  {Warburton}(2015)}]{Wardell:2015ada}%
  \BibitemOpen
  \bibfield  {author} {\bibinfo {author} {\bibfnamefont {B.}~\bibnamefont
  {Wardell}}\ and\ \bibinfo {author} {\bibfnamefont {N.}~\bibnamefont
  {Warburton}},\ }\href {https://doi.org/10.1103/PhysRevD.92.084019} {\bibfield
   {journal} {\bibinfo  {journal} {Phys. Rev. D}\ }\textbf {\bibinfo {volume}
  {92}},\ \bibinfo {pages} {084019} (\bibinfo {year} {2015})},\ \Eprint
  {https://arxiv.org/abs/1505.07841} {arXiv:1505.07841 [gr-qc]} \BibitemShut
  {NoStop}%
\bibitem [{\citenamefont {Barack}\ \emph {et~al.}(2008)\citenamefont {Barack},
  \citenamefont {Ori},\ and\ \citenamefont {Sago}}]{Barack:2008ms}%
  \BibitemOpen
  \bibfield  {author} {\bibinfo {author} {\bibfnamefont {L.}~\bibnamefont
  {Barack}}, \bibinfo {author} {\bibfnamefont {A.}~\bibnamefont {Ori}},\ and\
  \bibinfo {author} {\bibfnamefont {N.}~\bibnamefont {Sago}},\ }\href
  {https://doi.org/10.1103/PhysRevD.78.084021} {\bibfield  {journal} {\bibinfo
  {journal} {Phys. Rev. D}\ }\textbf {\bibinfo {volume} {78}},\ \bibinfo
  {pages} {084021} (\bibinfo {year} {2008})},\ \Eprint
  {https://arxiv.org/abs/0808.2315} {arXiv:0808.2315 [gr-qc]} \BibitemShut
  {NoStop}%
\bibitem [{\citenamefont {Warburton}\ and\ \citenamefont
  {Barack}(2011)}]{Warburton:2011hp}%
  \BibitemOpen
  \bibfield  {author} {\bibinfo {author} {\bibfnamefont {N.}~\bibnamefont
  {Warburton}}\ and\ \bibinfo {author} {\bibfnamefont {L.}~\bibnamefont
  {Barack}},\ }\href {https://doi.org/10.1103/PhysRevD.83.124038} {\bibfield
  {journal} {\bibinfo  {journal} {Phys. Rev. D}\ }\textbf {\bibinfo {volume}
  {83}},\ \bibinfo {pages} {124038} (\bibinfo {year} {2011})},\ \Eprint
  {https://arxiv.org/abs/1103.0287} {arXiv:1103.0287 [gr-qc]} \BibitemShut
  {NoStop}%
\bibitem [{\citenamefont {Nasipak}\ \emph {et~al.}(2019)\citenamefont
  {Nasipak}, \citenamefont {Osburn},\ and\ \citenamefont
  {Evans}}]{Nasipak:2019hxh}%
  \BibitemOpen
  \bibfield  {author} {\bibinfo {author} {\bibfnamefont {Z.}~\bibnamefont
  {Nasipak}}, \bibinfo {author} {\bibfnamefont {T.}~\bibnamefont {Osburn}},\
  and\ \bibinfo {author} {\bibfnamefont {C.~R.}\ \bibnamefont {Evans}},\ }\href
  {https://doi.org/10.1103/PhysRevD.100.064008} {\bibfield  {journal} {\bibinfo
   {journal} {Phys. Rev. D}\ }\textbf {\bibinfo {volume} {100}},\ \bibinfo
  {pages} {064008} (\bibinfo {year} {2019})},\ \Eprint
  {https://arxiv.org/abs/1905.13237} {arXiv:1905.13237 [gr-qc]} \BibitemShut
  {NoStop}%
\bibitem [{\citenamefont {Akcay}\ \emph {et~al.}(2013)\citenamefont {Akcay},
  \citenamefont {Warburton},\ and\ \citenamefont {Barack}}]{Akcay:2013wfa}%
  \BibitemOpen
  \bibfield  {author} {\bibinfo {author} {\bibfnamefont {S.}~\bibnamefont
  {Akcay}}, \bibinfo {author} {\bibfnamefont {N.}~\bibnamefont {Warburton}},\
  and\ \bibinfo {author} {\bibfnamefont {L.}~\bibnamefont {Barack}},\ }\href
  {https://doi.org/10.1103/PhysRevD.88.104009} {\bibfield  {journal} {\bibinfo
  {journal} {Phys. Rev. D}\ }\textbf {\bibinfo {volume} {88}},\ \bibinfo
  {pages} {104009} (\bibinfo {year} {2013})},\ \Eprint
  {https://arxiv.org/abs/1308.5223} {arXiv:1308.5223 [gr-qc]} \BibitemShut
  {NoStop}%
\bibitem [{\citenamefont {Osburn}\ \emph {et~al.}(2014)\citenamefont {Osburn},
  \citenamefont {Forseth}, \citenamefont {Evans},\ and\ \citenamefont
  {Hopper}}]{Osburn:2014hoa}%
  \BibitemOpen
  \bibfield  {author} {\bibinfo {author} {\bibfnamefont {T.}~\bibnamefont
  {Osburn}}, \bibinfo {author} {\bibfnamefont {E.}~\bibnamefont {Forseth}},
  \bibinfo {author} {\bibfnamefont {C.~R.}\ \bibnamefont {Evans}},\ and\
  \bibinfo {author} {\bibfnamefont {S.}~\bibnamefont {Hopper}},\ }\href
  {https://doi.org/10.1103/PhysRevD.90.104031} {\bibfield  {journal} {\bibinfo
  {journal} {Phys. Rev. D}\ }\textbf {\bibinfo {volume} {90}},\ \bibinfo
  {pages} {104031} (\bibinfo {year} {2014})},\ \Eprint
  {https://arxiv.org/abs/1409.4419} {arXiv:1409.4419 [gr-qc]} \BibitemShut
  {NoStop}%
\bibitem [{\citenamefont {van~de Meent}\ and\ \citenamefont
  {Shah}(2015)}]{vandeMeent:2015lxa}%
  \BibitemOpen
  \bibfield  {author} {\bibinfo {author} {\bibfnamefont {M.}~\bibnamefont
  {van~de Meent}}\ and\ \bibinfo {author} {\bibfnamefont {A.~G.}\ \bibnamefont
  {Shah}},\ }\href {https://doi.org/10.1103/PhysRevD.92.064025} {\bibfield
  {journal} {\bibinfo  {journal} {Phys. Rev. D}\ }\textbf {\bibinfo {volume}
  {92}},\ \bibinfo {pages} {064025} (\bibinfo {year} {2015})},\ \Eprint
  {https://arxiv.org/abs/1506.04755} {arXiv:1506.04755 [gr-qc]} \BibitemShut
  {NoStop}%
\bibitem [{\citenamefont {Hopper}\ and\ \citenamefont
  {Evans}(2013)}]{Hopper:2012ty}%
  \BibitemOpen
  \bibfield  {author} {\bibinfo {author} {\bibfnamefont {S.}~\bibnamefont
  {Hopper}}\ and\ \bibinfo {author} {\bibfnamefont {C.~R.}\ \bibnamefont
  {Evans}},\ }\href {https://doi.org/10.1103/PhysRevD.87.064008} {\bibfield
  {journal} {\bibinfo  {journal} {Phys. Rev. D}\ }\textbf {\bibinfo {volume}
  {87}},\ \bibinfo {pages} {064008} (\bibinfo {year} {2013})},\ \Eprint
  {https://arxiv.org/abs/1210.7969} {arXiv:1210.7969 [gr-qc]} \BibitemShut
  {NoStop}%
\bibitem [{\citenamefont {Cutler}\ \emph {et~al.}(1994)\citenamefont {Cutler},
  \citenamefont {Kennefick},\ and\ \citenamefont {Poisson}}]{Cutler:1994pb}%
  \BibitemOpen
  \bibfield  {author} {\bibinfo {author} {\bibfnamefont {C.}~\bibnamefont
  {Cutler}}, \bibinfo {author} {\bibfnamefont {D.}~\bibnamefont {Kennefick}},\
  and\ \bibinfo {author} {\bibfnamefont {E.}~\bibnamefont {Poisson}},\ }\href
  {https://doi.org/10.1103/PhysRevD.50.3816} {\bibfield  {journal} {\bibinfo
  {journal} {Phys. Rev. D}\ }\textbf {\bibinfo {volume} {50}},\ \bibinfo
  {pages} {3816} (\bibinfo {year} {1994})}\BibitemShut {NoStop}%
\bibitem [{\citenamefont {Darwin}(1959)}]{Darwin:1959}%
  \BibitemOpen
  \bibfield  {author} {\bibinfo {author} {\bibfnamefont {C.~G.}\ \bibnamefont
  {Darwin}},\ }\href {https://doi.org/10.1098/rspa.1959.0015} {\bibfield
  {journal} {\bibinfo  {journal} {Proceedings of the Royal Society of London.
  Series A. Mathematical and Physical Sciences}\ }\textbf {\bibinfo {volume}
  {249}},\ \bibinfo {pages} {180} (\bibinfo {year} {1959})}\BibitemShut
  {NoStop}%
\bibitem [{\citenamefont {Zwillinger}(2014)}]{Zwillinger2014}%
  \BibitemOpen
  \bibfield  {author} {\bibinfo {author} {\bibfnamefont {D.}~\bibnamefont
  {Zwillinger}},\ }\href {https://books.google.co.uk/books?id=NjnLAwAAQBAJ}
  {\emph {\bibinfo {title} {{Table of Integrals, Series, and Products}}}}\
  (\bibinfo  {publisher} {Elsevier Science},\ \bibinfo {year}
  {2014})\BibitemShut {NoStop}%
\bibitem [{\citenamefont {Burko}(2000)}]{Burko:2000xx}%
  \BibitemOpen
  \bibfield  {author} {\bibinfo {author} {\bibfnamefont {L.~M.}\ \bibnamefont
  {Burko}},\ }\href {https://doi.org/10.1103/PhysRevLett.84.4529} {\bibfield
  {journal} {\bibinfo  {journal} {Phys. Rev. Lett.}\ }\textbf {\bibinfo
  {volume} {84}},\ \bibinfo {pages} {4529} (\bibinfo {year} {2000})},\ \Eprint
  {https://arxiv.org/abs/gr-qc/0003074} {arXiv:gr-qc/0003074} \BibitemShut
  {NoStop}%
\bibitem [{\citenamefont {Diaz-Rivera}\ \emph {et~al.}(2004)\citenamefont
  {Diaz-Rivera}, \citenamefont {Messaritaki}, \citenamefont {Whiting},\ and\
  \citenamefont {Detweiler}}]{Diaz-Rivera:2004nim}%
  \BibitemOpen
  \bibfield  {author} {\bibinfo {author} {\bibfnamefont {L.~M.}\ \bibnamefont
  {Diaz-Rivera}}, \bibinfo {author} {\bibfnamefont {E.}~\bibnamefont
  {Messaritaki}}, \bibinfo {author} {\bibfnamefont {B.~F.}\ \bibnamefont
  {Whiting}},\ and\ \bibinfo {author} {\bibfnamefont {S.~L.}\ \bibnamefont
  {Detweiler}},\ }\href {https://doi.org/10.1103/PhysRevD.70.124018} {\bibfield
   {journal} {\bibinfo  {journal} {Phys. Rev. D}\ }\textbf {\bibinfo {volume}
  {70}},\ \bibinfo {pages} {124018} (\bibinfo {year} {2004})},\ \Eprint
  {https://arxiv.org/abs/gr-qc/0410011} {arXiv:gr-qc/0410011} \BibitemShut
  {NoStop}%
\bibitem [{\citenamefont {Warburton}\ and\ \citenamefont
  {Barack}(2010)}]{Warburton:2010eq}%
  \BibitemOpen
  \bibfield  {author} {\bibinfo {author} {\bibfnamefont {N.}~\bibnamefont
  {Warburton}}\ and\ \bibinfo {author} {\bibfnamefont {L.}~\bibnamefont
  {Barack}},\ }\href {https://doi.org/10.1103/PhysRevD.81.084039} {\bibfield
  {journal} {\bibinfo  {journal} {Phys. Rev. D}\ }\textbf {\bibinfo {volume}
  {81}},\ \bibinfo {pages} {084039} (\bibinfo {year} {2010})},\ \Eprint
  {https://arxiv.org/abs/1003.1860} {arXiv:1003.1860 [gr-qc]} \BibitemShut
  {NoStop}%
\bibitem [{\citenamefont {Warburton}(2015)}]{Warburton:2014bya}%
  \BibitemOpen
  \bibfield  {author} {\bibinfo {author} {\bibfnamefont {N.}~\bibnamefont
  {Warburton}},\ }\href {https://doi.org/10.1103/PhysRevD.91.024045} {\bibfield
   {journal} {\bibinfo  {journal} {Phys. Rev. D}\ }\textbf {\bibinfo {volume}
  {91}},\ \bibinfo {pages} {024045} (\bibinfo {year} {2015})},\ \Eprint
  {https://arxiv.org/abs/1408.2885} {arXiv:1408.2885 [gr-qc]} \BibitemShut
  {NoStop}%
\bibitem [{\citenamefont {Nasipak}\ and\ \citenamefont
  {Evans}(2021)}]{Nasipak:2021qfu}%
  \BibitemOpen
  \bibfield  {author} {\bibinfo {author} {\bibfnamefont {Z.}~\bibnamefont
  {Nasipak}}\ and\ \bibinfo {author} {\bibfnamefont {C.~R.}\ \bibnamefont
  {Evans}},\ }\href {https://doi.org/10.1103/PhysRevD.104.084011} {\bibfield
  {journal} {\bibinfo  {journal} {Phys. Rev. D}\ }\textbf {\bibinfo {volume}
  {104}},\ \bibinfo {pages} {084011} (\bibinfo {year} {2021})},\ \Eprint
  {https://arxiv.org/abs/2105.15188} {arXiv:2105.15188 [gr-qc]} \BibitemShut
  {NoStop}%
\bibitem [{\citenamefont {Panosso~Macedo}\ \emph {et~al.}(2022)\citenamefont
  {Panosso~Macedo}, \citenamefont {Leather}, \citenamefont {Warburton},
  \citenamefont {Wardell},\ and\ \citenamefont
  {Zengino\u{g}lu}}]{PanossoMacedo:2022fdi}%
  \BibitemOpen
  \bibfield  {author} {\bibinfo {author} {\bibfnamefont {R.}~\bibnamefont
  {Panosso~Macedo}}, \bibinfo {author} {\bibfnamefont {B.}~\bibnamefont
  {Leather}}, \bibinfo {author} {\bibfnamefont {N.}~\bibnamefont {Warburton}},
  \bibinfo {author} {\bibfnamefont {B.}~\bibnamefont {Wardell}},\ and\ \bibinfo
  {author} {\bibfnamefont {A.}~\bibnamefont {Zengino\u{g}lu}},\ }\href
  {https://doi.org/10.1103/PhysRevD.105.104033} {\bibfield  {journal} {\bibinfo
   {journal} {Phys. Rev. D}\ }\textbf {\bibinfo {volume} {105}},\ \bibinfo
  {pages} {104033} (\bibinfo {year} {2022})},\ \Eprint
  {https://arxiv.org/abs/2202.01794} {arXiv:2202.01794 [gr-qc]} \BibitemShut
  {NoStop}%
\bibitem [{\citenamefont {Barack}\ and\ \citenamefont
  {Sago}(2007)}]{Barack:2007tm}%
  \BibitemOpen
  \bibfield  {author} {\bibinfo {author} {\bibfnamefont {L.}~\bibnamefont
  {Barack}}\ and\ \bibinfo {author} {\bibfnamefont {N.}~\bibnamefont {Sago}},\
  }\href {https://doi.org/10.1103/PhysRevD.75.064021} {\bibfield  {journal}
  {\bibinfo  {journal} {Phys. Rev. D}\ }\textbf {\bibinfo {volume} {75}},\
  \bibinfo {pages} {064021} (\bibinfo {year} {2007})},\ \Eprint
  {https://arxiv.org/abs/gr-qc/0701069} {arXiv:gr-qc/0701069} \BibitemShut
  {NoStop}%
\bibitem [{\citenamefont {Barack}\ and\ \citenamefont
  {Sago}(2010)}]{Barack:2010tm}%
  \BibitemOpen
  \bibfield  {author} {\bibinfo {author} {\bibfnamefont {L.}~\bibnamefont
  {Barack}}\ and\ \bibinfo {author} {\bibfnamefont {N.}~\bibnamefont {Sago}},\
  }\href {https://doi.org/10.1103/PhysRevD.81.084021} {\bibfield  {journal}
  {\bibinfo  {journal} {Phys. Rev. D}\ }\textbf {\bibinfo {volume} {81}},\
  \bibinfo {pages} {084021} (\bibinfo {year} {2010})},\ \Eprint
  {https://arxiv.org/abs/1002.2386} {arXiv:1002.2386 [gr-qc]} \BibitemShut
  {NoStop}%
\bibitem [{\citenamefont {Barack}\ and\ \citenamefont
  {Ori}(2001)}]{Barack:2001ph}%
  \BibitemOpen
  \bibfield  {author} {\bibinfo {author} {\bibfnamefont {L.}~\bibnamefont
  {Barack}}\ and\ \bibinfo {author} {\bibfnamefont {A.}~\bibnamefont {Ori}},\
  }\href {https://doi.org/10.1103/PhysRevD.64.124003} {\bibfield  {journal}
  {\bibinfo  {journal} {Phys. Rev. D}\ }\textbf {\bibinfo {volume} {64}},\
  \bibinfo {pages} {124003} (\bibinfo {year} {2001})},\ \Eprint
  {https://arxiv.org/abs/gr-qc/0107056} {arXiv:gr-qc/0107056} \BibitemShut
  {NoStop}%
\bibitem [{\citenamefont {Pound}\ \emph {et~al.}(2014)\citenamefont {Pound},
  \citenamefont {Merlin},\ and\ \citenamefont {Barack}}]{Pound:2013faa}%
  \BibitemOpen
  \bibfield  {author} {\bibinfo {author} {\bibfnamefont {A.}~\bibnamefont
  {Pound}}, \bibinfo {author} {\bibfnamefont {C.}~\bibnamefont {Merlin}},\ and\
  \bibinfo {author} {\bibfnamefont {L.}~\bibnamefont {Barack}},\ }\href
  {https://doi.org/10.1103/PhysRevD.89.024009} {\bibfield  {journal} {\bibinfo
  {journal} {Phys. Rev. D}\ }\textbf {\bibinfo {volume} {89}},\ \bibinfo
  {pages} {024009} (\bibinfo {year} {2014})},\ \Eprint
  {https://arxiv.org/abs/1310.1513} {arXiv:1310.1513 [gr-qc]} \BibitemShut
  {NoStop}%
\bibitem [{\citenamefont {Haas}(2007)}]{Haas:2007kz}%
  \BibitemOpen
  \bibfield  {author} {\bibinfo {author} {\bibfnamefont {R.}~\bibnamefont
  {Haas}},\ }\href {https://doi.org/10.1103/PhysRevD.75.124011} {\bibfield
  {journal} {\bibinfo  {journal} {Phys. Rev. D}\ }\textbf {\bibinfo {volume}
  {75}},\ \bibinfo {pages} {124011} (\bibinfo {year} {2007})},\ \Eprint
  {https://arxiv.org/abs/0704.0797} {arXiv:0704.0797 [gr-qc]} \BibitemShut
  {NoStop}%
\bibitem [{\citenamefont {Barack}\ and\ \citenamefont
  {Long}(2022)}]{Barack:2022pde}%
  \BibitemOpen
  \bibfield  {author} {\bibinfo {author} {\bibfnamefont {L.}~\bibnamefont
  {Barack}}\ and\ \bibinfo {author} {\bibfnamefont {O.}~\bibnamefont {Long}},\
  }\href {https://doi.org/10.1103/PhysRevD.106.104031} {\bibfield  {journal}
  {\bibinfo  {journal} {Phys. Rev. D}\ }\textbf {\bibinfo {volume} {106}},\
  \bibinfo {pages} {104031} (\bibinfo {year} {2022})},\ \Eprint
  {https://arxiv.org/abs/2209.03740} {arXiv:2209.03740 [gr-qc]} \BibitemShut
  {NoStop}%
\bibitem [{\citenamefont {Poisson}\ \emph {et~al.}(2011)\citenamefont
  {Poisson}, \citenamefont {Pound},\ and\ \citenamefont
  {Vega}}]{Poisson:2011nh}%
  \BibitemOpen
  \bibfield  {author} {\bibinfo {author} {\bibfnamefont {E.}~\bibnamefont
  {Poisson}}, \bibinfo {author} {\bibfnamefont {A.}~\bibnamefont {Pound}},\
  and\ \bibinfo {author} {\bibfnamefont {I.}~\bibnamefont {Vega}},\ }\href
  {https://doi.org/10.12942/lrr-2011-7} {\bibfield  {journal} {\bibinfo
  {journal} {Living Rev. Rel.}\ }\textbf {\bibinfo {volume} {14}},\ \bibinfo
  {pages} {7} (\bibinfo {year} {2011})},\ \Eprint
  {https://arxiv.org/abs/1102.0529} {arXiv:1102.0529 [gr-qc]} \BibitemShut
  {NoStop}%
\bibitem [{\citenamefont {Detweiler}\ and\ \citenamefont
  {Whiting}(2003)}]{Detweiler:2003}%
  \BibitemOpen
  \bibfield  {author} {\bibinfo {author} {\bibfnamefont {S.}~\bibnamefont
  {Detweiler}}\ and\ \bibinfo {author} {\bibfnamefont {B.~F.}\ \bibnamefont
  {Whiting}},\ }\href {https://doi.org/10.1103/PhysRevD.67.024025} {\bibfield
  {journal} {\bibinfo  {journal} {Phys. Rev. D}\ }\textbf {\bibinfo {volume}
  {67}},\ \bibinfo {pages} {024025} (\bibinfo {year} {2003})}\BibitemShut
  {NoStop}%
\bibitem [{\citenamefont {Heffernan}\ \emph {et~al.}(2012)\citenamefont
  {Heffernan}, \citenamefont {Ottewill},\ and\ \citenamefont
  {Wardell}}]{Heffernan:2012su}%
  \BibitemOpen
  \bibfield  {author} {\bibinfo {author} {\bibfnamefont {A.}~\bibnamefont
  {Heffernan}}, \bibinfo {author} {\bibfnamefont {A.}~\bibnamefont
  {Ottewill}},\ and\ \bibinfo {author} {\bibfnamefont {B.}~\bibnamefont
  {Wardell}},\ }\href {https://doi.org/10.1103/PhysRevD.86.104023} {\bibfield
  {journal} {\bibinfo  {journal} {Phys. Rev. D}\ }\textbf {\bibinfo {volume}
  {86}},\ \bibinfo {pages} {104023} (\bibinfo {year} {2012})},\ \Eprint
  {https://arxiv.org/abs/1204.0794} {arXiv:1204.0794 [gr-qc]} \BibitemShut
  {NoStop}%
\bibitem [{\citenamefont {Barack}(2001)}]{Barack:2001bw}%
  \BibitemOpen
  \bibfield  {author} {\bibinfo {author} {\bibfnamefont {L.}~\bibnamefont
  {Barack}},\ }\href {https://doi.org/10.1103/PhysRevD.64.084021} {\bibfield
  {journal} {\bibinfo  {journal} {Phys. Rev.}\ }\textbf {\bibinfo {volume}
  {D64}},\ \bibinfo {pages} {084021} (\bibinfo {year} {2001})},\ \Eprint
  {https://arxiv.org/abs/gr-qc/0105040} {arXiv:gr-qc/0105040 [gr-qc]}
  \BibitemShut {NoStop}%
\bibitem [{\citenamefont {Heffernan}\ \emph {et~al.}(2018)\citenamefont
  {Heffernan}, \citenamefont {Ottewill}, \citenamefont {Warburton},
  \citenamefont {Wardell},\ and\ \citenamefont {Diener}}]{Heffernan:2017cad}%
  \BibitemOpen
  \bibfield  {author} {\bibinfo {author} {\bibfnamefont {A.}~\bibnamefont
  {Heffernan}}, \bibinfo {author} {\bibfnamefont {A.~C.}\ \bibnamefont
  {Ottewill}}, \bibinfo {author} {\bibfnamefont {N.}~\bibnamefont {Warburton}},
  \bibinfo {author} {\bibfnamefont {B.}~\bibnamefont {Wardell}},\ and\ \bibinfo
  {author} {\bibfnamefont {P.}~\bibnamefont {Diener}},\ }\href
  {https://doi.org/10.1088/1361-6382/aad420} {\bibfield  {journal} {\bibinfo
  {journal} {Class. Quant. Grav.}\ }\textbf {\bibinfo {volume} {35}},\ \bibinfo
  {pages} {194001} (\bibinfo {year} {2018})},\ \Eprint
  {https://arxiv.org/abs/1712.01098} {arXiv:1712.01098 [gr-qc]} \BibitemShut
  {NoStop}%
\bibitem [{\citenamefont {Barack}(2019)}]{Barack:2019}%
  \BibitemOpen
  \bibfield  {author} {\bibinfo {author} {\bibfnamefont {L.}~\bibnamefont
  {Barack}},\ }\href@noop {} {\bibinfo {title} {Worldtube method for a
  frequency-domain puncture implementation}},\ \bibinfo {howpublished}
  {Personal Communication} (\bibinfo {year} {2019})\BibitemShut {NoStop}%
\bibitem [{sup()}]{supplemental_material}%
  \BibitemOpen
  \href@noop {} {}\bibinfo {note} {See the supplemental material at for a
  Mathematica notebook containing the high-order puncture.}\BibitemShut {Stop}%
\bibitem [{BHP()}]{BHPToolkit}%
  \BibitemOpen
  \href@noop {} {\bibinfo {title} {{Black Hole Perturbation Toolkit}}},\
  \bibinfo {howpublished}
  {(\href{http://bhptoolkit.org/}{bhptoolkit.org})}\BibitemShut {NoStop}%
\bibitem [{\citenamefont {Hopper}(2018)}]{Hopper:2017iyq}%
  \BibitemOpen
  \bibfield  {author} {\bibinfo {author} {\bibfnamefont {S.}~\bibnamefont
  {Hopper}},\ }\href {https://doi.org/10.1103/PhysRevD.97.064007} {\bibfield
  {journal} {\bibinfo  {journal} {Phys. Rev. D}\ }\textbf {\bibinfo {volume}
  {97}},\ \bibinfo {pages} {064007} (\bibinfo {year} {2018})},\ \Eprint
  {https://arxiv.org/abs/1706.05455} {arXiv:1706.05455 [gr-qc]} \BibitemShut
  {NoStop}%
\bibitem [{\citenamefont {Whittall}\ and\ \citenamefont
  {Barack}(2023)}]{Whittall:2023xjp}%
  \BibitemOpen
  \bibfield  {author} {\bibinfo {author} {\bibfnamefont {C.}~\bibnamefont
  {Whittall}}\ and\ \bibinfo {author} {\bibfnamefont {L.}~\bibnamefont
  {Barack}},\ }\href {https://doi.org/10.1103/PhysRevD.108.064017} {\bibfield
  {journal} {\bibinfo  {journal} {Phys. Rev. D}\ }\textbf {\bibinfo {volume}
  {108}},\ \bibinfo {pages} {064017} (\bibinfo {year} {2023})},\ \Eprint
  {https://arxiv.org/abs/2305.09724} {arXiv:2305.09724 [gr-qc]} \BibitemShut
  {NoStop}%
\bibitem [{\citenamefont {Hopper}\ and\ \citenamefont
  {Evans}(2010)}]{Hopper:2010uv}%
  \BibitemOpen
  \bibfield  {author} {\bibinfo {author} {\bibfnamefont {S.}~\bibnamefont
  {Hopper}}\ and\ \bibinfo {author} {\bibfnamefont {C.~R.}\ \bibnamefont
  {Evans}},\ }\href {https://doi.org/10.1103/PhysRevD.82.084010} {\bibfield
  {journal} {\bibinfo  {journal} {Phys. Rev. D}\ }\textbf {\bibinfo {volume}
  {82}},\ \bibinfo {pages} {084010} (\bibinfo {year} {2010})},\ \Eprint
  {https://arxiv.org/abs/1006.4907} {arXiv:1006.4907 [gr-qc]} \BibitemShut
  {NoStop}%
\bibitem [{\citenamefont {Glampedakis}\ and\ \citenamefont
  {Kennefick}(2002)}]{Glampedakis:2002ya}%
  \BibitemOpen
  \bibfield  {author} {\bibinfo {author} {\bibfnamefont {K.}~\bibnamefont
  {Glampedakis}}\ and\ \bibinfo {author} {\bibfnamefont {D.}~\bibnamefont
  {Kennefick}},\ }\href {https://doi.org/10.1103/PhysRevD.66.044002} {\bibfield
   {journal} {\bibinfo  {journal} {Phys. Rev. D}\ }\textbf {\bibinfo {volume}
  {66}},\ \bibinfo {pages} {044002} (\bibinfo {year} {2002})},\ \Eprint
  {https://arxiv.org/abs/gr-qc/0203086} {arXiv:gr-qc/0203086} \BibitemShut
  {NoStop}%
\bibitem [{\citenamefont {Dolan}\ \emph {et~al.}(2021)\citenamefont {Dolan},
  \citenamefont {Kavanagh},\ and\ \citenamefont {Wardell}}]{Dolan:2021ijg}%
  \BibitemOpen
  \bibfield  {author} {\bibinfo {author} {\bibfnamefont {S.~R.}\ \bibnamefont
  {Dolan}}, \bibinfo {author} {\bibfnamefont {C.}~\bibnamefont {Kavanagh}},\
  and\ \bibinfo {author} {\bibfnamefont {B.}~\bibnamefont {Wardell}},\
  }\href@noop {} {\  (\bibinfo {year} {2021})},\ \Eprint
  {https://arxiv.org/abs/2108.06344} {arXiv:2108.06344 [gr-qc]} \BibitemShut
  {NoStop}%
\bibitem [{\citenamefont {Spiers}\ \emph {et~al.}(2023)\citenamefont {Spiers},
  \citenamefont {Pound},\ and\ \citenamefont {Moxon}}]{Spiers:2023cip}%
  \BibitemOpen
  \bibfield  {author} {\bibinfo {author} {\bibfnamefont {A.}~\bibnamefont
  {Spiers}}, \bibinfo {author} {\bibfnamefont {A.}~\bibnamefont {Pound}},\ and\
  \bibinfo {author} {\bibfnamefont {J.}~\bibnamefont {Moxon}},\ }\href@noop {}
  {\  (\bibinfo {year} {2023})},\ \Eprint {https://arxiv.org/abs/2305.19332}
  {arXiv:2305.19332 [gr-qc]} \BibitemShut {NoStop}%
\end{thebibliography}%
\end{document}